%% file: main.tex
\RequirePackage{rotating}
\documentclass[acmsmall]{acmart}

\usepackage{rotating}  %

\AtBeginDocument{%
  \providecommand\BibTeX{{%
    \normalfont B\kern-0.5em{\scshape i\kern-0.25em b}\kern-0.8em\TeX}}}

\setcopyright{none}
\acmDOI{10.1145/3478026}

\acmJournal{TWEB}

\begin{document}

\title{A Large-scale Empirical Analysis of Browser Fingerprints Properties for Web Authentication}

\author{Nampoina Andriamilanto}
\orcid{0000-0002-0224-5664}
\email{nampoina.andriamilanto@b-com.com}
\affiliation{%
  \institution{Institute of Research and Technology b$<>$com}
  \streetaddress{1219 avenue Champs Blancs}
  \city{Cesson-Sévigné}
  \country{France}
  \postcode{35510}
}
\affiliation{%
  \institution{Univ Rennes, CNRS, IRISA}
  \streetaddress{263 avenue du général Leclerc}
  \city{Rennes}
  \country{France}
  \postcode{35000}
}

\author{Tristan Allard}
\orcid{0000-0002-2777-0027}
\email{tristan.allard@irisa.fr}
\affiliation{%
  \institution{Univ Rennes, CNRS, IRISA}
  \streetaddress{263 avenue du général Leclerc}
  \city{Rennes}
  \country{France}
  \postcode{35000}
}

\author{Ga\"etan Le Guelvouit}
\email{gaetan.leguelvouit@b-com.com}
\affiliation{%
  \institution{Institute of Research and Technology b$<>$com}
  \streetaddress{1219 avenue Champs Blancs}
  \city{Cesson-Sévigné}
  \country{France}
  \postcode{35510}
}

\author{Alexandre Garel}
\email{alexandre.garel@b-com.com}
\authornote{The author participated to the fingerprint collection and analysis when working at the institution, but is not anymore affiliated to it.}
\affiliation{%
  \institution{Institute of Research and Technology b$<>$com}
  \streetaddress{1219 avenue Champs Blancs}
  \city{Cesson-Sévigné}
  \country{France}
  \postcode{35510}
}

\input{1-abstract}

\ccsdesc[500]{Security and privacy~Multi-factor authentication}
\ccsdesc[300]{Information systems~Browsers}

\keywords{browser fingerprinting, web authentication, multi-factor authentication}

\maketitle

\input{2-introduction}
\input{3-dataset}
\input{4-authentication-factor-properties}
\input{5-results}
\input{6-statistical-analysis}
\input{7-discussion}
\input{8-related-work}
\input{9-conclusion}

\input{10-acknowledgments}

\bibliographystyle{ACM-Reference-Format}
\bibliography{bibliography}

\input{11-appendix}

\end{document}

%% file: 1-abstract.tex
\begin{abstract}
  Modern browsers give access to several attributes that can be collected to form a browser fingerprint.
  Although browser fingerprints have primarily been studied as a web tracking tool, they can contribute to improve the current state of web security by augmenting web authentication mechanisms.
  In this paper, we investigate the adequacy of browser fingerprints for web authentication.
  We make the link between the digital fingerprints that distinguish browsers, and the biological fingerprints that distinguish Humans, to evaluate browser fingerprints according to properties inspired by biometric authentication factors.
  These properties include their distinctiveness, their stability through time, their collection time, their size, and the accuracy of a simple verification mechanism.
  We assess these properties on a large-scale dataset of $4,145,408$~fingerprints composed of $216$~attributes and collected from $1,989,365$~browsers.
  We show that, by time-partitioning our dataset, more than $81.3$\% of our fingerprints are shared by a single browser.
  Although browser fingerprints are known to evolve, an average of $91$\% of the attributes of our fingerprints stay identical between two observations, even when separated by nearly $6$~months.
  About their performance, we show that our fingerprints weigh a dozen of kilobytes and take a few seconds to collect.
  Finally, by processing a simple verification mechanism, we show that it achieves an equal error rate of $0.61$\%.
  We enrich our results with the analysis of the correlation between the attributes and their contribution to the evaluated properties.
  We conclude that our browser fingerprints carry the promise to strengthen web authentication mechanisms.
\end{abstract}

\makeatletter{
  \renewcommand*{\@makefnmark}{}
  \footnotetext{
    This paper is a major extension (more than $50$\% of new material) of work originally presented in~\cite{AAL21}.
  }
  \makeatother
}

%% file: 2-introduction.tex
\section{Introduction}
  Web authentication widely relies on the use of identifier-password pairs defined by the end user.
  The password authentication factor is easy to use and to deploy, but has been shown to suffer from severe security flaws when used without any additional factor.
  Real-life users indeed use common passwords~\cite{TroyHuntPasswordReuse}, which paves the way to brute-force or guessing attacks~\cite{BON12}.
  Moreover, they tend to use similar passwords across different websites~\cite{DBCBW14}, which increases the impact of successful attacks.
  Phishing attacks are also a major threat to the use of passwords.
  Over the course of a year, Thomas et al.~\cite{TLZBRIMCEMMPB17} achieved to retrieve $12.4$~million credentials stolen by phishing kits.
  These flaws bring the need for supplementary security layers, primarily through multi-factor authentication~\cite{BHVS15}, such that each additional factor provides an \emph{additional security barrier}.
  However, this usually comes at the cost of \emph{usability} (i.e., users have to remember, possess, or do something) and \emph{deployability} (i.e., implementers have to deploy dedicated hardware or software, teach users how to use them, and maintain the deployed solution).

\begin{figure}
    \centering
    \includegraphics[width=0.7\columnwidth]{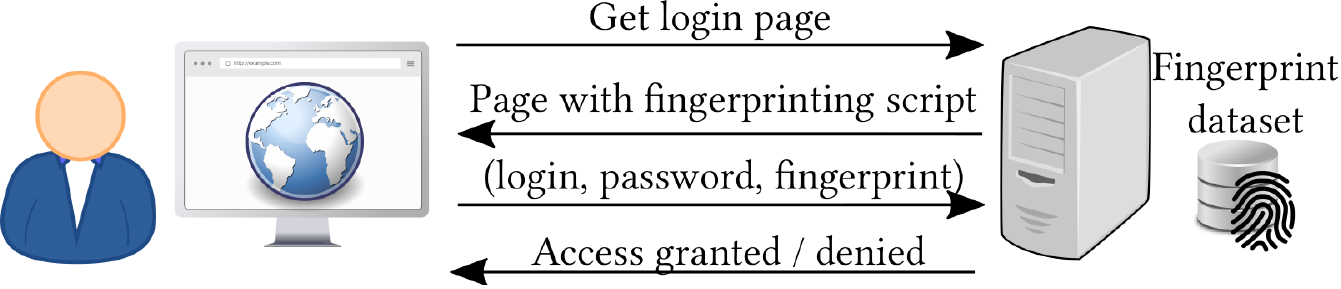}
    \caption{
      A simplified web authentication mechanism that relies on browser fingerprinting.
    }
    \label{fig:auth-mechanism-scheme}
    \Description[
      A simplified web authentication mechanism that relies on browser fingerprinting.
    ]{
      A simplified web authentication mechanism that relies on browser fingerprinting.
    }
  \end{figure}

  In the meantime, \emph{browser fingerprinting}~\cite{LBBA20} gains more and more attention.
  The seminal Panopticlick study~\cite{ECK10} is the first work to highlight the possibility to build a \emph{browser fingerprint} by collecting attributes from a browser (e.g., the \texttt{userAgent} property of the \texttt{navigator} Java\-Script object).
  In addition to being widely used for web tracking purposes~\cite{AEEJND14, EN16, IES21} (raising legal, ethical, and technical issues), browser fingerprinting is already used as an additional web authentication factor \emph{in real-life}.
  The browser fingerprints constitute a supplementary factor that is verified at login with the other factors, as depicted in Figure~\ref{fig:auth-mechanism-scheme} (see Section~\ref{sec:browser-fingerprinting-based-authentication-mechanism} for an example of an authentication mechanism that relies on browser fingerprints).
  Browser fingerprints are indeed a good \emph{candidate} as an additional web authentication factor thanks to their distinctive power, their frictionless deployment (e.g., no additional software or hardware to install), and their usability (no secret to remember, no additional object to possess, and no supplementary action to carry out).
  As a result, companies like MicroFocus~\cite{MicroFocusDFP} or SecureAuth~\cite{SecureAuthDFP} include this technique in their authentication mechanisms.

  However, to the best of our knowledge, no large-scale study rigorously evaluates the adequacy of browser fingerprints as an additional web authentication factor.
  On the one hand, most works about the use of browser fingerprints for authentication concentrate on the design of the authentication mechanism~\cite{UMFHSW13, PJ15, GSPJ16, SPJ17, LABN19, REKP19, LC20}.
  On the other hand, the large-scale empirical studies on browser fingerprints focus on their effectiveness as a web tracking tool~\cite{ECK10, LRB16, GLB18, PRGB20}.
  Such a mismatch between the understanding of browser fingerprints for authentication -- currently poor -- and their ongoing adoption in real-life is a serious harm to the security of web users.
  The lack of documentation from the existing authentication tools (e.g., about the used attributes, about the distinctiveness and the stability of the resulting fingerprints) only adds up to the current state of ignorance, all this whereas security-by-obscurity directly contradicts the most fundamental security principles.
  Moreover, the distinctiveness of browser fingerprints that can be achieved when considering a wide-surface of fingerprinting attributes on a large population is, to the best of our knowledge, unknown.
  On the one hand, the studies that analyze browser fingerprints in a large-scale (more than $100,000$ fingerprints) consider fewer than $35$ attributes~\cite{ECK10, LRB16, VLRR18, GLB18, LC20}.
  This underestimates the distinctiveness of the fingerprints (e.g., \cite{GLB18} and \cite{LC20} report a unique fingerprint rate of about $35$\%), as it increases the chances for browsers to share the same fingerprint.
  All this whereas more than a hundred attributes are accessible.
  On the other hand, the studies that consider more than fifty attributes either work on less than two thousands users~\cite{KZW15, PRGB20}, or do not analyze the resulting fingerprints at all~\cite{ALM18}.
  The current knowledge about the hundreds of accessible attributes (e.g., their stability, their collection time, their correlation) is, to the best of our knowledge, also incomplete.
  Indeed, previous studies either consider few attributes~\cite{ECK10, MS12, FE15, BMPT16, LRB16, SVS17, GLB18, QF19} or focus on a single aspect of them (e.g., their evolution~\cite{VLRR18, LC20}).

  \textbf{Our contributions.}
  We conduct the first \emph{large-scale data-centric empirical study} of the \emph{fundamental properties of browser fingerprints} when used as an additional web authentication factor.
  We base our findings on an in-depth analysis of a real-life fingerprint dataset collected over a period of $6$~months, that contains $4,145,408$~fingerprints composed of $216$~attributes.
  In particular, our dataset includes nine \emph{dynamic attributes} of three types, which values depend on instructions provided by the fingerprinter: five HTML5 canvases~\cite{BMPT16}, three audio fingerprinting methods~\cite{QF19}, and a WebGL canvas~\cite{MS12}.
  The dynamic attributes are used within state-of-the-art web authentication mechanisms to mitigate replay attacks~\cite{REKP19, LABN19}.
  Each dynamic attribute has been studied singularly, but their fundamental properties have not yet been studied simultaneously on the same browser population.
  To the best of our knowledge, no related work considers a dataset of this scale, in terms of both fingerprints and attributes, together with various dynamic attributes.
  We formalize, and assess on our dataset, the properties necessary for paving the way to elaborate browser fingerprinting authentication mechanisms.
  We make the link between the digital fingerprints that distinguish browsers, and the biological fingerprints that distinguish Humans, to evaluate browser fingerprints according to properties inspired by biometric authentication factors~\cite{MMJP09, ZD04, GLMPSS05}.
  The properties aim at characterizing the adequacy and the practicability of browser fingerprints, independently of their use within future authentication mechanisms.
  In particular, we measure the size of the browser anonymity sets through time, the proportion of identical attributes between two observations of the fingerprint of a browser, the collection time of the fingerprints, their size, the loss of efficacy between device types, and the accuracy of a simple illustrative verification mechanism.
  To comprehend the results obtained on the complete fingerprints, we include an in-depth study of the contribution of the attributes to the properties of the fingerprints.
  Moreover, we discuss the correlation between the attributes, make a focus on the contribution of the dynamic attributes, and provide the exhaustive list of the attributes together with their properties.
  To the best of our knowledge, no previous work analyzed browser fingerprinting attributes at this scale, in terms of the number of attributes, the number of fingerprints, and the variety of properties (e.g., stability, collection time).

  In a nutshell, we make the following contributions:
  \begin{enumerate}
    \item We formalize the fundamental properties that browser fingerprints should provide to be usable and practical as a web authentication factor.
    \item About their adequacy, we show that
      (1) considering a wide surface of $216$~fingerprinting attributes on our large population provides a proportion of unique fingerprints -- also called unicity rate -- of $81.8$\% on our complete dataset,
      (2) by time-partitioning our dataset, the unicity rates are stable on the long term at around $81.3$\%, and $94.7$\% of our fingerprints are shared by $8$~browsers or fewer,
      (3) on average, a fingerprint has more than $91$\% of identical attributes between two observations, even when separated by nearly $6$~months,
      (4) our mobile browsers lack distinctiveness, as they show a unicity rate of $42$\%.
    \item About their practicability, we show that
      (1) the generated fingerprints weigh a dozen of kilobytes,
      (2) they are collected within seconds,
      (3) the accuracy of a simple illustrative verification mechanism is close to perfect, as it achieves an equal error rate of $0.61$\%.
    \item We enrich our results with
      (1) a precise analysis of the contribution of each attribute
        (a) to the distinctiveness, and show that $10$\% of the attributes provide a normalized entropy higher than $0.25$,
        (b) to the stability, and show that $85$\% of the attributes stay identical for $99$\% of the consecutive fingerprints coming from the same browser,
        (c) to the collection time, and show that only $33$ attributes take more than $5$ms to collect,
        (d) to the fingerprint size, and show that only $20$ attributes weigh more than $100$~bytes,
      (2) a discussion about the correlation of the attributes, and show that only $49$ attributes can completely be inferred when knowing another attribute,
      (3) a focus on the properties of the nine dynamic attributes.
    \item We provide an in-depth description of our methodology and our dataset, with the goal of making our results reproducible.
      In particular, we include the exhaustive list of the collected attributes together with their properties, and a detailed description of the preprocessing of the fingerprints.
  \end{enumerate}

  This paper is a major extension of work originally presented in~\cite{AAL21}, and brings more than $50$\% of new material.
  Specifically, we clarify the obtained results by discussing them further, we evaluate the accuracy of a simple illustrative verification mechanism, we highlight the contribution of the attributes to the properties, and we provide a comprehensive list of the attributes, together with their properties and their concrete implementation.
  In a summary, we make the following additions:
  \begin{enumerate}
    \item Section~\ref{sec:dataset} gains a description of the studied browser population.
      This includes the share of the families of browser and operating system, and insights about the bias towards French browsers.
      We also add a description of the data preprocessing step, and a comparison between our dataset and the dataset of previous studies.
    \item Section~\ref{sec:authentication-factor-properties} gains the accuracy of a simple illustrative verification mechanism as an additional performance property.
    \item Section~\ref{sec:results} gets the results further discussed, and gains the results of the accuracy of the simple illustrative verification mechanism.
    \item Section~\ref{sec:attribute-wise-analysis} is entirely new.
      It discusses the contribution of the attributes to the fingerprint properties, the correlation between the attributes, and the properties of the dynamic attributes.
    \item Section~\ref{sec:discussion} is entirely new.
      It describes how browser fingerprints can be integrated in a web authentication mechanism and discusses the attacks that are possible on such mechanism.
    \item Section~\ref{sec:related-works} is entirely new.
      It provides related works about the use of browser fingerprinting for authentication.
    \item The appendices are entirely new.
      Appendix~\ref{app:attributes} describes the concrete implementation of the studied attributes for reproducibility.
      Appendix~\ref{app:keywords} provides the keywords used to classify the fingerprints.
      Appendix~\ref{app:anomalous-collection-times} discusses a side effect encountered by our fingerprinting script that results in anomalous high collection time of some fingerprints or attributes.
      Appendix~\ref{app:advanced-verification-mechanism} discusses a more complex verification mechanism that relies on distance functions on the attributes to compare fingerprints.
      Appendix~\ref{app:attribute-list-and-property} provides the complete list of the attributes with their properties (e.g., number of distinct values, stability).
  \end{enumerate}

  The rest of the paper is organized as follows.
  Section~\ref{sec:dataset} describes the dataset analyzed in this study.
  Section~\ref{sec:authentication-factor-properties} presents and formalizes the properties evaluated in the analysis.
  Section~\ref{sec:results} presents the experimental results.
  Section~\ref{sec:attribute-wise-analysis} breaks down the analysis to the attributes to comprehend the results on the complete fingerprints.
  Section~\ref{sec:discussion} describes how browser fingerprinting can contribute to an authentication mechanism and discusses the attacks that are possible on such mechanism.
  Section~\ref{sec:related-works} positions this study with the related works.
  Finally, Section~\ref{sec:conclusion} concludes.

%% file: 3-dataset.tex
\section{Dataset}
\label{sec:dataset}
  In this section, we describe the browser fingerprint dataset that is analyzed in this study.
  First, we present the conditions of the collection and describe the precautions taken to protect the privacy of the experimenters.
  Then, we detail the preprocessing steps to cleanse the raw dataset.
  Finally, we describe the working dataset and compare it with the large-scale datasets of previous studies.

  \subsection{Fingerprints collection}
    To study the properties of browser fingerprints on a real-world browser population, we launched an experiment in collaboration with the authors of the Hiding in the Crowd study~\cite{GLB18}, together with an industrial partner that controls one of the top $15$ French websites according to the site ranking service Alexa~\cite{TopFrenchAlexa}.
    The authors of the Hiding in the Crowd study only consider the $17$ attributes of their previous work~\cite{LRB16} and focus on the issue of web tracking.
    On the contrary, we consider in this work more than one order of magnitude more attributes -- $216$ attributes -- and focus on the use of browser fingerprinting as an additional web authentication factor.
    Our dataset contains more fingerprints than~\cite{GLB18} because two browsers that show different fingerprints for a given set of attributes can come to the same fingerprint if a subset of these attributes is considered.
    As a result, the authors of~\cite{GLB18} remove more duplicated fingerprints due to the higher chances for a browser to present the same fingerprint for $17$ attributes than for $216$ attributes.

    \subsubsection{Experiment}
      The experiment consisted in integrating a fingerprinting script on two general audience web pages that are controlled by our industrial partner, which subjects are political news and weather forecast.
      The script was active between December $7$, $2016$, and June $7$, $2017$.
      It fingerprinted the visitors who consented to the use of cookies in compliance with the European directives $2002$/$58$/CE~\cite{EropeanDirective200258EC} and $2009$/$136$/CE~\cite{EropeanDirective2009136EC}.
      To differentiate two browsers in future analysis, we assigned them a unique identifier (UID) as a $6$-months cookie, which was sent alongside fingerprints.
      Similarly to previous studies~\cite{ECK10, LRB16}, we coped with the issue of cookie deletion by storing a one-way hash of the IP address computed by a secure cryptographic hash function.
      We refer the interested reader to Section~\ref{sec:privacy-concerns} for more details on the measures taken to protect the privacy of the experimenters.

    \subsubsection{Browser fingerprinting attributes}
      The fingerprinting script used in the experiment includes $216$ attributes divided in $200$ JavaScript properties, together with their collection time, and $16$ HTTP header fields.
      In particular, they include three types of dynamic attributes that comprise five HTML5 canvases~\cite{BMPT16}, three audio fingerprinting methods~\cite{QF19}, and a WebGL canvas~\cite{MS12}.

      We sought to evaluate the properties of browser fingerprints when considering as many attributes as possible, to estimate more precisely what can really be achieved.
      We compiled the attributes from previous studies and open source projects.
      If the value of an attribute is not accessible, a flag explaining the reason is stored instead, as it is still exploitable information.
      Indeed, two browsers can be distinguished if they behave differently on the inaccessibility of an attribute (e.g., returning the value \texttt{undefined} is different from throwing an exception).
      We also configure a timeout after which the fingerprint is sent without waiting for every attribute to be collected.
      The attributes that were not collected are set to a specific flag.
      The complete list of attributes and their properties is available in Appendix~\ref{app:attribute-list-and-property}.

      Client-side attributes only consist of JavaScript properties.
      No plugins (e.g., Flash, Silverlight) are used due to their removal and replacement by HTML5 functionalities~\cite{EndOfFlash}.
      Moreover, the fingerprinting script collects the HTTP headers from requests sent by JavaScript, hence the dataset contains no fingerprint of browsers having JavaScript disabled.

  \begin{sidewaystable}
    \centering
    \caption{
      Comparison of the share of browser and operating system families between the studies Panopticlick (PTC), AmIUnique (AIU), Hiding in the Crowd (HitC), the average share of each family computed by StatCounter for the worldwide and French browser populations between January 2017 and June 2017, and this study.
      The symbol - denotes missing information.
      The families are ordered from the most common to the least common in this study.
      The shares for the AIU and the HitC studies both come from the HitC paper in which they are provided for the desktop and the mobile browsers.
      We calculated and provide here the shares for the complete browser population for comparability.
      We stress that the classification methodology can vary between the populations that are compared, and refer to Appendix~\ref{app:keywords} for a description of our methodology.
    }
    \label{tab:experiments-browser-os-family}
    \begin{tabular}{lcccccc}
      \toprule
                          & PTC~\cite{ECK10}   & AIU~\cite{LRB16}
                          & HitC~\cite{GLB18}
                          & Worldwide~\cite{Worldwide2017BrowserShareStatCounter, Worldwide2017OperatingSystemShareStatCounter}
                          & France~\cite{France2017BrowserShareStatCounter, France2017OperatingSystemShareStatCounter}
                          & \textbf{This study} \\
      \midrule
        Starting date     & 01/2010 & 11/2014 & 12/2016 & 01/2017 & 01/2017
                          & \textbf{12/2016}  \\
        Duration          & 3 weeks & 3-4 months & 6 months & 6 months
                          & 6 months & \textbf{6 months} \\
      \midrule
        Firefox           & 0.568 & 0.437 & -     & 0.064 & 0.193
                          & \textbf{0.268} \\
        Chrome            & 0.142 & 0.398 & -     & 0.531 & 0.478
                          & \textbf{0.265} \\
        Internet Explorer & 0.126 & 0.043 & -     & 0.046 & 0.065
                          & \textbf{0.260} \\
        Edge              & -     & -     & -     & 0.017 & 0.040
                          & \textbf{0.078} \\
        Safari            & 0.077 & 0.079 & -     & 0.144 & 0.193
                          & \textbf{0.064} \\
        Samsung Internet  & -     & -     & -     & 0.033 & 0.029
                          & \textbf{0.048} \\
        Others            & 0.088 & 0.042 & -     & 0.165 & 0.002
                          & \textbf{0.017} \\
      \hline
        Windows-based     & -     & 0.568 & 0.831 & 0.375 & 0.523
                          & \textbf{0.821} \\
        \hspace{4mm}Windows 10        & -     & -     & -     & -     & -
                          & \textbf{0.338} \\
        \hspace{4mm}Windows 7         & -     & -     & -     & -     & -
                          & \textbf{0.312} \\
        \hspace{4mm}Other Windows     & -     & -     & -     & -     & -
                          & \textbf{0.171} \\
        Android           & -     & 0.061 & 0.087 & 0.385 & 0.189
                          & \textbf{0.091} \\
        Mac OS X          & -     & 0.133 & 0.048 & 0.051 & 0.120
                          & \textbf{0.051} \\
        iOS               & -     & 0.047 & 0.023 & 0.132 & 0.138
                          & \textbf{0.026} \\
        Linux-based       & -     & 0.150 & 0.008 & 0.008 & 0.015
                          & \textbf{0.008} \\
        Others            & -     & 0.041 & 0.003 & 0.049 & 0.015
                          & \textbf{0.003} \\
      \bottomrule
    \end{tabular}
  \end{sidewaystable}

  \subsection{Browser population bias}
  \label{sec:browser-population-bias}
    The previously presented datasets were collected through dedicated websites and are biased towards privacy-aware and technically-skilled persons~\cite{ECK10, LRB16, PRGB20}.
    Our dataset is more general audience oriented and is not biased towards this type of population.
    Nevertheless, the website audience is mainly French-speaking users, resulting in biases that we discuss below.
    We emphasize that the obtained browser fingerprints -- and accordingly, their properties -- depend on the browser population.
    A verifier that seeks to use browser fingerprints as an authentication factor should analyze the fingerprints of a browser population that is as close as possible to the target population.
    Some browser populations can show inadequate properties, like the standardized browsers of a university which lack distinctiveness~\cite{AND20}.
    In the following analysis (see Section~\ref{sec:results}), we evaluate our properties on the complete browser population and on the subpopulations of the browsers running on desktop and mobile devices.
    A finer-grained analysis -- down to a given browser family in a given version -- can be done by sampling the population on the wanted subset.
    Studying the fingerprints at this level is out of the scope of this paper.
    We let such analysis as future works.

    We match the \texttt{userAgent} JavaScript property with manually selected keywords (see Appendix~\ref{app:keywords}) to infer the operating system and browser family of our browser population.
    Table~\ref{tab:experiments-browser-os-family} compares the share of each operating system and browser family between the previous works that include such statistics~\cite{ECK10, LRB16, GLB18}, the average share of each family between January 2017 and June 2017 measured by StatCounter for the worldwide~\cite{Worldwide2017BrowserShareStatCounter, Worldwide2017OperatingSystemShareStatCounter} and French browser populations~\cite{France2017BrowserShareStatCounter, France2017OperatingSystemShareStatCounter}, and our browser population.
    We stress that the browser population that visits a website varies through time as the browser market evolves~\cite{DAT19} and depends on the website (e.g., only a subset of the worldwide browser population is expected to visit a local news website).
    Among our browser population, $82.1$\% run on a Windows operating system, resulting in a high proportion of Internet Explorer ($26$\%) and Edge ($7.8$\%) browsers.
    This can be explained by the population visiting the website being less technically savvy, hence using more common web environments (e.g., a Firefox on a Windows operating system) than technical environments (e.g., Linux-based operating systems).

    Among the desktop browsers, the most common browsers are Firefox ($31.4$\%), Internet Explorer ($28.3$\%), and Chrome ($26.2$\%).
    Our dataset contains more Firefox browsers and less Chrome browsers than the French population observed by StatCounter~\cite{France2017BrowserShareStatCounter}.
    According to previous studies, the fingerprints of Firefox and Chrome browsers are among the most distinctive~\cite{AL17, LC20}.
    Firefox and Chrome both include automated updates, which results in higher instability of the fingerprints collected from these browsers~\cite{LC20}.
    Internet Explorer and Edge supported badly WebRTC at the time of the experiment~\cite{AL17, CanIUseCreateDataChannel}.
    On these browsers, no information could have been collected using WebRTC, which reduces the distinctiveness of this attribute on these browsers.
    However, Edge is considered as one of the browsers that provide highly distinctive fingerprints~\cite{AL17, LC20}.

    The vast majority of our mobile browsers runs on an Android platform ($84.4$\%), followed by Windows Phone ($8.8$\%), and iOS ($5.2$\%).
    The browsers running on Android devices tend to be more distinguishable than the ones running on iOS~\cite{LRB16}, due to the plurality of device vendors and models that embark the Android operating system.
    The browser fingerprints of iOS devices show more updates than these of other mobile operating systems~\cite{LC20}, which can result in increased instability of the fingerprints of iOS devices.
    The most common mobile browser is Samsung Internet ($45.1$\%), followed by Chrome ($38.5$\%), Internet Explorer mobile ($8.4$\%), and Safari ($6.7$\%).
    Smartphones embark a default browser, which can explain this distribution that includes the four default browsers\footnote{
      Samsung Internet is the default browser on some Android devices manufactured by Samsung.
    } of the major operating systems.
    Using another browser than the default browser results in an increased distinctiveness as the combination of operating system and browser becomes less common~\cite{LC20}.
    Although Chrome is one of the default browser of Android devices, it provides highly distinctive information in its UserAgent~\cite{LRB16}.
    As a result, Firefox and Chrome are known as the two most distinctive mobile browsers~\cite{LC20}.

    The contextual attributes related to the time zone or to the configured language are less distinctive in our dataset than in previous studies (see Section~\ref{sec:attributes-distinctiveness}).
    For example, the normalized entropy of the \texttt{Timezone} JavaScript property is of only $0.008$, against $0.161$ for the Panopticlick study~\cite{ECK10}, and $0.198$ for the AmIUnique study~\cite{LRB16}.
    These attributes also tend towards the typical French values: $98.48$\% of the browsers have a \texttt{Timezone} value of $-1$, $98.59$\% of the browsers have the daylight saving time enabled, and \texttt{fr} is present in $98.15$\% of the values of the \texttt{Accept-Language} HTTP header.

    The browsers of our dataset mostly belong to general audience French users.
    Counter-intuitively, considering an international population may not reduce the distinctiveness.
    Indeed, we can expect foreign users to have a combination of contextual attributes (e.g., the time zone, the configured languages) different from the French users, making them distinguishable even if the remaining attributes have identical values.
    Forging new browser fingerprints from a dataset to realistically simulate a different browser population is a difficult task.
    First, it requires to know the distribution of the values that the target population would show for each attribute.
    The value of some attributes can be inferred easily (e.g., the typical time zone of a given location), but the typical value that other attributes would present is harder to infer (e.g., the textual value of the UserAgent of a new browser family).
    Second, although we could sample the value of each individual attribute from their hypothetical distribution, the value of the attributes should be picked with respect to the correlations that occur between them.
    When forging a browser fingerprint, the lack of consideration of the correlations between the attributes leads to unrealistic fingerprints~\cite{ECK10, BRO12, TJM15, AM20}.

  \subsection{Privacy concerns}
  \label{sec:privacy-concerns}
    The browser fingerprints are sensitive due to their identification capacity.
    We complied with the European directives $2002$/$58$/CE~\cite{EropeanDirective200258EC} and $2009$/$136$/CE~\cite{EropeanDirective2009136EC} in effect at the time of the experiment, and took additional measures to protect the participating users.
    First, the script was set on two web pages of a single domain in a first-party manner, hence providing no extra information about the browsing activity of the users.
    The content of the web pages are generic, hence they do not leak any information about the interests of the users.
    Second, we restricted the collection to the users having consented to cookies, as required by the European directives $2002$/$58$/CE and $2009$/$136$/CE.
    Third, a unique identifier (UID) was set as a cookie with a $6$-months lifetime, corresponding to the duration of the experiment.
    Fourth, we deleted the fingerprints for which the \texttt{cookieEnabled} property was not set to \texttt{true}.
    Finally, we hashed the IP addresses by passing them through the HMAC-SHA256 algorithm using a key that we threw afterward.
    It was done using the \texttt{secret} and the \texttt{hmac} libraries of Python$3.6$.
    These one-way hashed IP addresses are only used for the UIDs resynchronization (see Section~\ref{sec:uid-resynchronization}) and are not used as an attribute in the working dataset.

  \subsection{Data preprocessing}
  \label{sec:data-preprocessing}
    Given the experimental aspect of browser fingerprints, and the scale of our collection, the raw dataset contains erroneous or irrelevant samples.
    That is why we perform several preprocessing steps before any analysis.
    The dataset is composed of entries in the form of $(f, b, t)$ tuples so that the fingerprint~$f$ was presented by the browser~$b$ at the given time~$t$.
    We talk here about entries (i.e., $(f, b, t)$ tuples) and not fingerprints (i.e., only $f$) to avoid confusion.
    The preprocessing is divided in four steps: the cleaning, the UIDs resynchronization, the deduplication, and the derivation of the extracted attributes.
    Initially, we have $8,205,416$ entries in the raw dataset.

    \subsubsection{Dataset cleaning}
      The dataset cleaning step filters out $70,460$ irrelevant entries, following the method described below.
      The fingerprinting script prepares, sends, and stores the entries in string format consisting of the attribute values separated by semicolons.
      We remove $769$~entries that have a wrong number of fields, mainly due to truncated or unrelated data (e.g., the body of a post request).
      We filter out $53,251$~entries that belong to robots, by checking whether blacklisted keywords are present in the \texttt{User-Agent} HTTP header (see Appendix~\ref{app:keywords} for the list of keywords).
      We reduce the entries that have multiple exact copies (down to the same moment of collection) to a single instance.
      Finally, we remove $18,591$~entries that have the cookies disabled, and $2,412$~entries that have a time of collection that falls outside the time window of the experiment.

    \subsubsection{Unique IDs resynchronization}
    \label{sec:uid-resynchronization}
      The resynchronization step replaces $181,676$ UIDs with a total of $116,708$ other UIDs, following the method described here.
      Each entry includes a UID which was stored and retrieved using the cookie mechanism of the browser.
      The cookies are considered an unreliable browser identification solution~\cite{LC20}, hence we undergo a cookie resynchronization step similarly to the Panopticlick study~\cite{ECK10}.
      We consider the entries that have the same (fingerprint, IP address hash) pair to belong to the same browser, even if they show different UIDs.
      We group the entries by their pair of (fingerprint, IP address hash) and assign the entries of each group a single UID taken from those observed for this group.
      The resynchronization is processed on all the groups at the exception of the groups that show interleaved UIDs for which we do not modify the UID of their entries.
      These groups contain entries that have the same (fingerprint, IP address hash) pair but which UID values are interleaved (e.g., $b_1$, $b_2$, then $b_1$ again).
      The interleaved UIDs are a good indicator that several genuine identical browsers operate on the same private network and share the same public IP address~\cite{ECK10}.

    \subsubsection{Deduplication}
    \label{sec:dataset-deduplication}
      The deduplication step constitutes the biggest cut in our dataset, and filters out $2,420,217$ entries.
      To avoid storing duplicates of the same fingerprint observed several times for a browser, the usual method is to ignore a fingerprint if it was already seen for a browser during the collection~\cite{ECK10, LRB16}.
      Our script collects the fingerprint on each visit, no matter if it was already seen for this browser or not.
      To stay consistent with common methodologies, we deduplicate the fingerprints offline.
      For each browser, we hold the first entry that contains a given fingerprint, and ignore the following entries if they also contain this fingerprint.
      This method takes the interleaved fingerprints into account, which are the fingerprints so that we observe $f_1$, $f_2$, then $f_1$ again.
      For example, if a browser~$b$ has the entries $\{(f_1, b, t_1), (f_2, b, t_2), (f_2, b, t_3), (f_1, b, t_4)\}$, we only hold the entries $\{(f_1, b, t_1), (f_2, b, t_2), (f_1, b, t_4)\}$ after the deduplication step.

      We hold the interleaved fingerprints to realistically simulate the state of the fingerprint of each browser through time.
      We find that $10.59$\% of our browsers showed at least one case of interleaved fingerprints.
      The interleaved fingerprints can come from attributes that switch between two values.
      An example is the screen size that changes when an external screen is plugged or unplugged.
      Previous studies discarded the fingerprints that were already encountered for a given browser~\cite{ECK10, LRB16, GLB18}, hiding the interleaved fingerprints.

    \subsubsection{Extracted attributes}
    \label{sec:extracted-attributes}
      We derive $46$ extracted attributes of two types from $9$ original attributes.
      First, we have the extracted attributes that are parts of an original attribute, like an original attribute that is composed of $28$ triplets of RGB (Red Green Blue) color values that we split in $28$ single attributes.
      Then, we have the extracted attributes that are derived from an original attribute, like the number of plugins derived from the list of plugins.
      The extracted attributes do not increase the distinctiveness as they come from an original attribute, and they are at most as distinctive as their original attribute.
      However, the extracted attributes can offer a higher stability than their original attribute, as the latter is impacted by any little change among the extracted attributes.
      For example, if exactly one of the $28$ RGB values changes between two fingerprint observations, the original attribute is counted as having changed, but only one of the extracted attributes will be.

  \begin{sidewaystable}
    \begin{minipage}{\textwidth}
      \centering
      \caption{
        Comparison between the datasets of the studies Panopticlick (PTC), AmIUnique (AIU), Hiding in the Crowd (HitC), Long-Term Observation (LTO), Who Touched My Browser Fingerprint (WTMBF), and this study.
        As the LTO study tests various attribute sets, we present the ranges of the results obtained by their attribute selection method.
        The symbols - denotes missing information and * denotes deduced information from the paper.
        The attributes comprise the derived attributes.
        The fingerprints are counted after data preprocessing.
      }
      \label{tab:dataset-comparison}
      \begin{tabular}{lcccccc}
        \toprule
                                 & PTC~\cite{ECK10}    & AIU~\cite{LRB16}
                                 & HitC~\cite{GLB18}   & LTO~\cite{PRGB20}
                                 & WTMBF~\cite{LC20}   & \textbf{This study} \\
        \midrule
          Collection period      & 01-02/2010          & 11/2014-02/2015
                                 & 12/2016-06/2017     & 02/2016-02/2019
                                 & 12/2017-07/2018
                                 & \textbf{12/2016-06/2017}                  \\
          Collection duration    & 3 weeks             & 3-4 months*
                                 & 6 months            & 3 years
                                 & 8 months            & \textbf{6 months}   \\
          Attributes             & 8                   & 17
                                 & 17                  & 305
                                 & 35                  & \textbf{262}        \\
          Browsers               & 455,604*            & 119,818*
                                 & -                   & -
                                 & 1,329,927           & \textbf{1,989,365}  \\
          Browsers seen multiple times & -                   & -
                                 & -                   & -
                                 & 661,827 (49.76\%)
                                 & \textbf{547,672 (27.53\%)}                \\
          Fingerprints           & 470,161             & 118,934
                                 & 2,067,942           & 88,088
                                 & 7,246,618           & \textbf{4,145,408}  \\
          Distinct fingerprints  & 409,296             & 142,023\footnote{
            This number is displayed in Figure~$11$ of~\cite{LRB16} as the number of distinct fingerprints, but it also corresponds to the number of raw fingerprints.
            Every fingerprint would be unique if the number of distinct and collected fingerprints are equal.
            Hence, we are not confident in this number, but it is the number provided by the authors.
          }
                                 & -                   & 9,822--16,541
                                 & 1,586,719           & \textbf{3,578,196}  \\
          \hline
          Ratio of desktop fingerprints & -            & 0.890*
                                 & 0.879               & 0.697*--0.707*
                                 & -                   & \textbf{0.805}      \\
          Ratio of mobile fingerprints & -             & 0.110*
                                 & 0.121               & 0.293--0.303
                                 & -                   & \textbf{0.134}      \\
          \hline
          Unicity of overall fingerprints & 0.836      & 0.894
                                 & 0.336               & 0.954--0.958
                                 & $\approx$ 0.350*    & \textbf{0.818}      \\
          Unicity of mobile fingerprints & -           & 0.810
                                 & 0.185               & 0.916--0.941
                                 & $\approx$ 0.350*    & \textbf{0.399}      \\
          Unicity of desktop fingerprints & -          & 0.900
                                 & 0.357               & 0.974--0.978
                                 & $\approx$ 0.350*    & \textbf{0.884}      \\
        \bottomrule
      \end{tabular}
    \end{minipage}
  \end{sidewaystable}

    \subsubsection{Working dataset}
    \label{sec:working-dataset}
      The working dataset obtained after preprocessing the raw dataset contains $5,714,738$ entries (comprising the identical fingerprints that are interleaved for each browser), with $4,145,408$ fingerprints (comprising no identical fingerprint for each browser), and $3,578,196$ distinct fingerprints.
      They are composed of $216$ original attributes and $46$ extracted attributes, for a total of $262$ attributes.
      The fingerprints come from $1,989,365$ browsers, $27.53$\% of which have multiple fingerprints.
      Table~\ref{tab:dataset-comparison} displays a comparison between the dataset of the studies Panopticlick~\cite{ECK10}, AmIUnique~\cite{LRB16}, Hiding in the Crowd~\cite{GLB18}, Long-Term Observation~\cite{PRGB20}, Who Touched My Browser Fingerprint~\cite{LC20}, and this study.

  \subsection{Comparison with previous studies}
  \label{sec:dataset-differences}
    We compare our working dataset with the datasets of previous studies in Table~\ref{tab:dataset-comparison}, notably by the unicity rate which is the proportion of the fingerprints that have been observed for a single browser.

    We have a slightly lower unicity rate compared to the studies Panopticlick (PTC)~\cite{ECK10} and AmIUnique (AIU)~\cite{LRB16}.
    Considering a larger browser population generally increases the chances for two browsers to share the same fingerprint, which can reduce the unicity rate.
    However, considering a wider surface of fingerprinting attributes usually reduces these chances, which can increase the unicity rate.
    The importance of these two effects depends on the browser population and on the attributes.
    The larger browser population and the larger surface of fingerprinting attributes can explain the slight decrease of the unicity rate compared to PTC and AIU.
    We also observe a lower unicity rate for the fingerprints of mobile browsers compared to the fingerprints of desktop browsers, confirming the findings of previous studies~\cite{ECK10, SPJ15, LRB16, GLB18}.
    However, we emphasize that our dataset and the datasets of these studies miss attributes that could improve the distinctiveness of the fingerprints of mobile browsers like the sensor API~\cite{W3CGenericSensorAPI} that is used in real-life~\cite{DABP18, MDIP19, IES21}.

    The authors of the Hiding in the Crowd (HitC) study~\cite{GLB18} worked on fingerprints collected from the same experiment and browser population as us.
    However, to stay consistent with their previous AIU study~\cite{LRB16}, they consider the same set of $17$~attributes.
    This explains our higher number of fingerprints, as two browsers that have different fingerprints for a given set of attributes can come to the same fingerprint if a subset of these attributes is considered.
    As a result, they remove more duplicated fingerprints than us due to the higher chances for a browser to present the same fingerprint for $17$~attributes than for $216$~attributes.
    Our unicity rate is also higher, being at $81.8$\% for the complete dataset against $33.6$\% for HitC~\cite{GLB18}.
    This is due to the larger set of considered attributes that distinguish browsers more efficiently, as each additional attribute can provide a way to distinguish browsers.
    Our little drops on the proportion of desktop and mobile browsers come from a finer-grained classification, as we have $4.8$\% of the browsers that are classified as belonging to tablets, smart TVs, or game consoles.

    The recent study about the evolution of fingerprints by Song Li and Yinzhi Cao~\cite{LC20} analyzes a large-scale dataset of $7,246,618$ fingerprints collected from a European website.
    They focus on the evolution of the fingerprints and their composing attributes.
    Their fingerprints show a low unicity rate of approximately $0.350$.
    The high number of collected fingerprints and the low number of used attributes can explain this lower unicity rate\footnote{
      The authors of~\cite{LC20} emphasize that the unicity rate is ``relatively low because many browsers visited their collection website more than once''.
      This can indicate that they set the identical fingerprints collected from the same browser in an anonymity set.
      As a result, this set is at least of size $2$ (non-unique) whereas the fingerprints may have been seen for this single browser only, which makes them actually unique.
    }.
    We remark that this unicity rate is close to the unicity rate of HitC~\cite{GLB18} which includes fewer attributes ($17$ against $35$ for~\cite{LC20}) and fewer fingerprints ($2,067,942$).
    The unicity rate of the mobile browsers of~\cite{LC20} is close to the unicity rate of their desktop browsers, contrary to previous studies~\cite{ECK10, SPJ15, LRB16, GLB18} and ours.
    This can be explained by the inclusion of new attributes that provide more distinctiveness to the mobile fingerprints.
    The attributes that they have in addition to the previous studies\footnote{
      The Panopticlick study~\cite{ECK10} leverages IP address hashes to resynchronize fingerprints (see Section~\ref{sec:uid-resynchronization}) but do not include any information related to the IP address in the fingerprints.
    } are the features that they infer from the IP address.
    Song Li and Yinzhi Cao~\cite{LC20} focus on the evolution of fingerprints and analyze what they call \textit{dynamics}.
    They define a dynamic as a representation of the evolutions between two different and consecutive fingerprints that belong to the same browser.
    They generate $960,853$ dynamics from their seven million fingerprints whereas we generate $3,725,373$ dynamics (see Section~\ref{sec:fingerprint-stability}) from our four million fingerprints.
    This difference is explained by our fingerprints being composed of more attributes, which increases the chances for two consecutive fingerprints to differ.
    Indeed, the change of any attribute between the consecutive fingerprints generates a dynamic, and the more attributes there is the more chances one of them changes between two evolutions\footnote{
      Would all the additional attributes never change, their addition would not induce more changes between the consecutive fingerprints.
      However, our dataset only contains $5$ attributes that never presented any change between the three million consecutive fingerprints (see Section~\ref{sec:attributes-stability}).
    }.

%% file: 4-authentication-factor-properties.tex
\section{Authentication factor properties}
\label{sec:authentication-factor-properties}
  Biometric authentication factors and browser fingerprints share strong similarities.
  They both work by extracting features from a unique entity, which is a person for the former and a browser for the latter, that can be used for identification or authentication.
  Although the entity is unique, the extracted features are a digital representation of the entity which can show imperfections (e.g., the fingerprints of two different persons can show similar representations).
  Previous studies~\cite{MMJP09, ZD04, GLMPSS05} identified the properties for a biometric characteristic to be \emph{usable}\footnote{
    Here, \emph{usable} refers to the adequacy of the characteristic to be used for authentication, rather than the ease of use by the users.
  } as an authentication factor, and the additional properties for a biometric authentication scheme to be \emph{practical}.
  We evaluate browser fingerprints according to these properties because of their similarity with biometric authentication factors.

  The four properties needed for a biometric characteristic to be \emph{usable} as an authentication factor are the following.
  \begin{itemize}
    \item \emph{Universality}: the characteristic should be present in everyone.
    \item \emph{Distinctiveness}: two distinct persons should have different characteristics.
    \item \emph{Permanence}: the same person should have the same characteristic over time. We rather use the term \emph{stability}.
    \item \emph{Collectibility}: the characteristic should be collectible and measurable.
  \end{itemize}

  The three properties that a biometric authentication scheme requires to be \emph{practical} are the following.
  \begin{itemize}
    \item \emph{Performance}: the scheme should be accurate, consume few resources, and be robust against environmental changes.
    \item \emph{Acceptability}: the users should accept to use the scheme in their daily lives.
    \item \emph{Circumvention}: it should be difficult for an attacker to deceive the scheme.
  \end{itemize}

  The properties that we study are the \emph{distinctiveness}, the \emph{stability}, and the \emph{performance}.
  We consider that the \emph{universality} and the \emph{collectibility} are satisfied, as the HTTP headers that are automatically sent by browsers constitute a fingerprint.
  However, we stress that a loss of distinctiveness occurs when no JavaScript attribute is available.
  About the \emph{circumvention}, we discuss the design of an authentication mechanism that leverages browser fingerprints and the possible attacks on such mechanism in Section~\ref{sec:discussion}.
  As for the \emph{acceptability}, we refer to~\cite{AND20} for a survey about the acceptability of an authentication mechanism that relies on browser fingerprinting, and emphasize that such mechanisms are already used in real-life~\cite{WLD19, MicroFocusDFP, SecureAuthDFP}.

  \subsection{Distinctiveness}
    To satisfy the \emph{distinctiveness} property, the browser fingerprints should distinguish two different browsers.
    The two extreme cases are every browser sharing the same fingerprint, which makes them indistinguishable from each other, and no two browsers sharing the same fingerprint, making every browser distinguishable.
    The distinctiveness falls between these extremes, depending on the attributes and the browser population.
    We consider the use of browser fingerprinting as an additional authentication factor.
    Hence, we do not require a perfect distinctiveness, as it is used in combination with other authentication factors to improve the overall security.

    The dataset entries are composed of a fingerprint, the source browser, and the moment of collection in the form of a Unix timestamp in milliseconds.
    We denote $B$ the domain of the unique identifiers, $F$ the domain of the fingerprints, and $T$ the domain of the timestamps.
    The fingerprint dataset is denoted $D$ and is formalized as:
    \begin{equation}
      D = \{ (f, b, t) \mid f \in F, b \in B, t \in T \}
    \end{equation}

    We use the size of the browser anonymity sets to quantify the distinctiveness, as the browsers that belong to the same anonymity set are indistinguishable.
    We denote $\mathcal{B}(f, \mathcal{D})$ the browsers that provided the fingerprint~$f$ in the dataset~$D$.
    It is formalized as:
    \begin{equation}
      \mathcal{B}(f, \mathcal{D}) = \{
        b \in B \mid
        \forall (g, b, t) \in \mathcal{D}, f = g
      \}
    \end{equation}

    We denote $\mathcal{A}(\epsilon, \mathcal{D})$ the fingerprints that have an anonymity set of size~$\epsilon$ (i.e., that are shared by $\epsilon$~browsers) in the dataset~$D$.
    It is formalized as:
    \begin{equation}
      \mathcal{A}(\epsilon, \mathcal{D}) = \{
        f \in F \mid
        \mathrm{card}(\mathcal{B}(f, \mathcal{D})) = \epsilon
      \}
    \end{equation}

    A common measure of the fingerprint distinctiveness is the \emph{unicity rate}~\cite{ECK10, LRB16, GLB18}, which is the proportion of the fingerprints that were observed for a single browser.
    We denote $\mathcal{U}(\mathcal{D})$ the unicity rate of the dataset $\mathcal{D}$, which is formalized as:
    \begin{equation}
      \mathcal{U}(\mathcal{D}) =
        \frac{
          \mathrm{card}(\mathcal{A}(1, \mathcal{D}))
        }{
          \mathrm{card}( \{ (f, b) \mid
                            \exists t \in T, (f, b, t) \in \mathcal{D}
                         \})
        }
    \end{equation}

    Previous studies measured the anonymity set sizes on the whole dataset~\cite{ECK10, LRB16, GLB18}.
    We measure the anonymity set sizes on the fingerprints currently in use by each browser, and not on their whole history.
    Two different browsers, sharing the same fingerprint on different time windows, are then distinguishable (e.g., a browser was updated before the other).
    Moreover, a browser that runs in a fancy web environment (e.g., having a custom font) and that has several fingerprints in the dataset (e.g., fifty), will bloat the proportion of unique fingerprints and bias the study (e.g., fifty fingerprints are unique whereas they come from a single browser).

    We evaluate the anonymity set sizes on the time-partitioned datasets composed of the last fingerprint seen for each browser at a given time.
    Let $\mathcal{S}_{\tau}(\mathcal{D})$ be the time-partitioned dataset originating from $\mathcal{D}$ that represents the state of the fingerprint of each browser after $\tau$~days.
    With $t_{\tau}$ the last timestamp of this day, $\mathcal{S}_{\tau}(\mathcal{D})$ is defined as:
    \begin{equation}
      \mathcal{S}_{\tau}(\mathcal{D}) = \{
        (f_i, b_j, t_k) \in \mathcal{D} \mid
        \forall (f_p, b_q, t_r) \in \mathcal{D},
        b_j = b_q,
        t_r \leq t_k \leq t_{\tau}
      \}
    \end{equation}

  \subsection{Stability}
    To satisfy the \emph{stability} property, the fingerprint of a browser should stay sufficiently similar between two observations to be recognizable.
    Browser fingerprints have the particularity of evolving through time, due to changes in the web environment like a software update or a user configuration.
    We measure the stability by the average similarity between the consecutive fingerprints of browsers, given the elapsed time between their observation.
    The two extreme cases are every browser holding the same fingerprint through its life, and the fingerprint changing completely with each observation.
    A lack of stability makes it harder to recognize the fingerprint of a browser between two observations.

    We denote $\mathcal{C}(\Delta, \mathcal{D})$ the function that provides the pairs of consecutive fingerprints of $\mathcal{D}$ that are separated by a time-lapse comprised in the $\Delta$ time range.
    It is formalized as:
    \begin{equation}
      \begin{aligned}
        \mathcal{C}(\Delta, \mathcal{D}) = \{ (f_i, f_p) \mid \,
          & \forall ((f_i, b_j, t_k), (f_p, b_q, t_r)) \in \mathcal{D}^2,
            b_j = b_q, t_k < t_r, (t_r - t_k) \in \Delta, \\
          & \nexists (f_c, b_d, t_e) \in \mathcal{D},
            b_d = b_j, f_c \neq f_i, f_c \neq f_p,
            t_k < t_e < t_r
        \}
      \end{aligned}
    \end{equation}

    We consider the Kronecker delta $\delta(x, y)$, being $1$ if $x$ equals $y$ and $0$ otherwise.
    We consider the set $\Omega$ of the $n$ used attributes.
    We denote $f[\omega]$ the value taken by the attribute~$\omega$ for the fingerprint~$f$.
    Let $\mathrm{sim}(f, g)$ be the proportion of identical attributes between the fingerprints $f$ and $g$, which is formalized as:
    \begin{equation}
      \mathrm{sim}(f, g) = \frac{1}{n}
      \sum_{\omega \in \Omega} \delta(f[\omega], g[\omega])
    \end{equation}

    We define the function $\mathrm{avsim}(\Delta, \mathcal{D})$ that provides the average similarity between the pairs of the consecutive fingerprints, for a given time range~$\Delta$ and a dataset~$D$.
    It is formalized as:
    \begin{equation}
      \mathrm{avsim}(\Delta, \mathcal{D}) =
        \frac{
          \sum_{(f, g) \in \mathcal{C}(\Delta, \mathcal{D})}
            \mathrm{sim}(f, g)
        }{
          \mathrm{card}(\mathcal{C}(\Delta, \mathcal{D}))
        }
    \end{equation}

  \subsection{Performance}
    We consider four aspects of the \emph{performance} of browser fingerprints for web authentication: their \emph{collection time}, their \emph{size} in memory, the \emph{loss of efficacy} between different device types, and the \emph{accuracy} of a simple illustrative verification mechanism.
    Browser fingerprinting can easily be deployed by adding a script on the authentication page, and by preparing servers to handle the reception, the storage, and the verification of the fingerprints.
    The users solely rely on their regular web browser and do not have to run any dedicated application, nor possess specific hardware, nor undergo a configuration step.
    The main additional load is on the supplementary consumption of memory and time resources.
    Moreover, the web environment differs between device types (e.g., mobile browsers have more limited functionalities than desktop browsers) and through time (e.g., modern browsers differ from the browsers from ten years ago), impacting the efficacy of browser fingerprinting.
    Finally, we evaluate the accuracy of a simple illustrative verification mechanism under fingerprints evolution.

    \subsubsection{Collection time}
      The browser fingerprints can be solely composed of passive attributes (e.g., HTTP headers) that are transmitted along with the communications with the server.
      In this case, the fingerprints are collected without the user perceiving any collection time, but major attributes are set aside.
      The client-side properties collected through JavaScript provide more distinctive attributes, at the cost of an additional collection time.
      We measure the collection time of the fingerprints considering only the JavaScript attributes, and ignore the HTTP headers that are transmitted passively.

    \subsubsection{Size}
      Browser fingerprinting consumes memory resources on the clients during the buffering of the fingerprints, on the wires during their sending, and on the servers during their storage.
      The memory consumption depends on the storage format of the fingerprints.
      For example, a canvas can be stored as an image encoded in a base64 string, or as a hash which is shorter.
      A trade-off has to be done between the quantity of information and the memory consumption.
      The less memory-consuming choice is to store the complete fingerprint as a single hash.
      However, the fingerprints evolve through time -- even more when they are composed of many attributes -- which results in the use of a hash of the complete fingerprint being impractical.
      Due to the unspecified size of the attributes (e.g., the specification of the \texttt{User-Agent} HTTP header does not define a size limit~\cite{RFC7231UserAgent}), we measure their \emph{size} given a fingerprint dataset.

    \subsubsection{Loss of efficacy}
      The \emph{loss of efficacy} is the loss of stability, distinctiveness, or performance of the fingerprints.
      It can occur either for a group of browsers (e.g., mobile browsers) or resulting from changes brought to web technologies.
      First, previous works showed differences in the properties of the fingerprints coming from mobile and desktop devices~\cite{SPJ15, LRB16, GLB18}, notably a limited distinctiveness for the mobile browsers.
      Following these findings, we compare the properties shown by the mobile and the desktop browsers.
      We match keywords on the \texttt{userAgent} JavaScript property (see Appendix~\ref{app:keywords}) to differentiate the browsers running on a desktop or a laptop (referred to as \emph{desktops}) from the browsers running on a mobile phone (referred to as \emph{mobiles}).
      Second, browser fingerprinting is closely dependent on the evolution of web technologies.
      As new technologies are integrated into browsers, new attributes are accessible and conversely for removal.
      Similarly, functionality updates can lead to a change in the fingerprint properties.
      For example, Kurtz et al.~\cite{KGBRF16} detected an iOS update by the sudden instability of an attribute that provides the iOS version.
      Following their finding, we verify whether the evolution of web technologies provokes major losses in the properties of the fingerprints.

    \subsubsection{Accuracy of a simple verification mechanism}
    \label{sec:verification-mechanism-accuracy}
      We evaluate the accuracy of a simple illustrative verification mechanism under the evolution of fingerprints.
      This mechanism counts the identical attributes between the presented and the stored fingerprint, and considers the evolution legitimate if this number is above a threshold~$\Theta$.
      The simplicity of this mechanism gives us an idea of the accuracy that can be easily achieved without having to engineer more complex rules.
      More elaborate mechanisms can obviously be designed (see Appendix~\ref{app:advanced-verification-mechanism}).

%% file: 5-results.tex
\section{Evaluation of browser fingerprints properties}
\label{sec:results}
  In this section, we evaluate the browser fingerprints of our dataset according to the properties of distinctiveness, stability, and performance.
  We present here the results on the complete fingerprints and let Section~\ref{sec:attribute-wise-analysis} provide insights on the contribution of the attributes to each property.
  We show that, by time-partitioning our dataset, our fingerprints provide a unicity rate of more than $81.3$\% which is stable through time.
  However, our fingerprints of mobile browsers are less distinctive than our fingerprints of desktop browsers\footnote{
    Our study is in line with previous studies~\cite{ECK10, SPJ15, LRB16, GLB18} about the fingerprints of mobile browsers showing less distinctiveness.
    However, we emphasize that our dataset and the datasets of these studies miss attributes that could improve the distinctiveness of the fingerprints of mobile browsers like the sensor API~\cite{W3CGenericSensorAPI} that is used in real-life~\cite{DABP18, MDIP19, IES21}.
  }, with a respective unicity rate of $42$\% against $84$\% considering the time-partitioned datasets.
  We also show that, on average, a fingerprint has more than $91$\% of identical attributes between two observations, even when they are separated by nearly $6$~months.
  About their performance, we show that our fingerprints weigh a dozen of kilobytes and are collected within seconds.
  The accuracy of the simple illustrative verification mechanism is close to perfect, as it achieves an equal error rate of $0.61$\%.
  This results from most of the consecutive fingerprints coming from a browser having at least $234$ identical attributes, whereas most of the fingerprints of different browsers have fewer.

  \subsection{Distinctiveness}

    \subsubsection{Overall distinctiveness}
\begin{figure}
        \centering
        \includegraphics[width=0.70\columnwidth]{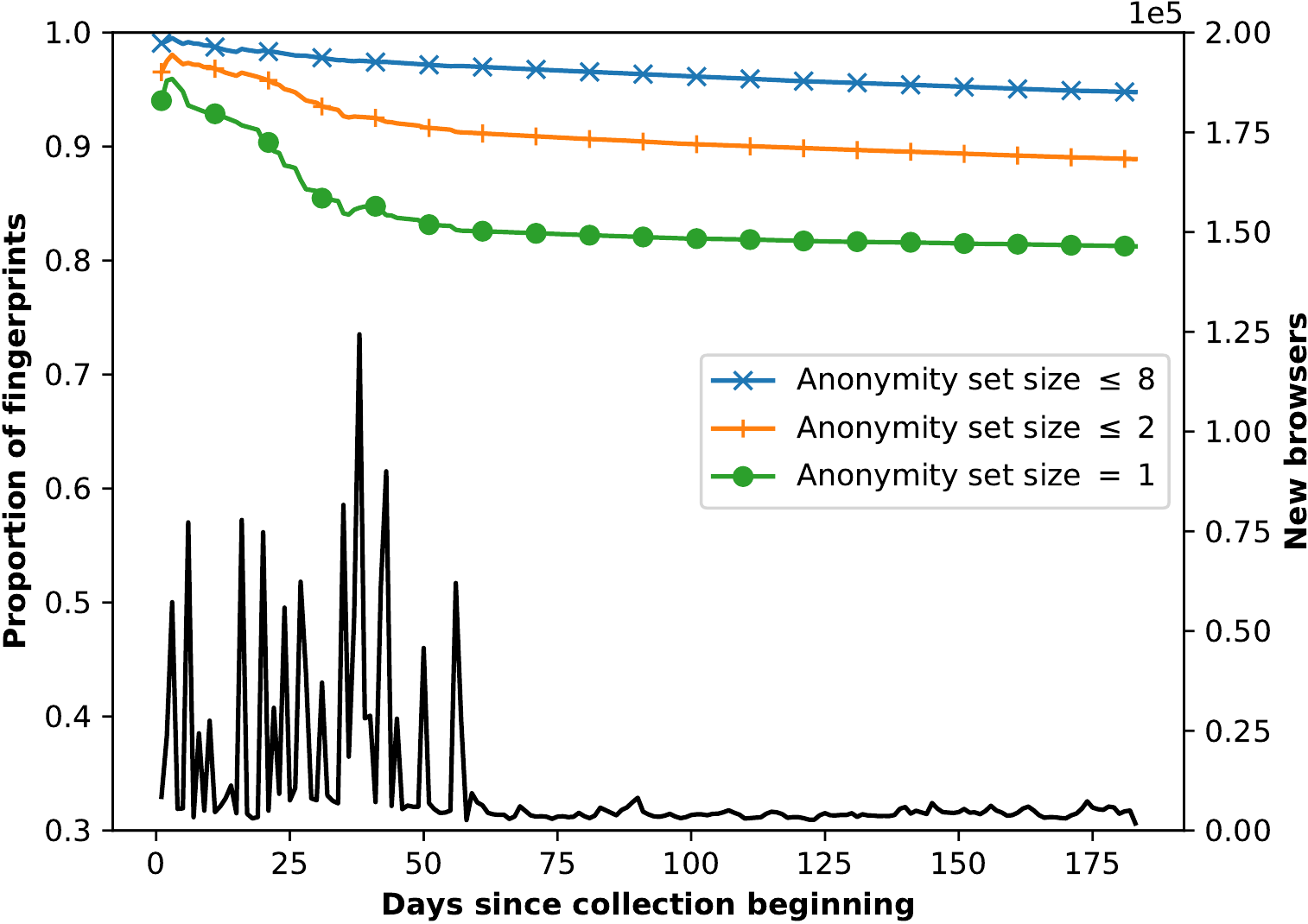}
        \caption{
          Anonymity set sizes and frequency of browser arrivals through the time-partitioned datasets obtained after each day.
          The new browsers are displayed in hundreds of thousands.
        }
        \label{fig:anonymity-sets-full}
        \Description[
          Anonymity set sizes and frequency of browser arrivals through the time-partitioned datasets obtained after each day.
        ]{
          Anonymity set sizes and frequency of browser arrivals through the time-partitioned datasets obtained after each day.
          On the long run, more than $81.3$\% of the fingerprints are shared by a single browser, and more than $94.7$\% are shared by at most eight browsers.
        }
      \end{figure}

      Figure~\ref{fig:anonymity-sets-full} presents the size of the anonymity sets alongside the frequency of browser arrival for the time-partitioned datasets.
      The time-partitioned datasets are designed so that each browser has the last fingerprint observed at the end of the $\tau$-th day.
      The overall fingerprints have a stable unicity rate of more than $81.3$\% for the partitioned-datasets, and more than $94.7$\% of the fingerprints are shared by at most $8$ browsers.
      The overall fingerprints comprise those collected from desktop and mobile browsers, but also those of tablets, game consoles, and smart TVs.
      The comparisons are done using fingerprint hashes, resulting in $4$ hash collisions which we deem negligible.

      The anonymity sets tend to grow as more browsers are encountered due to the higher chances of collision.
      However, the fingerprints tend to stay in small anonymity sets, as can be seen by the growth of the anonymity sets of size $2$ being more important than the growth of the anonymity sets of size $8$ or higher.
      The unicity rate of the time-partitioned datasets ($81.3$\%) is lower than the unicity rate of the complete dataset ($81.8$\%).
      This is due to browsers having multiple unique fingerprints in the complete dataset, which typically occurs when a browser having a unique web environment is fingerprinted multiple times.
      Considering the time-partitioned datasets removes this over-counting effect.

      New browsers are encountered continually.
      However, starting from the $60$th day, the arrival frequency stabilizes around $5,000$ new browsers per day.
      Before this stabilization, the arrival frequency is variable and has major spikes that seem to correspond to events that happened in France.
      These events could lead to more visits and explain these spikes.
      For example, the spike on the $38$th day corresponds to a live political debate on TV, and the spike on the $43$rd correlates with the announcement of a cold snap.

    \subsubsection{Distinctiveness of desktop and mobile browsers}
\begin{figure}
        \centering
        \includegraphics[width=0.70\columnwidth]{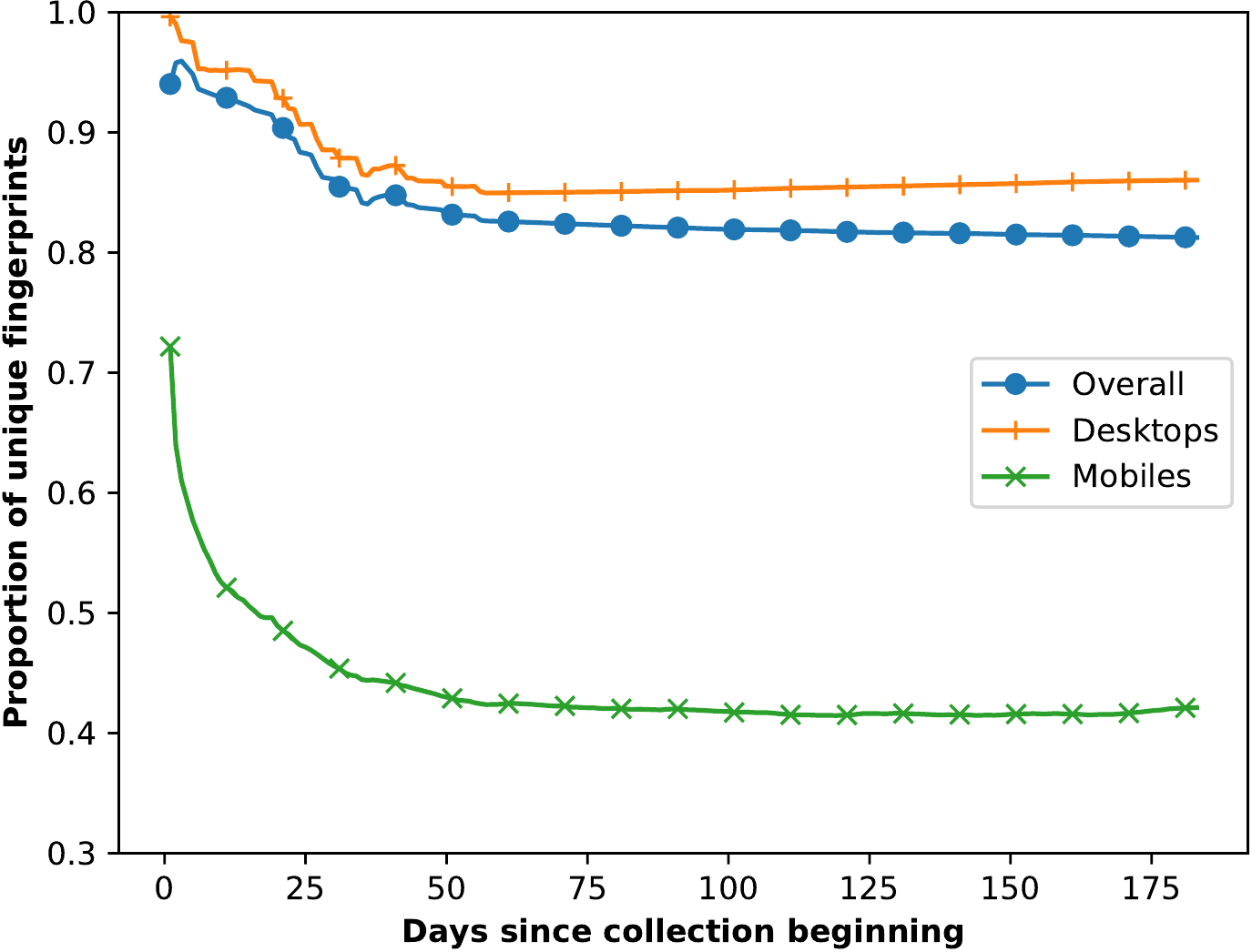}
        \caption{
          Unicity rate for the overall, the mobile, and the desktop browsers, through the time-partitioned datasets obtained after each day.
        }
        \label{fig:anonymity-sets-web-environment}
        \Description[
          Unicity rate for the overall, the mobile, and the desktop browsers, through the time-partitioned datasets obtained after each day.
        ]{
          Unicity rate for the overall, the mobile, and the desktop browsers, through the time-partitioned datasets obtained after each day.
          The fingerprints of the mobile browsers are more uniform than the fingerprints of the desktop browsers, with a unicity rate of approximately $42$\% against $84$\% on the long run.
        }
      \end{figure}

      Figure~\ref{fig:anonymity-sets-web-environment} presents the unicity rate through the time-partitioned datasets for the overall, the mobile, and the desktop browsers.
      The fingerprints of mobile browsers are more uniform than those of desktop browsers, with a unicity rate of approximately $42$\% against $84$\% considering the time-partitioned datasets.
      The unicity rate of the desktop browsers slightly increases by $1.04$~points from the $60$th to the $183$th day, from $84.99$\% to $86.03$\%.
      On the contrary, the unicity rate of the mobile browsers slightly decreases by $0.29$~points on the same period, from $42.42$\% to $42.13$\%.
      These results are in line with previous studies~\cite{ECK10, SPJ15, LRB16, GLB18} that also observed a reduction of the distinctiveness among the fingerprints of mobile browsers compared to those of desktop browsers.
      However, we acknowledge that our dataset misses attributes that could improve the distinctiveness of the fingerprints of mobile browsers.
      A good candidate attribute is the sensor API~\cite{W3CGenericSensorAPI} that is already used by real-life fingerprinters~\cite{DABP18, MDIP19, IES21}.

  \subsection{Stability}
  \label{sec:fingerprint-stability}
\begin{figure}
      \centering
      \includegraphics[width=0.70\columnwidth]{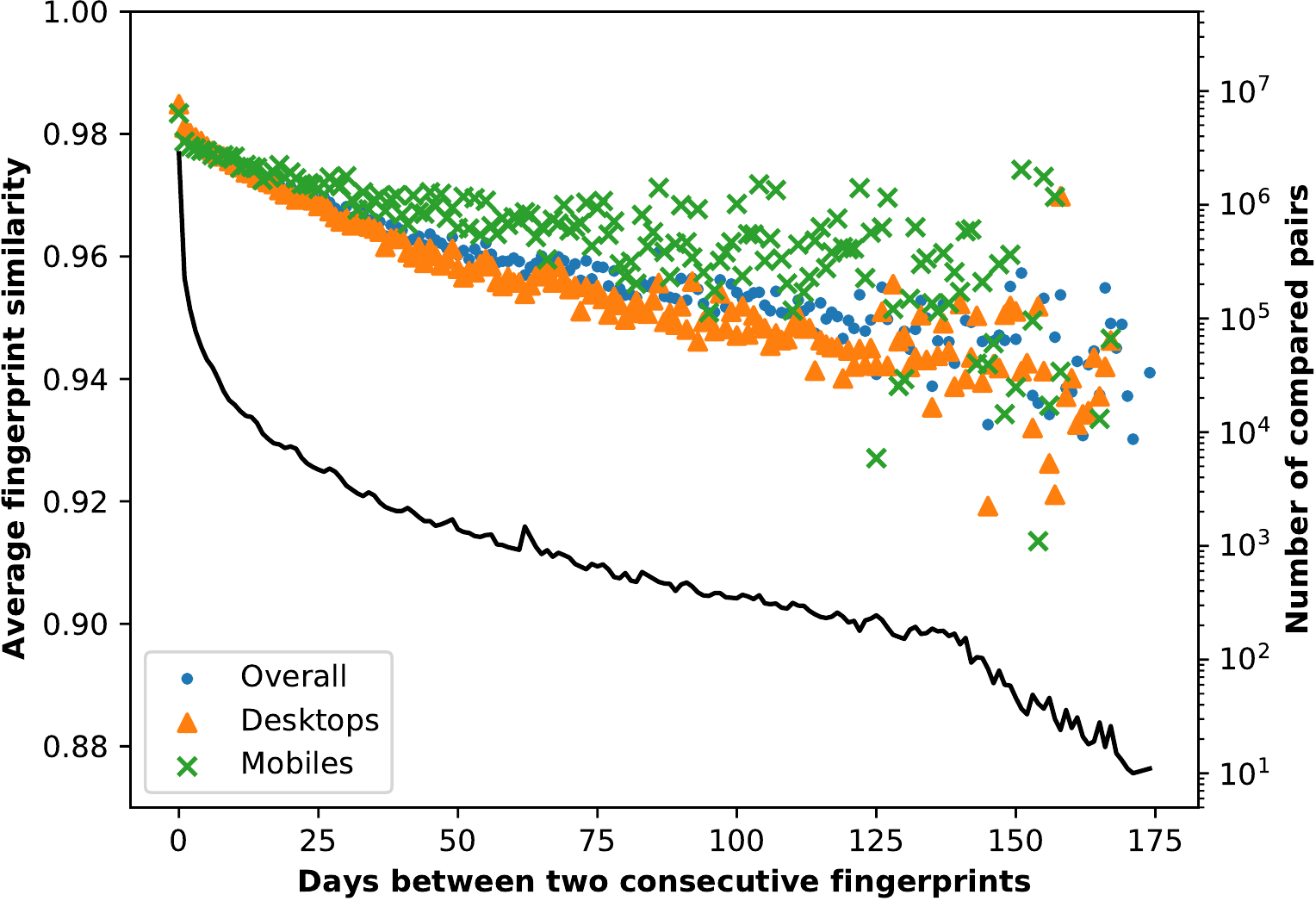}
      \caption{
        Average similarity between the pairs of consecutive fingerprints as a function of the time difference, with the number of compared pairs, for the overall, the mobile, and the desktop browsers.
      }
      \label{fig:stability}
      \Description[
        Average similarity between the pairs of consecutive fingerprints as a function of the time difference, with the number of compared pairs, for the overall, the mobile, and the desktop browsers.
      ]{
        Average similarity between the pairs of consecutive fingerprints as a function of the time difference, with the number of compared pairs, for the overall, the mobile, and the desktop browsers.
        On average, a fingerprint has more than $91$\% of its attributes unchanged even after nearly $6$~months.
        The fingerprints of the mobile browsers are generally more stable than the fingerprints of the desktop browsers.
      }
    \end{figure}

    Figure~\ref{fig:stability} displays the average similarity between the pairs of consecutive fingerprints as a function of the time difference, together with the number of compared pairs for each time difference.
    The ranges~$\Delta$ are expressed in days, so that the day~$d$ on the x-axis represents the fingerprints that are separated by ${\Delta = [d; d+1[}$~days.
    We ignore the comparisons of the time ranges that have less than $10$~pairs to have samples of sufficient size without putting too many comparisons aside.
    We also ignore the bogus comparisons that have a time difference higher than the limit of our experiment ($182$~days).
    These two sets of ignored comparisons account for less than $0.03$\% of each group.
    The results are obtained by comparing a total of $3,725,373$~pairs of consecutive fingerprints, which include $2,912,860$~pairs for the desktop browsers and $594,591$~pairs for the mobile browsers.
    Two consecutive fingerprints are necessarily different as we remove the duplicated consecutive fingerprints (see Section~\ref{sec:dataset-deduplication}).
    Considering ${(f_1, f_2, f_3)}$ the fingerprints collected for a browser that are ordered by the time of collection.
    The resulting set of consecutive fingerprints is $\{(f_1, f_2), (f_2, f_3)\}$.
    Our stability results are a lower bound, as the consecutive fingerprints are necessarily different (i.e., their similarity is strictly lower than~$1$).

    Our fingerprints are stable, as on average more than $91$\% of the attributes are expected to not change, considering up to $174$ elapsed days (nearly $6$~months) between two observations.
    In a web authentication context, we assume that the users would connect more frequently or would accept to undergo the account recovery process (see Section~\ref{sec:browser-fingerprinting-based-authentication-mechanism}).
    The fingerprints of mobile browsers are generally more stable than those of desktop browsers, as suggests their respective similarity curve.
    Few attributes of our script are highly instable.
    They are discussed in Section~\ref{sec:attributes-stability}.
    Getting rid of these attributes could reduce the distinctiveness of the fingerprints, but would improve their stability.

  \subsection{Performance}

    \subsubsection{Time consumption}
    \label{sec:fingerprints-collection-time}
\begin{figure}
        \centering
        \includegraphics[width=0.70\columnwidth]{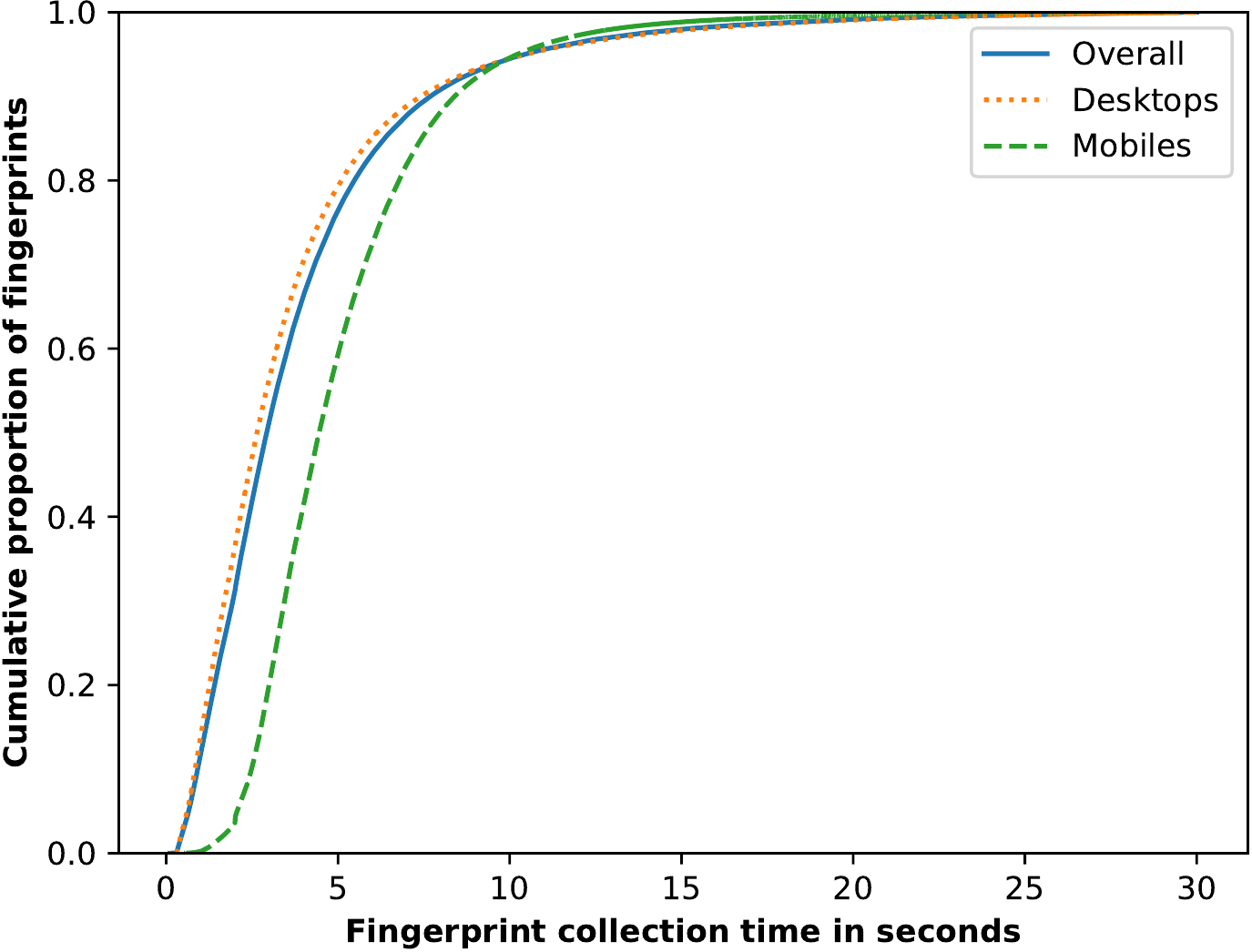}
        \caption{
          Cumulative distribution of the collection time of the fingerprints in seconds.
          We consider the fingerprints that take more than $30$~seconds to collect as outliers, which account for less than $1$\% of each group.
        }
        \label{fig:fps-collection-time}
        \Description[
          Cumulative distribution of the collection time of the fingerprints in seconds.
          We consider the fingerprints that take more than $30$~seconds to collect as outliers, which account for less than $1$\% of each group.
        ]{
          Cumulative distribution of the collection time of the fingerprints in seconds.
          We consider the fingerprints that take more than $30$~seconds to collect as outliers, which account for less than $1$\% of each group.
          Half of our fingerprints are collected in less than $2.92$~seconds and the majority ($95$\%) in less than $10.42$~seconds.
        }
      \end{figure}

      Figure~\ref{fig:fps-collection-time} displays the cumulative distribution of the collection time of our fingerprints in seconds with the outliers removed.
      We measure the collection time by the difference between two timestamps, one recorded at the starting of the script and the other one just before sending the fingerprint.
      Some fingerprints take a long time to collect that span from several hours to days.
      We limit the results to the fingerprints that take $30$~seconds or less to collect and consider the higher values as outliers.
      The outliers account for less than $1$\% of each group and are discussed in Appendix~\ref{app:anomalous-collection-times}.

      Half of our fingerprints are collected in less than $2.92$~seconds, and the majority ($95$\%) in less than $10.42$~seconds.
      The time to collect the fingerprints is lower for the desktop browsers  than for the mobile browsers.
      Half of the fingerprints of the desktop browsers are collected in less than $2.64$~seconds, and the majority ($95$\%) in less than $10.45$~seconds.
      These numbers are respectively of $4.44$~seconds and $10.16$~seconds for the mobile browsers.
      The median collection time of our fingerprints is less than the estimated median time taken by web pages to load completely~\cite{WebPageMedianLoadingTime}, which is $6.5$~seconds for the desktop browsers and $17.9$~seconds for the mobile browsers at the date of May~$1$, $2021$.
      Mobile devices generally have less computing power than desktop devices, which can explain the longer collection time together with the throttling of inactive tabs.
      The collection time of the fingerprints of mobile browsers has less variance than those of desktop browsers.
      This can be explained by the former having more uniform computing power than the latter.
      This is supported by the presence in our dataset of desktop browsers running on old systems like Windows~Vista or Windows~XP.

      Our script takes several seconds to collect the attributes that compose the fingerprints.
      However, we stress that this script is purely experimental and was developed to collect many attributes to get closer to what a fingerprinter may achieve in real-life.
      Although collecting the fingerprint induces an additional collection time on the authentication page, such page is expected to be lightweight and fast to load.
      Moreover, the time to collect the fingerprint can be reduced by ways that we describe below.
      The attributes that are longer to collect and that are less distinctive can be removed without a major loss of distinctiveness~\cite{AAL20}.
      For example, our method to detect an ad-blocker waits a few seconds for a simulated advertisement to be removed, but only provides a Boolean value.
      We discuss the longest attributes to collect in Section~\ref{sec:attributes-collection-time}.
      The attributes can also be collected in parallel or in the background.
      In their recent work, Song Li and Yinzhi Cao~\cite{LC20} collect their attributes in parallel and manage to collect the fingerprints of desktop and mobile browsers within one second.
      Our script can also be updated to leverage the most advanced web technologies, like the OffscreenCanvas API~\cite{OffscreenCanvas} that migrates the generation of the canvases off the main thread to another thread.
      More generally, we can use the Service Workers API~\cite{ServiceWorkers} to collect the attributes concurrently in the background to reduce the perceived collection time.

    \subsubsection{Memory consumption}
\begin{figure}
        \centering
        \includegraphics[width=0.70\columnwidth]{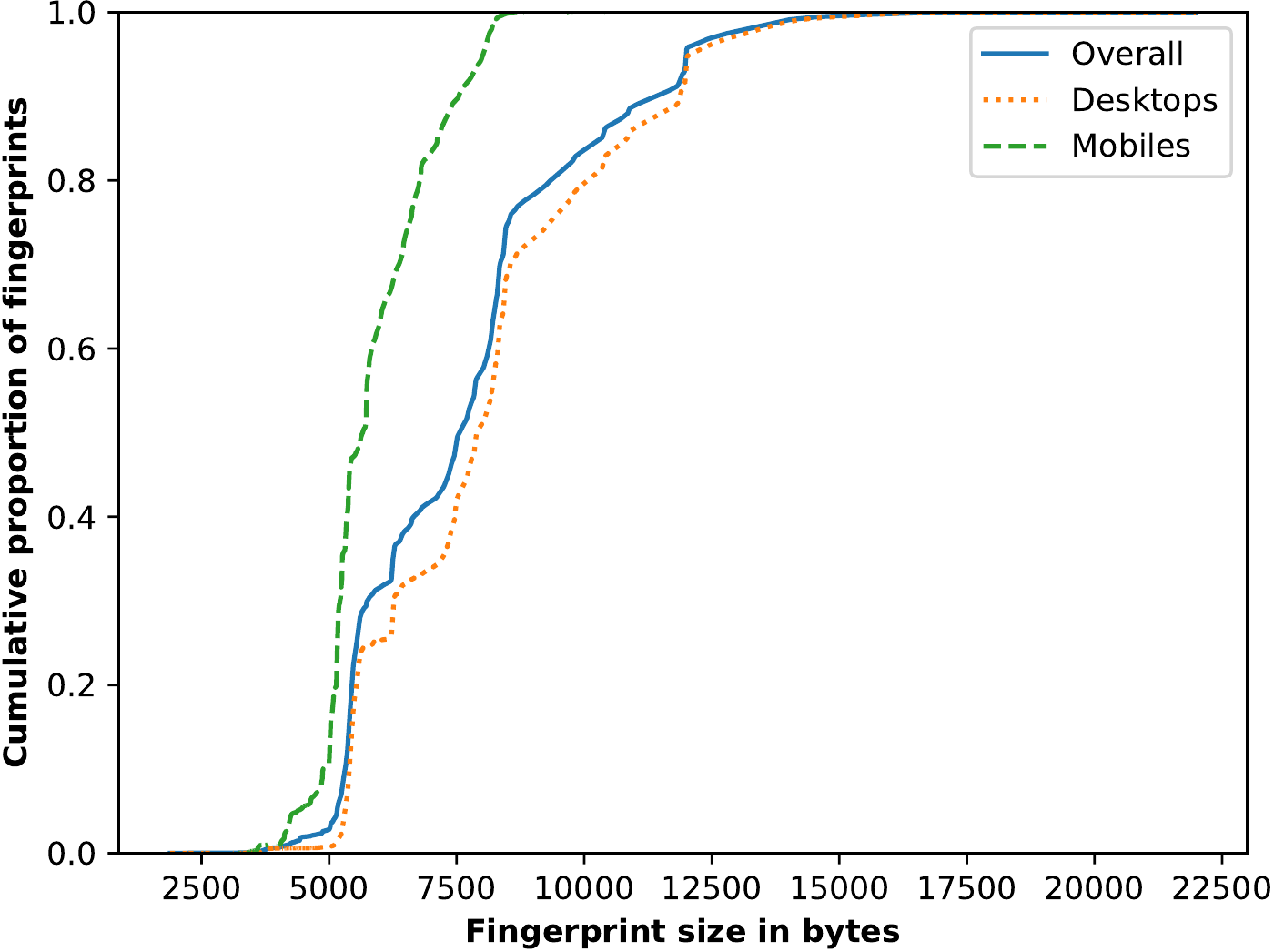}
        \caption{
          Cumulative distribution of the fingerprint size in bytes, for the overall, the mobile, and the desktop browsers.
          We ignore one outlier collected from a desktop browser that shows an extremely higher size.
        }
        \label{fig:fps-size-cdf}
        \Description[
          Cumulative distribution of the fingerprint size in bytes, for the overall, the mobile, and the desktop browsers.
          We ignore one outlier collected from a desktop browser that shows an extremely higher size.
        ]{
          Cumulative distribution of the fingerprint size in bytes, for the overall, the mobile, and the desktop browsers.
          We ignore one outlier collected from a desktop browser that shows an extremely higher size.
          Half of our fingerprints take less than $7,550$~bytes, $95$\% less than $12$~kilobytes, and all of them less than $22$~kilobytes.
        }
      \end{figure}

      Figure~\ref{fig:fps-size-cdf} displays the cumulative distribution of the size of our fingerprints in bytes with the outliers removed.
      Our fingerprints are encoded in UTF-8 using only ASCII characters and the canvases are stored as sha256 hashes.
      One character then takes one byte and the results can be expressed in both units.
      The fingerprint sizes comprise the value of the $262$~attributes without the metadata fields (e.g., the UID, the timestamp).
      The average fingerprint size is of ${\mu=7,692}$~bytes, and the standard deviation is of ${\sigma=2,294}$.
      We remove $1$~fingerprint from a desktop browser considered an outlier due to its size being greater than ${\mu + 15 \cdot \sigma}$.

      The memory consumption takes place on three components: on the client during the buffering of the fingerprints, on the wire during their sending, and on the server during their storage.
      Half of our fingerprints take less than $7,550$~bytes, $95$\% less than $12$~kilobytes, and all of them less than $22$~kilobytes.
      This is negligible given the current storage and bandwidth capacities.
      We observe a difference between the fingerprints of mobile and desktop browsers, with $95$\% of the fingerprints weighing respectively less than $8,020$~bytes and $12,082$~bytes.
      This is due to heavy attributes being lighter on mobile browsers, like the list of plugins or mime types that are most of the time empty.
      We discuss the heaviest attributes in Section~\ref{sec:attributes-size}.

    \subsubsection{Accuracy of the simple verification mechanism}
    \label{sec:accuracy-of-the-simple-verification-mechanism}
      The accuracy of the simple illustrative verification mechanism is measured according to the following methodology.
      First, we split our dataset in \emph{six samples}, one for each month.
      We assume that a user would spend at most one month between two connections, and otherwise would accept to undergo a heavier fingerprint update process.
      Two sets are afterward extracted from each sample.
      They are composed of pairs of compared fingerprints that we call \emph{comparisons}.
      The \emph{same-browser} comparisons are composed of the consecutive fingerprints of each browser, and the \emph{different-browsers} comparisons are composed of two randomly picked fingerprints of different browsers.
      After constituting the same-browser comparisons for each month, we sample the different-browsers comparisons to have the same size as the same-browser comparisons.
      The month sampling also helps the different-browsers comparisons to be realistic by pairing fingerprints that are separated by at most one month.
      Both the two sets of comparisons contain a total of $3,467,289$ comparisons.

\begin{figure*}
        \minipage{0.49\columnwidth}
          \centering
          \includegraphics[width=0.9\columnwidth]{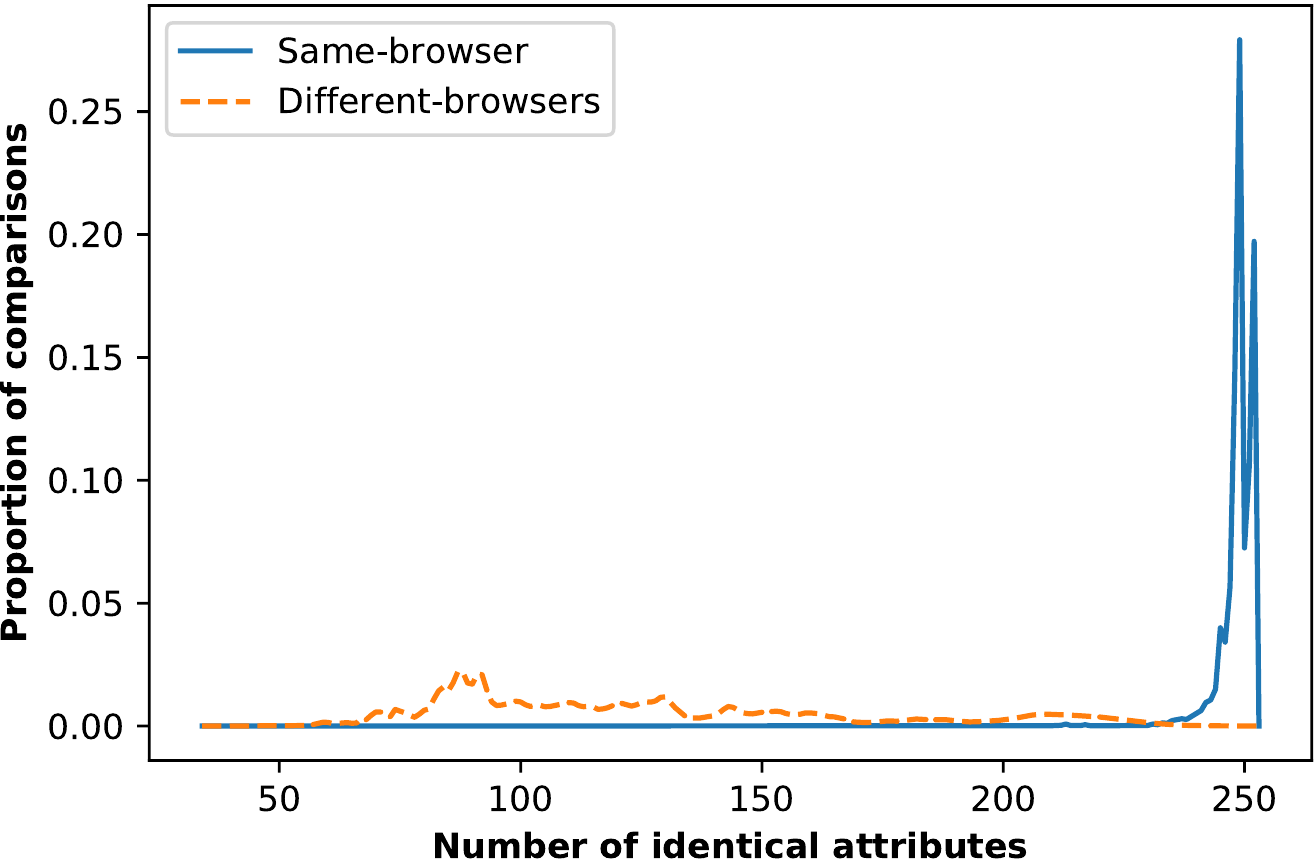}
          \caption{
            The number of identical attributes between the same-browser comparisons and the different-browsers comparisons.
          }
          \label{fig:identical-attributes-distribution-complete}
          \Description[
            The number of identical attributes between the same-browser comparisons and the different-browsers comparisons.
          ]{
            The number of identical attributes between the same-browser comparisons and the different-browsers comparisons.
            The two sets of comparisons are well separated, as $99.05$\% of the same-browser comparisons have at least $234$ identical attributes, and $99.68$\% of the different-browsers comparisons have fewer.
          }
        \endminipage
        \hfill
        \minipage{0.49\columnwidth}
          \centering
          \includegraphics[width=0.9\columnwidth]{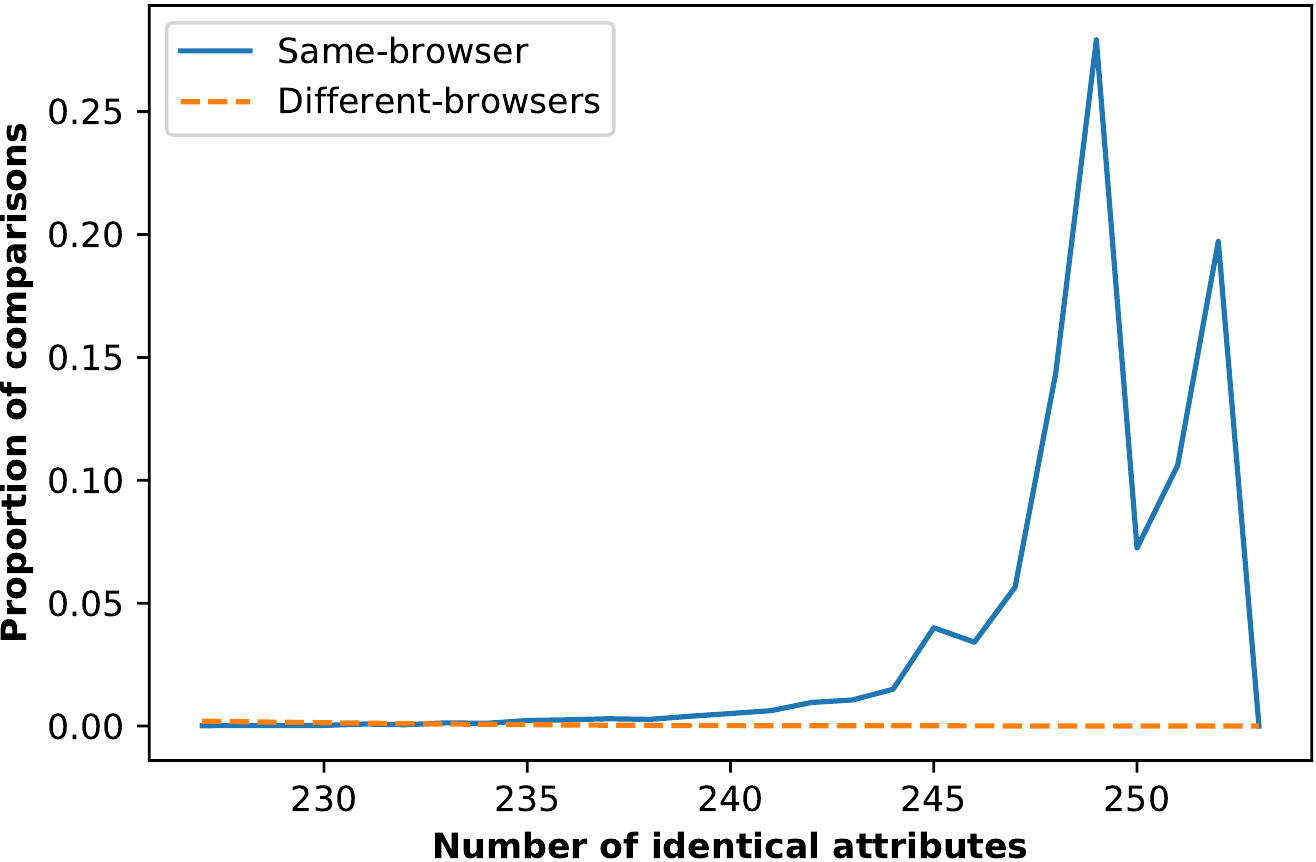}
          \caption{
            The number of identical attributes between the same-browser comparisons and the different-browsers comparisons, starting from $227$ attributes.
          }
          \label{fig:identical-attributes-distribution-zoom}
          \Description[
            The number of identical attributes between the same-browser comparisons and the different-browsers comparisons, starting from $227$ attributes.
          ]{
            The number of identical attributes between the same-browser comparisons and the different-browsers comparisons, starting from $227$ attributes.
            The two sets of comparisons are well separated, as $99.05$\% of the same-browser comparisons have at least $234$ identical attributes, and $99.68$\% of the different-browsers comparisons have fewer.
          }
        \endminipage
      \end{figure*}

      Figure~\ref{fig:identical-attributes-distribution-complete} displays the distribution of the identical attributes between the same-browser comparisons and the different-browsers comparisons, starting from $34$ identical attributes as there are no observed value below.
      Figure~\ref{fig:identical-attributes-distribution-zoom} presents a focus that starts from $227$ identical attributes, below which there are less than $0.005$ of the same-browser comparisons.
      We can observe that the two sets of comparisons are well separated, as $99.05$\% of the same-browser comparisons have at least $234$ identical attributes, and $99.68$\% of the different-browsers comparisons have fewer.
      The different-browsers comparisons have generally a fewer, and a more diverse, number of identical attributes compared to the same-browser comparisons.
      The different-browsers comparisons have between $34$ and $253$ identical attributes, with an average of $127.41$ attributes and a standard deviation of $44.06$ attributes.
      The same-browser comparisons have between $72$ and $252$ identical attributes, with an average of $248.64$ attributes and a standard deviation of $3.91$ attributes.

\begin{figure}
        \centering
        \includegraphics[width=0.70\columnwidth]{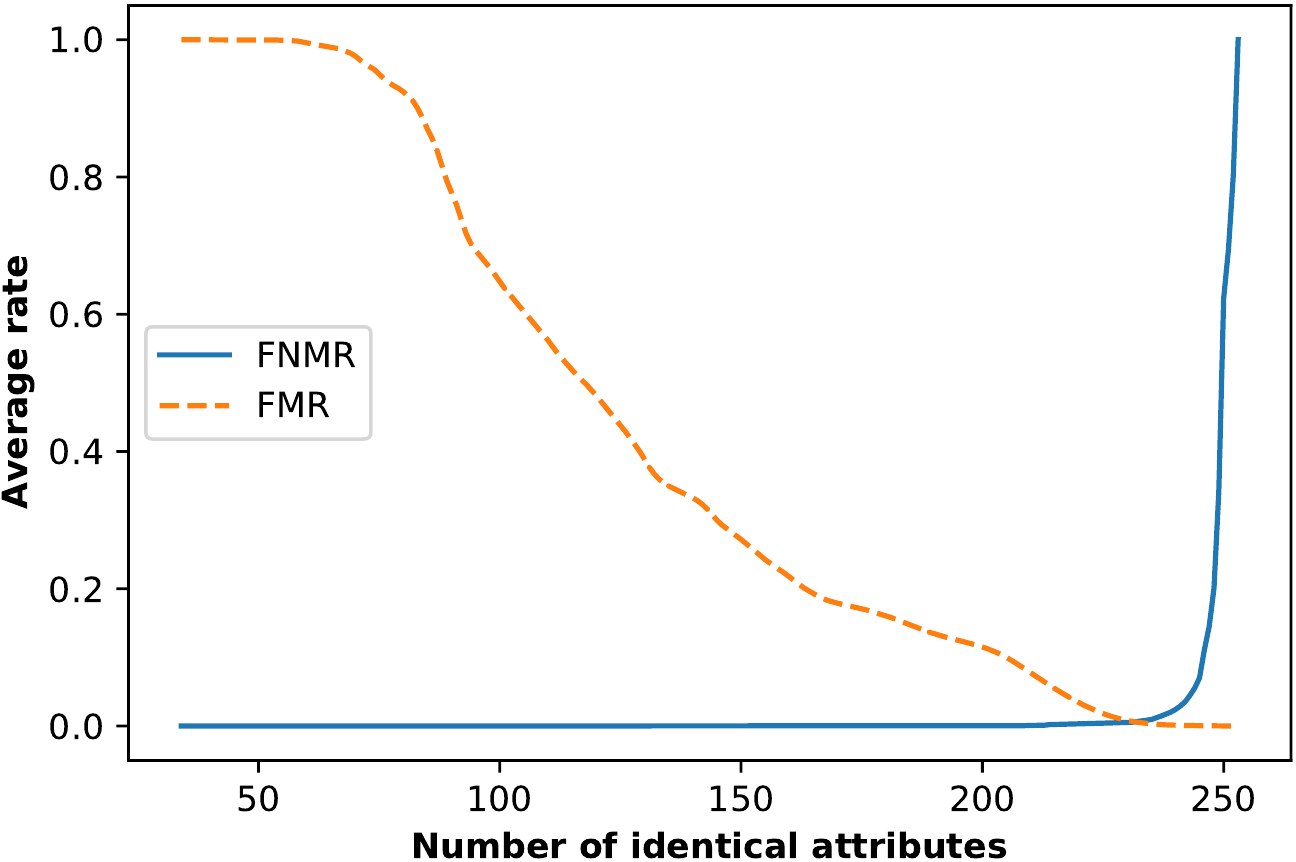}
        \caption{
          False match rate (FMR) and false non-match rate (FNMR) given the required number of identical attributes, averaged among the month six samples.
        }
        \label{fig:identical-fmr-fnmr-distribution}
        \Description[
          False match rate (FMR) and false non-match rate (FNMR) given the required number of identical attributes, averaged among the month six samples.
        ]{
          False match rate (FMR) and false non-match rate (FNMR) given the required number of identical attributes, averaged among the month six samples.
          The equal error rate, which is the rate where both the FMR and the FNMR are equal, is of $0.61$\% and is achieved for $232$ identical attributes.
        }
      \end{figure}

      Figure~\ref{fig:identical-fmr-fnmr-distribution} displays the false match rate (FMR) which is the proportion of the same-browser comparisons that are classified as different-browsers comparisons, and the false non-match rate (FNMR) which is the inverse.
      The displayed results are the average for each number of identical attributes among the six month-samples.
      As there are few same-browser comparisons that have less than $234$ identical attributes, the FNMR is null until this value.
      However, after exceeding this threshold, the FNMR increases as the same-browser comparisons begin to be classified as different-browsers comparisons.
      The equal error rate, which is the rate where both the FMR and the FNMR are equal, is of $0.61$\% and is achieved for $232$ identical attributes.
      Although the verification mechanism does not have a perfect accuracy of $100$\%, this is acceptable.
      Indeed, a user getting his browser unrecognized can undergo the fallback authentication process (see Section~\ref{sec:browser-fingerprinting-based-authentication-mechanism}).
      Moreover, we consider the use of browser fingerprinting as an additional authentication factor, hence the other factors can prevent a falsely recognized browser.
      Both these events are expected to rarely occur as can be seen by the low equal error rate.

      These results are tied to the distinctiveness and the stability of the fingerprints.
      Indeed, as more than $94.7$\% of the fingerprints are shared by less than $8$~browsers, two random fingerprints have little chances to match.
      Moreover, more than $244$ attributes ($96.64$\%) are identical between the consecutive fingerprints of a browser\footnote{
        On average, more than $90$\% of the attributes stay identical between the consecutive fingerprints, but these attributes are not always the same.
        We do not restrict the pairs of consecutive fingerprints to the subset of the identical attributes as these attributes differ between these pairs.
      }, on average and when separated by less than $31$~days.
      This is consistent with the $248.64$~identical attributes on average among the same-browser comparisons\footnote{
        The difference is due to the fingerprints of the same-browser comparisons belonging to the same month, whereas the consecutive fingerprints can overlap over two consecutive months.
      }.

  \subsection{Conclusion}
    About the distinctiveness, and considering the time-partitioned datasets, our fingerprints provide a unicity rate of more than $81.3$\% which is stable on the long-run.
    Moreover, more than $94.7$\% of our fingerprints are shared by at most $8$~browsers.
    About the stability, and on average, a fingerprint has more than $91$\% of its attributes that stay identical between two observations, even when they are separated by nearly $6$~months.
    About the performance, the majority ($95$\%) of our fingerprints weigh less than $12$~kilobytes and are collected in less than $10.42$~seconds.
    We do not remark any significant loss in the properties offered by our fingerprints through the $6$~months of our experiment.
    However, we fall to the same conclusion as previous studies~\cite{ECK10, SPJ15, LRB16, GLB18} about the fingerprints of mobile browsers lacking distinctiveness.
    Their unicity rate in the time-partitioned datasets falls down to approximately $42$\%.
    We remark that, in our dataset, the consecutive fingerprints of a browser have at least $234$ identical attributes, whereas the majority of the fingerprints of different browsers have fewer.
    This results in our simple verification mechanism achieving an equal error rate of $0.61$\%.

%% file: 6-statistical-analysis.tex
\section{Attribute-wise analysis}
\label{sec:attribute-wise-analysis}
  In this section, we discuss the contribution of the attributes to the properties of the fingerprints.
  Then, we show that most of the attributes are correlated with at least another one (i.e., they provide less than $1$~bit of entropy when the other one is known).
  Finally, we focus on the properties of the dynamic attributes (e.g., their collection time).
  We refer the reader to Appendix~\ref{app:attributes} for more information about the implementation of each attribute.

  \subsection{Contribution of particular attributes}
    In this section, we discuss the contribution of the attributes to the fingerprint properties of distinctiveness, stability, and performance.
    We express the stability of the attributes as the \emph{sameness rate}, which is the proportion of the consecutive fingerprints where the value of the attribute stays identical.
    Appendix~\ref{app:attribute-list-and-property} provides the exhaustive list of the attributes and their properties.

    \subsubsection{Attributes distinct values}
\begin{figure}
        \centering
        \includegraphics[width=0.70\columnwidth]{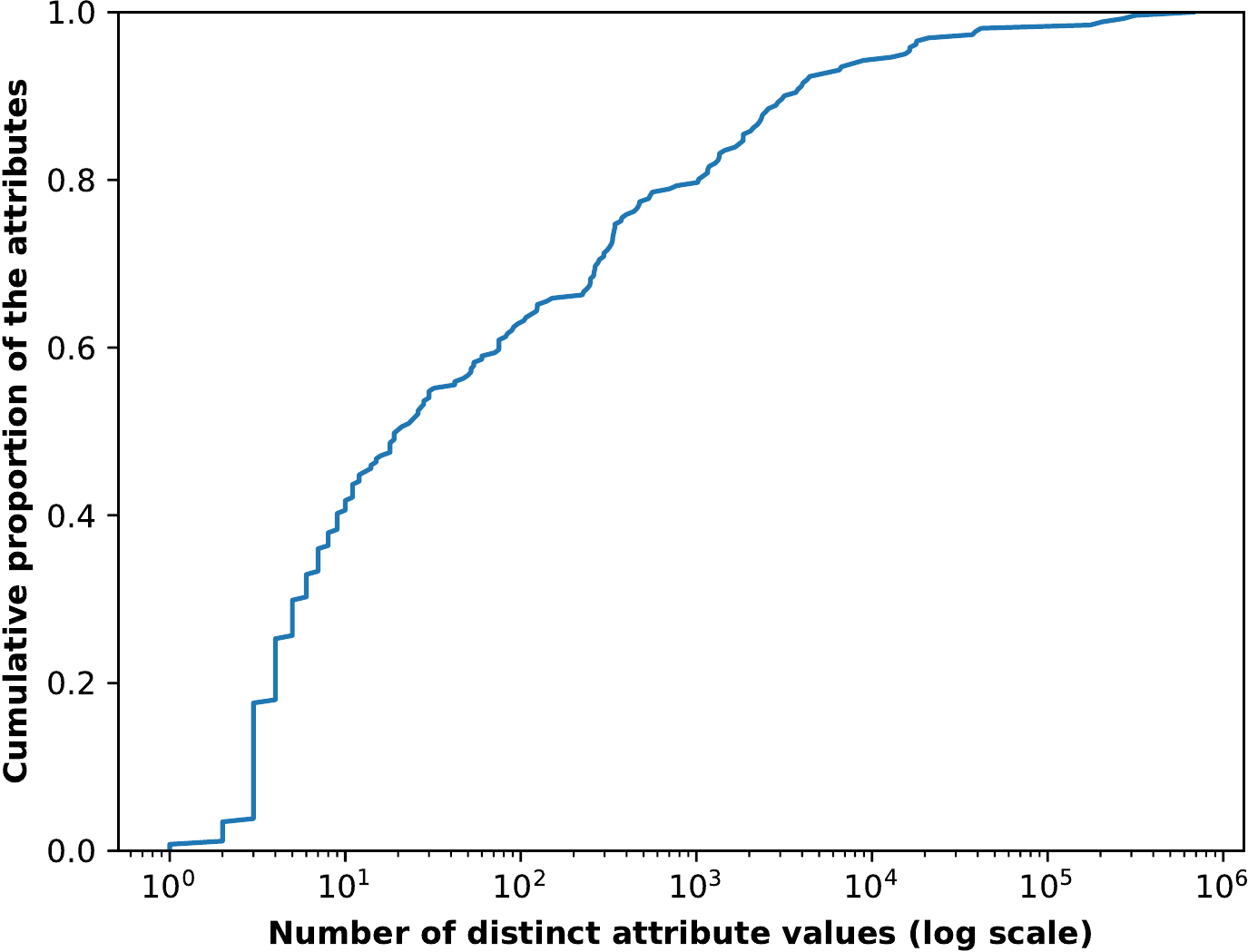}
        \caption{
          Cumulative distribution of the number of distinct values of the attributes in logarithmic scale.
        }
        \label{fig:cdf-distinct-values}
        \Description[
          Cumulative distribution of the number of distinct values of the attributes in logarithmic scale.
        ]{
          Cumulative distribution of the number of distinct values of the attributes in logarithmic scale.
          We have $42$\% of the attributes that have at most $10$ distinct values, $63$\% that have at most $100$ distinct values, and $79$\% that have at most $1,000$ distinct values.
          Only $5$~attributes have more than $100,000$ distinct values.
        }
      \end{figure}

      Figure~\ref{fig:cdf-distinct-values} depicts the cumulative distribution of the number of distinct values among the attributes.
      We have $42$\% of the attributes that have at most $10$ distinct values, $63$\% that have at most $100$ distinct values, and $79$\% that have at most $1,000$ distinct values.
      Only $5$~attributes have more than $100,000$ distinct values.
      They are the WebRTC fingerprinting method ($671,254$ values), the list of plugins ($314,518$ values), the custom canvas in the PNG format ($269,874$ values) and in the JPEG format ($205,005$ values), together with the list of mime types ($174,876$ values).

      The values of the attributes are of different nature, impacting the distinct values that can be observed among different population or time ranges.
      Some attributes have a \emph{fixed number of values}, like the Boolean attributes or the categorical attributes that have a fixed set of possibilities.
      Other attributes are composed of \emph{elements having a fixed set} of possibilities, but their combination provides a high number of values.
      It is the case for the attributes that are related to languages that typically are a combination of language identifiers (e.g., fr) that can be weighted (e.g., ${q=0.80}$).
      These attributes also comprise the list of fonts that is composed of Boolean values that indicates the presence of a font.
      Other attributes are \emph{integers} or \emph{floating-point numbers}, resulting in a large set of possible values.
      This category comprises any size-related attribute (e.g., the screen width and height) and our audio fingerprinting method which value is a floating-point number with a high precision of more than $8$~digits.
      Other attributes are \emph{textual information} that can include version numbers, which results in a high number of distinct values.
      Moreover, as new values appear through time (e.g., new versions, new software components), the set of the observed values is expected to grow over the observations.
      Examples are the UserAgent or the list of plugins which are composed of the name and version of hardware and software components.

      \begin{table}
        \begin{minipage}{\textwidth}
          \caption{
            Comparison of the distinct attribute values between the studies AmIUnique (AIU), Hiding in the Crowd (HitC), Who Touched My Browser Fingerprint (WTMBF), and this study.
            The attributes are composed of the five attributes showing the most distinct values for this study and the attributes reported by the previous studies that are in this study.
            The attributes are ranked according to this study.
            The symbol - denotes missing information and * denotes the attributes that are not collected exactly the same way (e.g., using different set of instructions or methods) among the studies.
          }
          \label{tab:distinct-attributes-comparison}
          \centering
          \begin{tabular}{lcccc}
            \toprule
              Attribute                 & AIU~\cite{LRB16} & HitC~\cite{GLB18}
                                        & WTMBF~\cite{LC20}
                                        & \textbf{This study}                 \\
            \midrule
              WebRTC                    & -       & -       & -
                                        & \textbf{671,254}                    \\
              List of plugins*          & 47,057  & 288,740 & 16,633
                                        & \textbf{314,518}                    \\
              Canvas (PNG)*             & 8,375   & 78,037  & 14,006
                                        & \textbf{269,874}                    \\
              Canvas (JPEG)             & -       & -       & -
                                        & \textbf{205,005}                    \\
              List of mime types        &  -      & -       & -
                                        & \textbf{174,876}                    \\
              User agent                & 11,237  & 19,775  & 41,060
                                        & \textbf{20,961}                     \\
              List of fonts*            & 36,202  & 17,372  & 115,128
                                        & \textbf{17,960}                     \\
              List of HTTP headers*     & 1,182   & 610     & 344
                                        & \textbf{5,394}                      \\
              WebGL renderer (unmasked) & 1,732   & 3,691   & 5,747
                                        & \textbf{3,786}                      \\
              Content language          & 4,694   & 2,739   & 14,214
                                        & \textbf{2,833}                      \\
              Pixel ratio               & -       & -       & 1,930
                                        & \textbf{2,035}                      \\
              WebGL canvas*             & -       & -       & 4,152
                                        & \textbf{1,158}                      \\
              CPU cores                 & -       & -       & 29
                                        & \textbf{60}                         \\
              Timezone                  & 55      & 60      & 38
                                        & \textbf{60}                         \\
              Platform                  & 187     & 32      & -
                                        & \textbf{32}                         \\
              Content encoding          & 42      & 30      & 26
                                        & \textbf{30}                         \\
              WebGL vendor (unmasked)   & 26      & 27      & 26
                                        & \textbf{27}                         \\
              Accept                    & 131     & 24      & 9
                                        & \textbf{26}                         \\
              Do Not Track              & 7       & 3       & -
                                        & \textbf{11}                         \\
              AdBlock*                  & 2       & 2       & -
                                        & \textbf{19}                         \\
              Color depth               & -       & -       & 6
                                        & \textbf{14}                         \\
              CPU class                 & -       & -       & 5
                                        & \textbf{6}                          \\
              Use of local storage      & 2       & 2       & 2
                                        & \textbf{4}                          \\
              Use of session storage    & 2      & 2       & -
                                        & \textbf{4}                          \\
              Cookies enabled           & 2       & 1       & 2
                                        & \textbf{1}                          \\
            \midrule
              Fingerprints              & 118,934    & 2,067,942    & 7,246,618
                                        & \textbf{4,145,408}                  \\
              Distinct fingerprints     & 142,023\footnote{
                This number is displayed in Figure~$11$ of~\cite{LRB16} as the number of distinct fingerprints, but it also corresponds to the number of raw fingerprints.
                Every fingerprint would be unique if the number of distinct and collected fingerprints are equal.
                Hence, we are not confident in this number, but it is the number provided by the authors.
              }                         & -            & 1,586,719
                                        & \textbf{3,578,196}                  \\
            \bottomrule
          \end{tabular}
        \end{minipage}
      \end{table}

      Table~\ref{tab:distinct-attributes-comparison} compares the number of distinct attributes of our dataset with the numbers reported by the studies AmIUnique~\cite{LRB16}, Hiding in the Crowd~\cite{GLB18}, and Who Touched My Browser Fingerprint~\cite{LC20}.
      We emphasize that the more fingerprints are observed, the more chances there is to observe new attribute values.
      We find two attributes missing from previous studies that show a high number of distinct values: the WebRTC fingerprinting method which includes numerous information about an instance of a current WebRTC connection, and the list of mime types.

      Three attributes have a higher number of distinct values in our study than in the comparison studies~\cite{LRB16, GLB18, LC20}: the list of plugins, the canvas, and the list of HTTP headers.
      For the list of plugins, this increase can be explained by the higher number of observed fingerprints compared to~\cite{LRB16} and by the higher proportion of desktop browsers\footnote{
        Due to a lack of personalization, the browsers that run on mobile devices tend to have fewer plugins than those that run on desktops or laptops~\cite{LRB16, GLB18}.
      } compared to~\cite{LC20}.
      The increase of distinct values of the canvas is due to the larger set of instructions that we use for the custom canvas compared to the comparison studies (see Section~\ref{sec:focus-dynamic-attributes}).
      The increase of distinct values of the HTTP headers list is explained by the comparison studies storing the name of the headers, whereas we store both the name and the value of the headers that we do not store in a dedicated attribute.
      Other attributes have a higher number of distinct attributes because we store a flag that indicates the type of error when one is encountered (e.g., the value was not collected in time, its collection raised an exception).
      For example, the support of the local storage has $4$ distinct values instead of $2$ for the previous studies~\cite{LRB16, GLB18}.
      Moreover, for the detection of an ad-blocker, we verify several modifications that a potential ad-blocker would make on a dummy advertisement (e.g., set its visibility to hidden) and store a list of whether each modification was done.

      Four attributes have more distinct values in the study of Laperdrix et al.~\cite{LRB16}: the list of fonts, the platform, and three HTTP headers that are accept, content language, and content encoding.
      The increase for the list of fonts is due to their usage of the Flash plugin that directly accesses the list of installed fonts.
      Due to the deprecation of plugins, we infer the presence of $66$ fonts through the size of text boxes~\cite{FE15}.
      The increase for the four other attributes can be explained by the diversity of the configurations of their browser population and by users modifying their browser fingerprint (e.g., they found users modifying the platform attribute and counted $5,426$ inconsistent fingerprints).

      Four attributes have more distinct values in the study of Song Li and Yinzhi Cao~\cite{LC20}: the user agent, the list of fonts, the content language, and the WebGL canvas.
      The increase of distinct values of the user agent can be explained by their higher proportion of mobile browsers, which tend to have more information in the user agent (e.g., the device model~\cite{LRB16}).
      The increase for the list of fonts can be explained by their higher number of fonts that are detected, which is of $96$ fonts compared to $66$ in our study.
      As for the content language, they collect their fingerprints from a European website and observe connections coming from $226$ countries.
      Although they remark that some connections are made through proxies, we can still suppose that the user population span over several countries.
      As a result, their dataset contains more diverse language configurations than our dataset which was collected from a French website.
      The increase of distinct values of the WebGL canvas can be explained by their set of instructions being more complex than ours (see Section~\ref{sec:focus-dynamic-attributes}).
      Unfortunately, they do not provide any example of generated images or description of the instructions they used.

    \begin{figure}
      \centering
      \includegraphics[width=0.70\columnwidth]{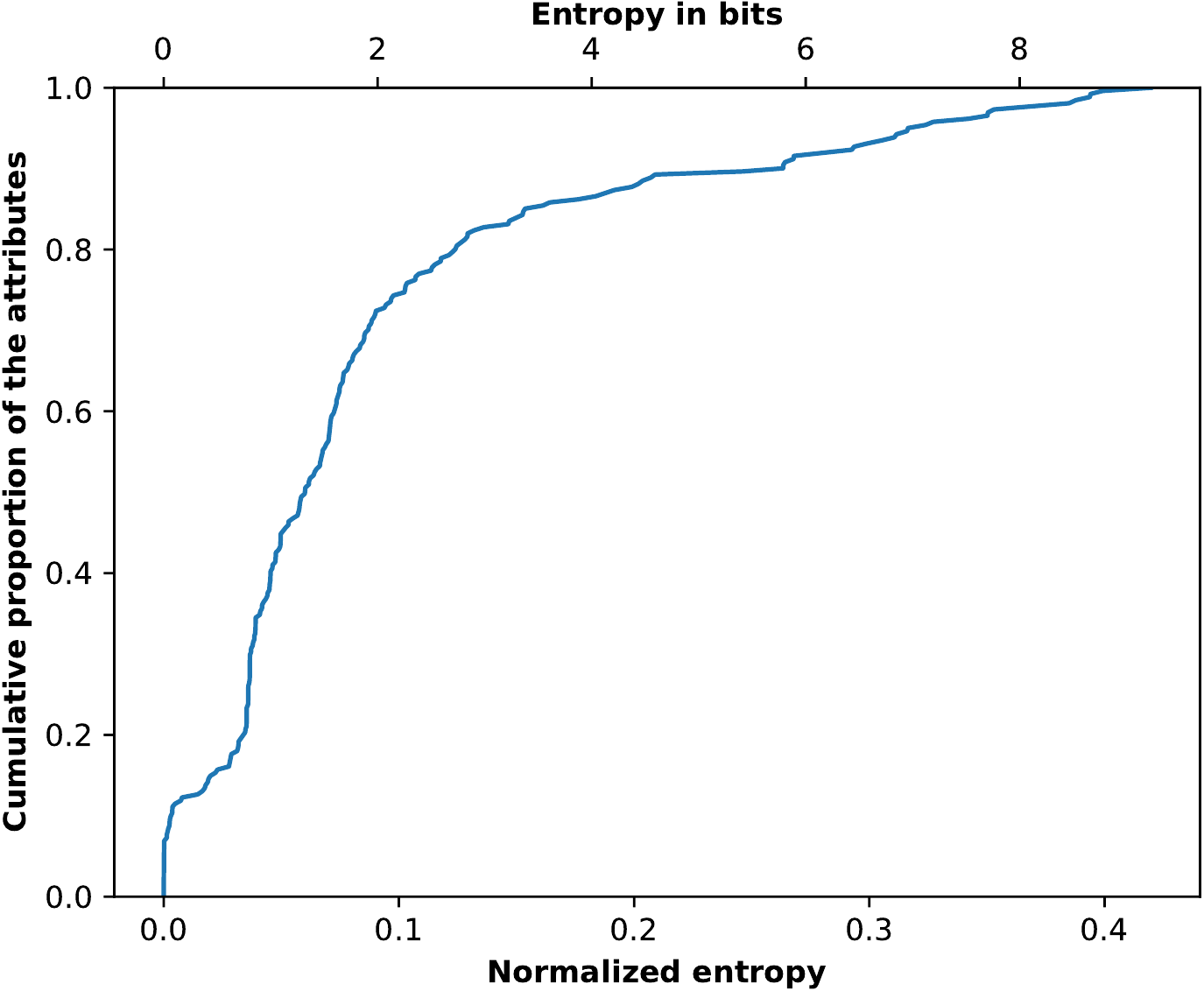}
      \caption{
        Cumulative distribution of the normalized entropy and entropy (in bits) among the attributes.
      }
      \label{fig:cdf-normalized-entropy}
      \Description[
        Cumulative distribution of the normalized entropy and entropy (in bits) among the attributes.
      ]{
        Cumulative distribution of the normalized entropy and entropy (in bits) among the attributes.
        We have $10$\% of the attributes that provide a low normalized entropy of less than~$0.003$, and another $10$\% that provide a normalized entropy between $0.25$ and $0.42$.
        The majority of the attributes ($80$\%) provide a normalized entropy comprised between $0.003$ and $0.25$.
      }
    \end{figure}

    \subsubsection{Attributes distinctiveness}
    \label{sec:attributes-distinctiveness}
      We measure the distinctiveness of the attributes as the \emph{normalized entropy} for comparability with previous studies.
      The normalized entropy was proposed by Laperdrix et al.~\cite{LRB16} to cope with the problem of comparing the entropy of attributes between fingerprint datasets of dissimilar sizes.
      The normalized entropy ${H_n(X)}$ is defined as the ratio of the entropy~${H(X)}$ of the attribute to the maximum entropy ${H_M = \log_2(N)}$, with $N$ being the number of fingerprints.
      The entropy can be retrieved as ${H(X) = H_n(X) * H_M}$.
      In our case, ${H_M = 21.983}$~bits and an entropy of $1$~bit is equivalent to a normalized entropy of $0.045$.

      Figure~\ref{fig:cdf-normalized-entropy} displays the cumulative distribution of the normalized entropy and entropy (in bits) among the attributes.
      We have $10$\% of the attributes that provide a low normalized entropy of less than~$0.003$, and another $10$\% that provide a normalized entropy between $0.25$ and $0.42$.
      The majority of the attributes ($80$\%) provide a normalized entropy comprised between $0.003$ and $0.25$.
      The most distinctive attributes of previous studies~\cite{ECK10, LRB16} also belong to the most distinctive attributes of our study.
      The three \emph{canvases} are among the $10$ most distinctive attributes.
      Our designed canvas in PNG has a normalized entropy of $0.420$, the canvas similar to the canvas presented in the Morellian study~\cite{LABN19} has a normalized entropy of $0.385$, and the canvas inspired by the AmIUnique study~\cite{LRB16} has a normalized entropy of $0.353$.
      The \emph{userAgent} collected from the JavaScript property is more distinctive than its HTTP header counterpart, as they respectively have a normalized entropy of $0.394$ and $0.350$.
      Finally, the \emph{list attributes} are also highly distinctive.
      The list of plugins has a normalized entropy of $0.394$, the list of supported mime types has a normalized entropy of $0.311$, and the list of fonts has a normalized entropy of $0.305$.

      \begin{table}
        \caption{
          Comparison of the normalized entropy between the studies Panopticlick (PTC), AmIUnique (AIU), Hiding in the Crowd (HitC), and this study.
          The attributes are ranked from the most to the least distinctive in this study.
          The symbol - denotes missing information and * denotes the attributes that are not collected exactly the same way (e.g., using different set of instructions or methods) among the studies.
        }
        \label{tab:normalized-entropy-comparison}
        \centering
        \begin{tabular}{lcccc}
          \toprule
            Attribute                 & PTC~\cite{ECK10} & AIU~\cite{LRB16}
                                      & HitC~\cite{GLB18}
                                      & \textbf{This study}                   \\
          \midrule
            Canvas (PNG)*             & -       & 0.491   & 0.407
                                      & \textbf{0.420}                        \\
            Canvas (JPEG)             & -       & -       & 0.391
                                      & \textbf{0.399}                        \\
            List of plugins*          & 0.817   & 0.656   & 0.452
                                      & \textbf{0.394}                        \\
            User agent                & 0.531   & 0.580   & 0.341
                                      & \textbf{0.350}                        \\
            List of fonts*            & 0.738   & 0.497   & 0.329
                                      & \textbf{0.305}                        \\
            WebGL renderer (unmasked) & -       & 0.202   & 0.264
                                      & \textbf{0.268}                        \\
            Content language          & -       & 0.351   & 0.129
                                      & \textbf{0.124}                        \\
            WebGL vendor (unmasked)   & -       & 0.127   & 0.109
                                      & \textbf{0.115}                        \\
            List of HTTP headers*     & -       & 0.249   & 0.085
                                      & \textbf{0.095}                        \\
            Do Not Track              & -       & 0.056   & 0.091
                                      & \textbf{0.085}                        \\
            Platform                  & -       & 0.137   & 0.057
                                      & \textbf{0.068}                        \\
            Accept                    & -       & 0.082   & 0.035
                                      & \textbf{0.028}                        \\
            Content encoding          & -       & 0.091   & 0.018
                                      & \textbf{0.019}                        \\
            Timezone                  & 0.161   & 0.198   & 0.008
                                      & \textbf{0.008}                        \\
            AdBlock*                  & -       & 0.059   & 0.002
                                      & \textbf{0.002}                        \\
            Use of local storage      & -       & 0.024   & 0.002
                                      & \textbf{0.001}                        \\
            Use of session storage    & -      & 0.024   & 0.002
                                      & \textbf{0.000}                        \\
            Cookies enabled           & 0.019   & 0.015   & 0.000
                                      & \textbf{0.000}                        \\
          \midrule
            Maximum entropy $H_M$     & 18.843  & 16.859  & 20.980
                                      & \textbf{21.983}                       \\
            Fingerprints              & 470,161 & 118,934 & 2,067,942
                                      & \textbf{4,145,408}                    \\
          \bottomrule
        \end{tabular}
      \end{table}

      Table~\ref{tab:normalized-entropy-comparison} compares the normalized entropy between the studies Panopticlick~\cite{ECK10}, AmIUnique~\cite{LRB16}, Hiding in the Crowd~\cite{GLB18}, and this study.
      The attributes are ranked from the most to the least distinctive in this study.
      The normalized entropy of our attributes are lower than what is reported in these studies, which can be explained by the following factors.
      First, the maximum entropy $H_M$ increases with the number of fingerprints.
      However, an attribute that has $n$~possibilities (e.g., a Boolean attribute has only two possible values) have a normalized entropy of at most ${\log_2(n)}$.
      As the number~$N$ of fingerprints increases, the normalized entropy decreases due to the entropy of the attribute being capped at ${\log_2(n)}$ whereas the ratio is to ${\log_2(N)}$.
      Second, contextual attributes are biased towards the French population (see Section~\ref{sec:browser-population-bias}) and provide a lower normalized entropy.
      For example, the time zone and the \texttt{Accept-Language} HTTP header (named content language in Table~\ref{tab:normalized-entropy-comparison}) provide a respective normalized entropy of $0.008$ and $0.124$, against $0.198$ and $0.351$ for the AmIUnique study.
      The third reason is the evolution of the web technologies since the Panopticlick and the AmIUnique study.
      For example, the list of plugins is less distinctive due to the replacement of plugins by HTML5 functionalities or extensions~\cite{EndOfFlash}.
      Another example is the list of fonts that was collected through plugins~\cite{ECK10, LRB16}, but now has to be inferred from the size of text boxes~\cite{FE15}.
      Although the $17$ attributes in common have a lower normalized entropy, we supplement them with more than a hundred attributes, resulting in the fingerprints showing a unicity rate above $80$\%.

      Interestingly, four attributes unreported by the previous large-scale studies~\cite{ECK10, LRB16, GLB18} are found to be highly distinctive.
      The \texttt{innerHeight} and \texttt{outerHeight} properties of the \texttt{window} JavaScript object, mentioned by~\cite{SLG19, VRRB20} but without any distinctiveness measure, have a respective normalized entropy of $0.388$ and $0.327$.
      The size of bounding boxes was used by~\cite{FE15} as a method of font fingerprinting.
      From the entropy reported in~\cite{FE15}, we obtain a normalized entropy of $0.761$ against $0.369$ in our dataset.
      To the best of our knowledge, no previous study use the width and the position of a newly created \texttt{div} element as a fingerprinting attribute.
      However, it is highly distinctive as it achieves a normalized entropy of $0.324$ in our dataset.
      All the mentioned attributes provide a sameness rate above $90$\%, except for the size of bounding boxes that goes down to $47$\%.
      However, when looking at its extracted parts, we observe that they have a sameness rate higher than $90$\% at the exception of the height of the first bounding box that has a sameness rate of $49$\%.
      This illustrates the necessity to break down some attributes to parts, as removing this part from the original attribute would drastically increase its sameness rate.

    \begin{figure}
      \centering
      \includegraphics[width=0.70\columnwidth]{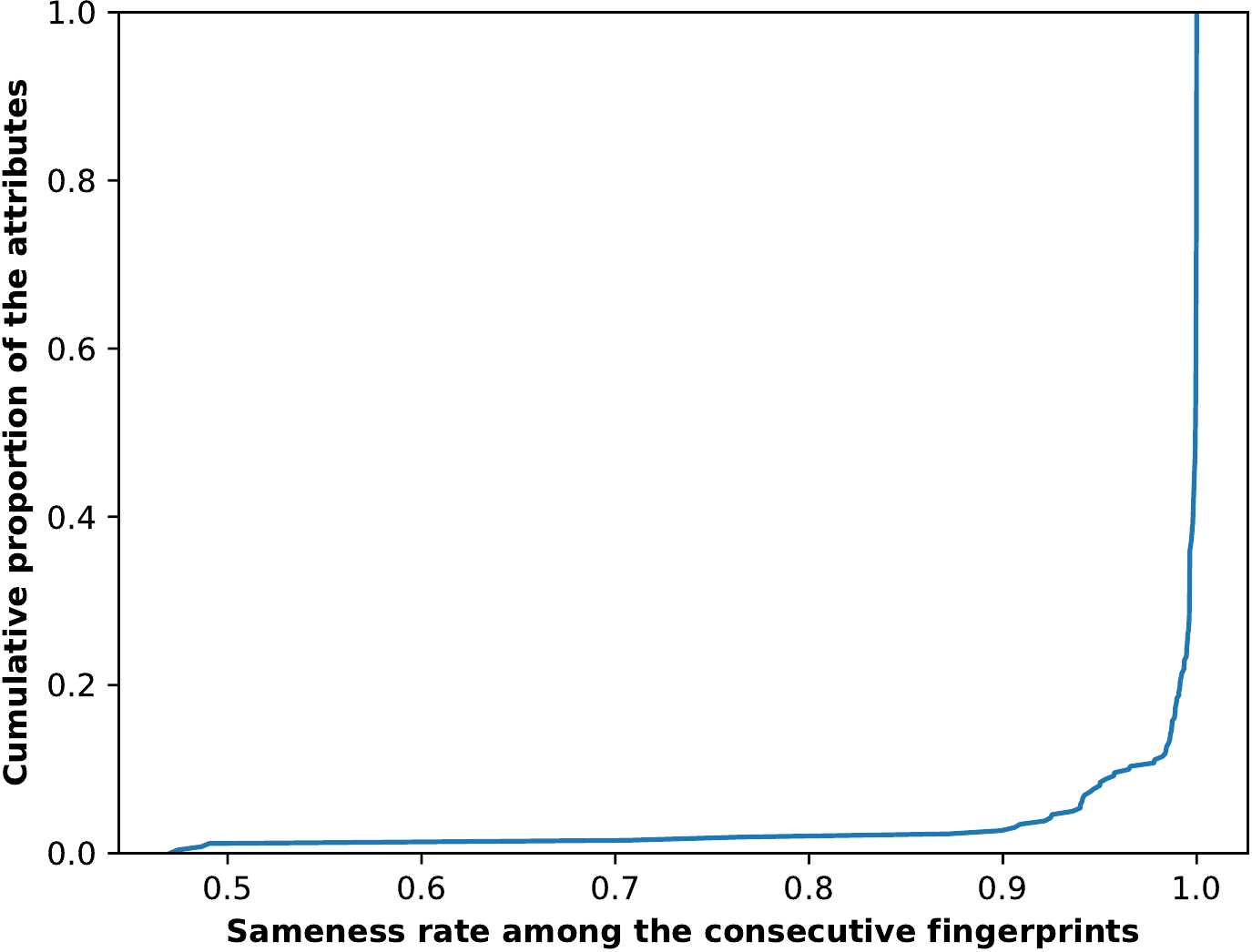}
      \caption{
        Cumulative distribution of the sameness rate of the attributes among the consecutive fingerprints.
      }
      \label{fig:cdf-sameness-rates}
      \Description[
        Cumulative distribution of the sameness rate of the attributes among the consecutive fingerprints.
      ]{
        Cumulative distribution of the sameness rate of the attributes among the consecutive fingerprints.
        Only $6$~attributes ($2.29$\%) provide a sameness rate below $85$\%, $5.7$\% of the attributes provide a sameness rate comprised in $[85, 95]$\%, $10.7$\% provide a sameness rate comprised in $[95, 99]$\%, and more than $80$\% of the attributes have a sameness rate above $99$\%.
      }
    \end{figure}

    \subsubsection{Attributes sameness rate}
    \label{sec:attributes-stability}
      Figure~\ref{fig:cdf-sameness-rates} displays the cumulative distribution of the sameness rate among the attributes.
      Only $6$~attributes ($2.29$\%) provide a sameness rate below $85$\%, $5.7$\% of the attributes provide a sameness rate comprised in $[85, 95]$\%, $10.7$\% provide a sameness rate comprised in $[95, 99]$\%, and more than $80$\% of the attributes have a sameness rate above $99$\%.
      Among the $3,725,373$ pairs of consecutive fingerprints, only $5$ attributes never change (i.e., they show a sameness rate of $100$\%).
      They comprise whether cookies are enabled, and four non-standard HTTP headers~\cite{UsefulXHeaders} that are \texttt{X-ATT-DeviceId}, \texttt{X-UIDH}, \texttt{X-WAP-Profile}, and \texttt{X-Network-Info}.
      These headers are typically added by mobile network operators to the communications of mobile devices~\cite{VSKP15}.
      The cookie enabling state is stable because the working dataset only contains fingerprints that have this functionality activated (see Section~\ref{sec:data-preprocessing}).
      As for the non-standard headers, the two first headers are always missing from the headers.
      The \texttt{X-Network-Info} header has three non-missing values that were seen on a single browser each.
      The \texttt{X-WAP-Profile} has three values that were seen on $37$ browsers and consist of URL links directing to an XML document that describes the device.
      The most common link is shared by $35$ browsers and concerns the Samsung B550H mobile phone~\cite{SamsungWAP}.

      The attributes that have a sameness rate below $85$\% are the size of bounding boxes, three extracted attributes derived from the bounding boxes, and two other attributes that we describe here.
      The instability of the size of bounding boxes is already explained in Section~\ref{sec:attributes-distinctiveness}.
      The \texttt{Cache-Control} HTTP header allows the browser to specify the cache policy used during requests.
      It is the second most instable attribute with a sameness rate of $70.63$\%.
      This header is instable because it is not always sent and some of its values contain the \texttt{max-age} parameter that can vary between requests.
      The third most instable attribute is the WebRTC fingerprinting method that has a sameness rate of $76.46$\%.
      This is due to three factors.
      First, the experimental state of this attribute and the bad support of the WebRTC API at the time of the experiment~\cite{AL17, CanIUseCreateDataChannel} results in it being unsupported, undefined, or erroneous for $75.23$\% of the observed fingerprints.
      Then, it contains local IP addresses\footnote{
        Our script hashes these local IP addresses in MD5 directly on the clients, see~\cite{TSYI15} for more information about how the WebRTC API gives access to this information.
      } which can change between two observations.
      Finally, it is composed by numerous pieces of information about an instance of a WebRTC connection, hence the change of any information modifies the value of the whole attribute.
      We reduce this effect by collecting these pieces of information from two connections and only keeping the identical values.

    \subsubsection{Attributes collection time}
    \label{sec:attributes-collection-time}
      Most of the attributes are HTTP headers or JavaScript properties that are collected sequentially and in a negligible amount of time.
      We call \emph{asynchronous attributes} the attributes that are collected asynchronously (i.e., in parallel).
      Only $33$ attributes have a median collection time~(MCT) higher than $5$ms.
      They are presented in Figure~\ref{fig:attributes-collection-time} and are thoroughly described in Appendix~\ref{app:attributes}.
      All these attributes are asynchronous at the exception of the canvases, hence the total collection time of the fingerprint is not their sum.
      These attributes can be separated in three classes: extension detection, browser component, and media related.
      Their high collection time can be explained by these attributes waiting for the web page to render, executing heavy processes (e.g., graphical computation), or being asynchronous attributes that are affected by the side effect described in Appendix~\ref{app:anomalous-collection-times}.
      We consider the fingerprints that take more than $30$~seconds to collect as outliers.
      We configured our fingerprinting script to collect the attributes in a given amount of time, after which the fingerprint is sent without waiting for all the attributes.
      In such case, the missing attribute do not have a collection time, and we ignore their collection time in the presented results.
      For all the attributes except five asynchronous attributes, these two types of outliers account for less than $1$\% of the overall, desktop, and mobile entries.
      Appendix~\ref{app:anomalous-collection-times} discusses these two types of outliers.

      \begin{figure}
        \centering
        \includegraphics[width=0.70\columnwidth]{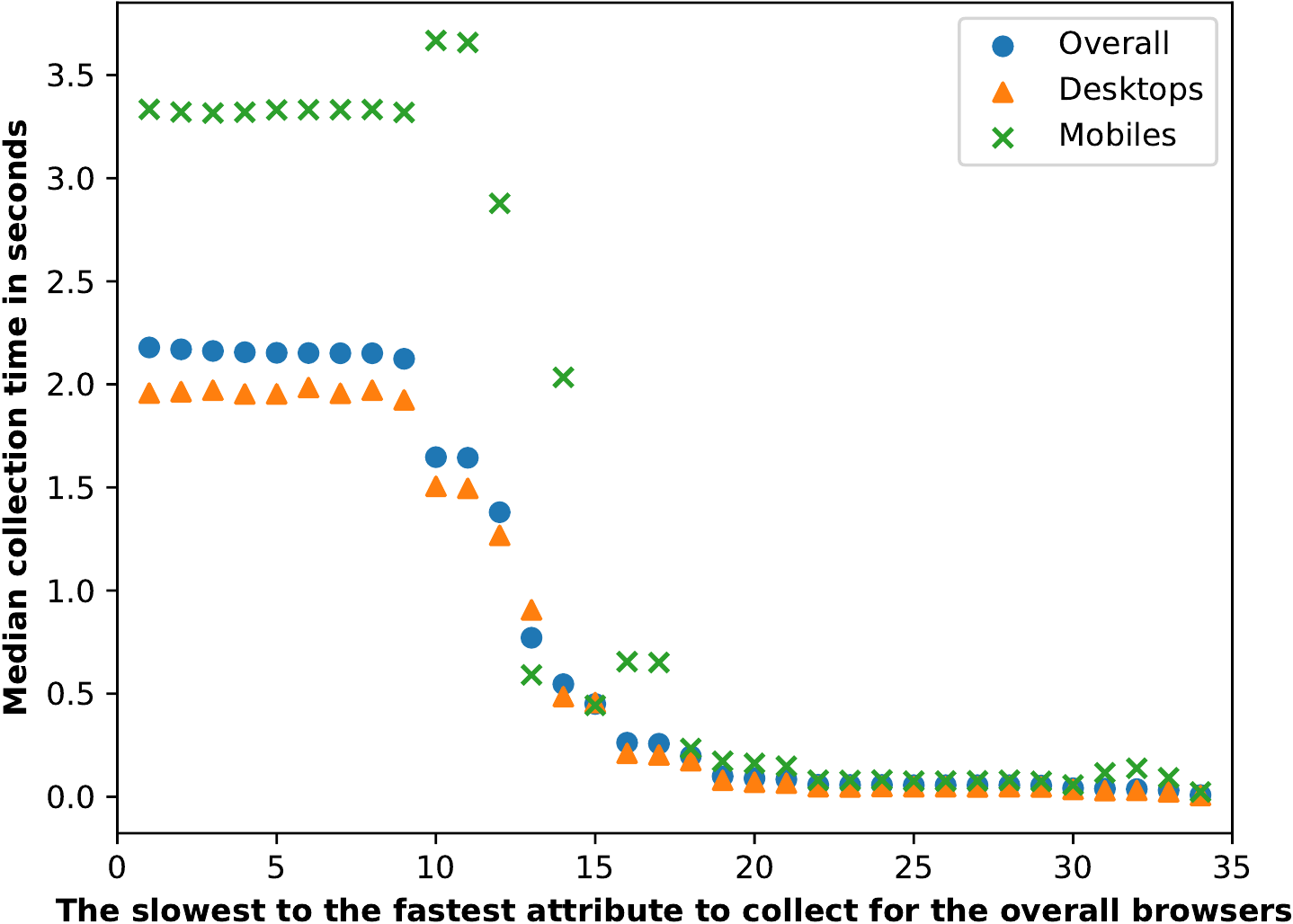}
        \caption{
          Median collection time in seconds of the $33$~attributes that have a median collection time higher than $5$ms.
          The attributes are ranked from the slowest to the fastest to collect for the overall browsers.
          We consider the fingerprints that take more than $30$~seconds to collect as outliers.
          The attributes that were not collected in time are not counted in these results.
          These two types of outliers account for less than $1$\% of the overall, desktop, and mobile entries, at the exception of the five asynchronous attributes discussed in Appendix~\ref{app:anomalous-collection-times}.
        }
        \label{fig:attributes-collection-time}
        \Description[
          Median collection time in seconds of the $33$~attributes that have a median collection time higher than $5$ms.
        ]{
          Median collection time in seconds of the $33$~attributes that have a median collection time higher than $5$ms.
          The attributes are ranked from the slowest to the fastest to collect for the overall browsers.
          We consider the fingerprints that take more than $30$~seconds to collect as outliers.
          The attributes that were not collected in time are not counted in these results.
          These two types of outliers account for less than $1$\% of the overall, desktop, and mobile entries, at the exception of the five asynchronous attributes discussed in Appendix~\ref{app:anomalous-collection-times}.
        }
      \end{figure}

      The first class of the attributes that take time to collect are the methods of \emph{extension detection}.
      All these attributes are asynchronous.
      The $9$ slowest attributes detect an extension by the changes it brings to the content of the web page~\cite{SN17}.
      They have a median collection time~(MCT) of approximately $2.2$~seconds due to the waiting time before checking the changes on the web page.
      There is a clear difference between desktop and mobile browsers that show a respective MCT of approximately $2$~seconds against $3.3$~seconds.
      The $22$nd to the $29$th slowest attributes detect an extension based on web-accessible resources~\cite{SVS17}, which consist in checking whether an image embarked in an extension is accessible or not.
      They have a MCT of around $56$ms, which is lower than the method that relies on detecting changes brought to the web page.

      The second class of the attributes that take time to collect infer the availability of \emph{browser components}.
      All these attributes are asynchronous.
      The list of speech synthesis voices is ranked $14$th with a MCT of $546$ms.
      This is due to this attribute being a list and the collection method that, for some browsers, requires to be done during an \texttt{onvoiceschanged} event which takes time to be triggered.
      The list of fonts and the inference of the default font are ranked $15$th and $19$th with a respective MCT of $450$ms and $99$ms.
      This is due to the detection method that measures the size of newly created text boxes~\cite{FE15}.
      The size of bounding boxes, the browser component colors, and the width and position of a newly created \texttt{div} element are ranked $18$th, $20$th, and $21$st, with a MCT ranging from $84$ms to $197$ms.
      This is due to their waiting for and manipulation of the web page that takes time.
      The WebRTC fingerprinting method is ranked $13$th with a MCT of $771$ms.
      This is due to the creation of two WebRTC connections that are needed to gather information about the WebRTC configuration.

      The third class of the attributes that take time to collect are those that generate a media file (e.g., a sound, an image) and are discussed in Section~\ref{sec:focus-dynamic-attributes}.
      The methods of advanced audio fingerprinting are asynchronous and ranked $10$th, $11$th, and $12$th.
      Our designed canvases are ranked $16$th and $17$th.
      The canvases inspired by the AmIUnique study~\cite{LRB16} are ranked $31$st and $33$rd.
      The canvas similar to the canvas of the Morellian study~\cite{LABN19} is ranked $32$nd.

    \begin{figure}
      \centering
      \includegraphics[width=0.70\columnwidth]{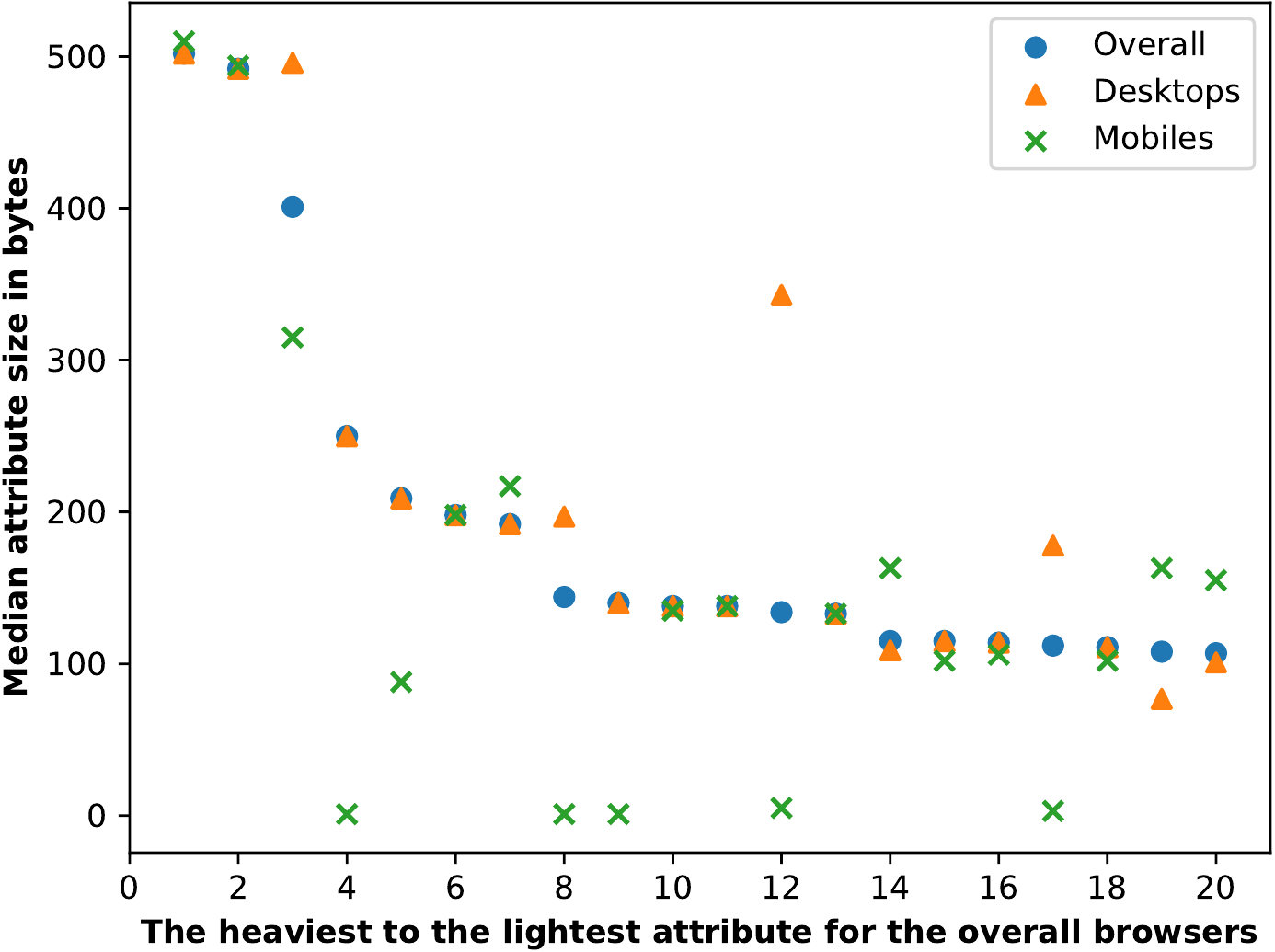}
      \caption{
        Median size of the $20$~heaviest attributes in bytes.
        The attributes are ranked from the heaviest to the lightest for the overall browsers.
      }
      \label{fig:attributes-size}
      \Description[
        Median size of the $20$~heaviest attributes in bytes.
      ]{
        Median size of the $20$~heaviest attributes in bytes.
        The attributes are ranked from the heaviest to the lightest for the overall browsers.
        We have $137$~attributes with a MS below $5$~bytes, $105$~attributes with a MS between $5$~bytes and $100$~bytes, and $20$~attributes with a MS above $100$~bytes.
      }
    \end{figure}

    \subsubsection{Attributes size}
    \label{sec:attributes-size}
      Most of our attributes have a negligible median size~(MS).
      We have $137$~attributes with a MS below $5$~bytes, $105$~attributes with a MS between $5$~bytes and $100$~bytes, and $20$~attributes with a MS above $100$~bytes.
      Figure~\ref{fig:attributes-size} displays the median size of the $20$~heaviest attributes in bytes, ranked from the heaviest to the lightest attribute for the overall browsers.
      We discuss below these $20$ heaviest attributes.

      The heaviest attributes are composed of \emph{list attributes} and \emph{verbose textual attributes}.
      The three heaviest attributes are list attributes.
      The list of the properties of the \texttt{navigator} object is the heaviest attribute with a MS of $502$~bytes, the list of colors of layout components is second with a MS of $492$~bytes, and the list of WebGL extensions is third with a MS of $401$~bytes.
      Examples of verbose textual attributes are the \texttt{appVersion} that is $18$th with a MS of $107$~bytes, and the \texttt{userAgent} JavaScript property that is $14$th with a MS of $115$~bytes, which is more verbose than its HTTP header counterpart that is $19$th with a MS of $108$~bytes.

      On mobile browsers, some list attributes are most of the time empty due to their lack of customization (e.g., plugins are mostly unsupported).
      They are the list of speech synthesis voices that is $4$th, the list of the constraints supported by the \texttt{mediaDevices} object that is $8$th, the list of plugins that is $12$th, and the list of supported mime types that is $17$th.
      On the contrary, the verbose attributes are slightly heavier on mobile browsers, which is explained by the presence of additional information like the device model.

    \subsection{Correlation between attributes}
    \label{sec:correlation-between-attributes}
      We can expect to have correlations occurring between the attributes when considering more than $200$~attributes.
      We provide here an overview of the correlation between the attributes, that include the nine dynamic attributes, and refer the reader to Appendix~\ref{app:attribute-list-and-property} for insight in the correlation of each attribute.

      For comparability with the results of attributes distinctiveness, we express the correlation by the conditional entropy ${\mathrm{H}(a_j | a_i)}$ of an attribute~$a_j$ when another attribute~$a_i$ is known, normalized to the maximum entropy $H_M$.
      We call this measure the \emph{normalized conditional entropy}.
      It is comprised between $0.0$ if knowing $a_i$ allows to completely infer $a_j$, and the normalized entropy of $a_j$ if knowing $a_i$ provides no information on the value of $a_j$ (i.e., they are independent).
      We denote $V_i$ the domain of the attribute~$a_i$ and $e^i_v$ the event that the attribute~$a_i$ takes the value~$v$.
      We consider the relative frequency~$p$ of the attribute values among the considered fingerprints.
      The measure of the conditional entropy ${\mathrm{H}(a_j | a_i)}$ of $a_j$ given $a_i$ is expressed as
      \begin{equation}
        \mathrm{H}(a_j | a_i) =
          - \sum_{v \in V_i, w \in V_j}
            p(e^i_v, e^j_w)
            \log
            \frac{
              p(e^i_v, e^j_w)
            }{
              p(e^i_v)
            }
      \end{equation}

      \begin{figure}
        \centering
        \includegraphics[width=0.70\columnwidth]{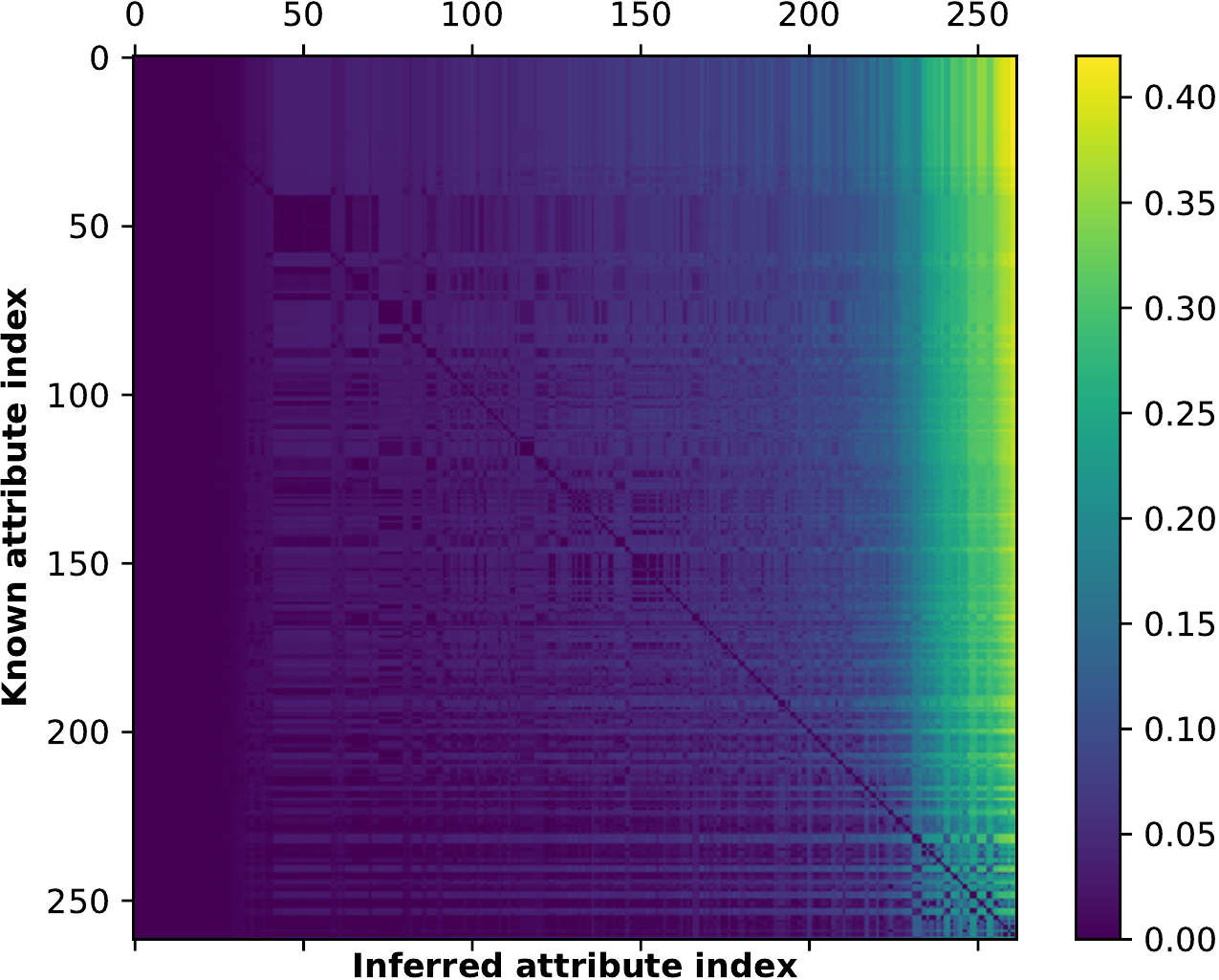}
        \caption{
          The normalized conditional entropy (NCE) of the attributes given the knowledge of another attribute.
          Given the known attribute of index $j$ on the vertical axis, the remaining NCE of the inferred attribute of index $i$ on the horizontal axis is read in the cell ($i$, $j$).
          The attributes are ordered by the average normalized conditional entropy and the indices match with Figure~\ref{fig:conditional-entropy}.
        }
        \label{fig:conditional-entropy-matrix}
        \Description[
          The normalized conditional entropy (NCE) of the attributes given the knowledge of another attribute.
          The attributes are ordered by the average normalized conditional entropy.
        ]{
          The normalized conditional entropy (NCE) of the attributes given the knowledge of another attribute.
          Given the known attribute of index $j$ on the vertical axis, the remaining NCE of the inferred attribute of index $i$ on the horizontal axis is read in the cell ($i$, $j$).
          The attributes are ordered by the average normalized conditional entropy and the indices match with Figure~\ref{fig:conditional-entropy}.
        }
      \end{figure}

      \begin{figure}
        \centering
        \includegraphics[width=0.70\columnwidth]{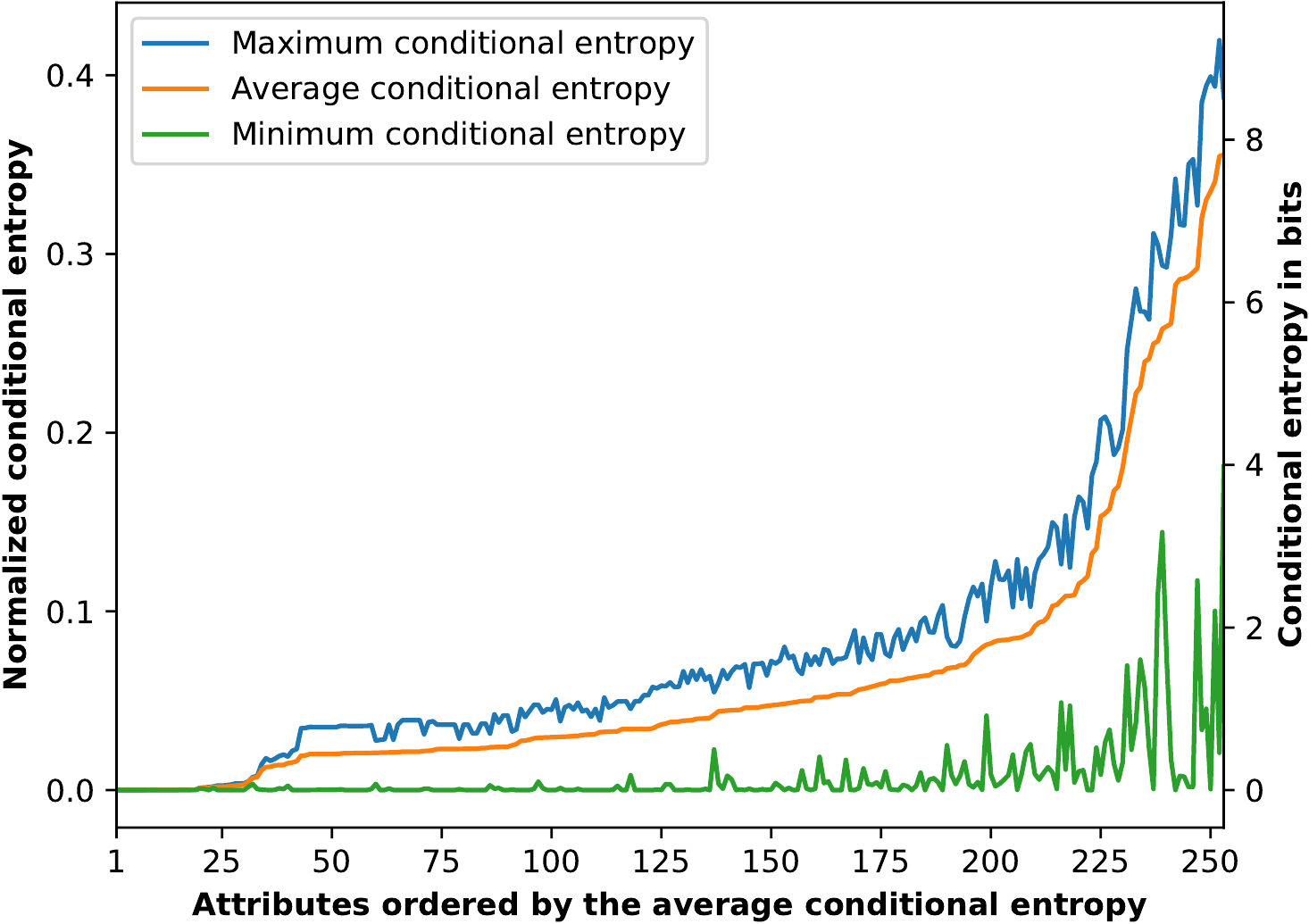}
        \caption{
          Minimum, average, and maximum normalized conditional entropy of an attribute when the value of another attribute is known.
          The attributes are ordered by the average normalized conditional entropy and the indices match with Figure~\ref{fig:conditional-entropy-matrix}.
        }
        \label{fig:conditional-entropy}
        \Description[
          Minimum, average, and maximum normalized conditional entropy of an attribute when the value of another attribute is known.
        ]{
          Minimum, average, and maximum normalized conditional entropy of an attribute when the value of another attribute is known.
          The attributes are ordered by the average normalized conditional entropy and the indices match with Figure~\ref{fig:conditional-entropy-matrix}.
          $19$~attributes have an average NCE of less than $10^{-3}$, $194$~attributes have an average NCE between $10^{-3}$ and $10^{-1}$, and $40$~attributes have an average NCE higher than $10^{-1}$.
        }
      \end{figure}

      Figure~\ref{fig:conditional-entropy-matrix} displays the normalized conditional entropy (NCE) of the attributes given the knowledge of another attribute and Figure~\ref{fig:conditional-entropy} displays the minimum, the average, and the maximum NCE of an attribute when the value of another attribute is known.
      In these two figures, the attributes are ordered identically by the average NCE.
      We recall that the maximum entropy is ${H_M = 21.983}$~bits and that a conditional entropy of $1$~bit is equivalent to a NCE of $0.045$.

      In Figure~\ref{fig:conditional-entropy}, we ignore the $9$ source attributes from which the extracted attributes are derived and the comparison of an attribute with itself.
      These cases are irrelevant as the extracted attributes are completely correlated with their source attribute, and an attribute is completely correlated with itself.
      We obtain a total of $253$~attributes.
      We stress that the maximum NCE of an attribute is always equal to the normalized entropy of this attribute.
      This is due to the cookie enabling state that is always \texttt{true}.
      As a result, it provides a null entropy and knowing its value does not provide any information on the value of the other attributes.
      We can see three parts in this figure.
      First, $19$~attributes have a low normalized entropy of less than $10^{-3}$ and their NCE when knowing another attribute is at most as much.
      Few different values have been observed for these attributes.
      Then, $194$~attributes have an average NCE between $10^{-3}$ and $10^{-1}$.
      The minimum normalized conditional entropy (MinNCE) of these attributes is near $0.0$, hence there exists another attribute that can be used to efficiently infer their value.
      Finally, $40$~attributes have an average NCE higher than $10^{-1}$ and generally have a strictly positive MinNCE.
      These attributes help to efficiently distinguish browsers and are less correlated to other attributes.

      The minimum normalized conditional entropy (MinNCE) is an interesting indicator of the efficiency to infer the value of an attribute if the value of another attribute is known.
      We have $49$~attributes that have a null MinNCE and can completely be inferred when another attribute is known.
      Moreover, $192$~attributes have a MinNCE comprised in the range $]0, 0.045]$, hence knowing the value of another attribute helps to infer their value but not completely.
      Finally, $12$~attributes have a MinNCE higher than $0.045$, which is equivalent to having a minimum conditional entropy higher than $1$~bit.
      They are displayed in Table~\ref{tab:less-correlated-attributes}.
      They consist of highly distinctive attributes that concern the screen or window size (e.g., W.screenX, W.innerHeight), browser components (e.g., the list of fonts or plugins), and verbose information about the browser (e.g., the UserAgent) or about external components (e.g., WebRTC fingerprinting).

      As the attributes tend to be correlated with each other, a subset of them can provide sufficient distinctiveness or the same distinctiveness as when using all the attributes.
      Several attribute selection methods were proposed by previous works.
      The most common methods weight the attributes (e.g., by the entropy) and successively pick the attribute of the highest weight until the obtained set reaches a threshold (e.g., on the number of attributes)~\cite{MEN11, KZW15, BRC16, HRA18, THS18}.
      Other selection methods leverage the correlation that occur between the attributes and pick the next attribute according to the attributes that are already selected~\cite{FE15, PRGB20, AAL20}.
      In particular, the authors of the FPSelect study~\cite{AAL20} apply three attribute selection methods to a sample of our dataset and compare the methods according to the distinctiveness and the usability of the resulting fingerprints.

      \begin{table}
        \caption{
          The attributes that have a minimum normalized conditional entropy (MinNCE) higher than $0.045$, which is equivalent to a minimum conditional entropy higher than $1$~bit.
          The minimum conditional entropy is in bits.
          W refers to the \texttt{window} JavaScript object, WG refers to an initialized WebGL context, and [...] denotes a truncated part.
        }
        \label{tab:less-correlated-attributes}

        \centering
        \begin{tabular}{lcc}
          \toprule
            \textbf{Attribute}                 & \textbf{MinNCE} &
            \textbf{Minimum conditional entropy (in bits)}                    \\
          \midrule
            W.innerHeight                      & 0.181         & 3.98         \\
            WebRTC fingerprinting              & 0.144         & 3.17         \\
            W.outerHeight                      & 0.117         & 2.58         \\
            List of fonts                      & 0.110         & 2.42         \\
            List of plugins                    & 0.100         & 2.21         \\
            W.outerWidth                       & 0.074         & 1.63         \\
            WebGL renderer (unmasked)          & 0.073         & 1.61         \\
            Height of first bounding box       & 0.070         & 1.53         \\
            S.availHeight                      & 0.058         & 1.27         \\
            W.screenY                          & 0.049         & 1.08         \\
            W.screenX                          & 0.047         & 1.04         \\
            userAgent JavaScript property      & 0.046         & 1.00         \\
          \bottomrule
        \end{tabular}
      \end{table}

  \subsection{Focus on dynamic attributes}
  \label{sec:focus-dynamic-attributes}
    Previous studies highlight the possibility to fingerprint browsers by rendering media files inside the browser, given instructions that are provided by the fingerprinter (e.g., WebGL canvas~\cite{MS12}, HTML5 canvas~\cite{BMPT16}, audio fingerprinting~\cite{QF19}).
    Later on, these attributes were integrated within challenge-response mechanisms~\cite{REKP19, LABN19} that mitigate replay attacks.
    We call these attributes the \emph{dynamic attributes} and include nine of them in our script: five HTML5 canvases, three audio fingerprinting methods, and a WebGL canvas.
    To the best of our knowledge, no study evaluates the properties of several dynamic attributes on a browser population, together with the evaluation of various set of instructions.
    In this section, we seek to fill this gap and focus on the properties offered by the nine dynamic attributes of our fingerprinting script.

    \begin{figure*}
      \setlength{\fboxsep}{0pt}  %
      \minipage{0.49\columnwidth}
        \centering
        \fbox{\includegraphics[width=0.9\columnwidth]{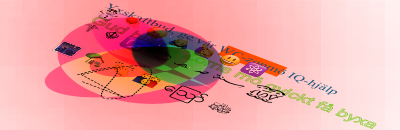}}
        \caption{
          A sample of our designed HTML5 canvas in PNG format.
        }
        \label{fig:our-canvas-png}
        \Description[
          A sample of our designed HTML5 canvas in PNG format.
        ]{
          A sample of our designed HTML5 canvas in PNG format.
          It includes several emojis, two strings including Swedish letters, many overlapping ellipses, a background with a color gradient, and a rotation of all these elements if the functionality is available.
        }
      \endminipage
      \hfill
      \minipage{0.49\columnwidth}
        \centering
        \fbox{\includegraphics[width=0.9\columnwidth]{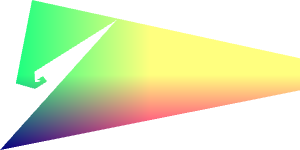}}
        \caption{
          A sample of our designed WebGL canvas in PNG format.
        }
        \label{fig:canvas-webgl-png}
        \Description[
          A sample of our designed WebGL canvas in PNG format.
        ]{
          A sample of our designed WebGL canvas in PNG format.
          It consists in sequential triangles with a color gradient.
        }
      \endminipage
    \end{figure*}

    \subsubsection{HTML5 canvas}
      The HTML5 canvas consists in asking the browser to draw an image using the canvas API within the two-dimensional context called \texttt{2d}.
      This method is already studied in several works, whether it is about its efficacy~\cite{BMPT16, LRB16, GLB18}, its use on the web~\cite{EN16, LFB17}, or the distinctiveness provided by various sets of instructions~\cite{LABN19}.
      However, to the best of our knowledge, no study evaluates the different properties (e.g., distinctiveness, stability, collection time) offered by various sets of complex instructions (i.e., mixes of texts, emojis, mathematical curves, drawn using different colors).
      We seek to fill this gap and evaluate the properties of five HTML5 canvases generated following three sets of instructions and in two image formats (PNG and JPEG).
      We name the canvases given the set of instructions and the format.
      We call \emph{AmIUnique canvas} the canvas generated by the set of instructions inspired by the AmIUnique\footnote{
        Contrarily to the AmIUnique study~\cite{LRB16}, we only have one sentence and one smiling face emoji instead of two.
        Also, we do not draw a colored rectangle and the sentence is drawn using a color gradient.
      } study~\cite{LRB16}.
      We extract it in PNG format only.
      We call \emph{Morellian canvas} the canvas generated by the set of instructions that is similar to the Morellian study~\cite{LABN19}.
      We extract it in PNG and JPEG format.
      We call \emph{custom canvas} the canvas generated by a set of instructions that we designed.
      We extract it in PNG and JPEG format.
      We generate the canvas using the API, then we collect the base64 representation of the image in PNG and JPEG format using the \texttt{toDataURL} function.
      The \texttt{quality} parameter of this function goes from $0.0$ to $1.0$ and allow us to control the level of compression of the JPEG versions.
      As we seek to compare the PNG canvases that are compressed without loss with their JPEG counterparts using a high level of compression, we set the \texttt{quality} to $0.1$ for the JPEG versions.
      An example of the custom canvas in PNG format is displayed in Figure~\ref{fig:our-canvas-png}.
      Examples of the AmIUnique and Morellian canvas are provided in Appendix~\ref{app:attributes}.

      We observe that canvases are less distinctive and more stable in JPEG format than in PNG format.
      For example, the custom canvas has a normalized entropy of $0.420$ when exported in PNG against $0.399$ for the JPEG version.
      However, the PNG version has a sameness rate of $92.16$\% against $93.59$\% for the JPEG version.
      These differences are due to distinct images in PNG format ending up the same after the lossy compression of the JPEG format.
      Acar et al.~\cite{AEEJND14} consider that a canvas generated in a lossy compression format as JPEG is not an indicator of a fingerprinting attempt.
      Although the JPEG version provides a lower distinctiveness than the PNG version, it is still highly distinctive when generated using a complex set of instructions.
      Indeed, it is the second most distinctive attribute among ours (see Section~\ref{sec:attributes-distinctiveness}).
      The time overhead induced by the additional extraction in JPEG format is negligible.
      For example, the custom canvas has a median collection time (MCT) increased by $5$ms for the JPEG version compared to the PNG version that has a MCT of $257$ms.
      We do not account the size differences as both formats are hashed and the resulting hashes weigh $64$~bytes.
      The PNG and the JPEG versions are also highly correlated, as knowing the value in the PNG format of the custom canvas leaves a normalized conditional entropy (NCE) of ${5.28 \times 10^{-4}}$ on the value of the JPEG format.
      On the opposite, knowing the value of the JPEG format provides a higher NCE of $0.021$ on the value of the PNG format.

      The properties of the three PNG canvases differ, with the custom canvas being the most distinctive.
      First, the Morellian canvas is an enhanced version of the AmIUnique canvas with additional curves.
      This enhancement provides an increase of the distinctiveness, with a normalized entropy of $0.385$ for the Morellian canvas against $0.353$ for the AmIUnique canvas.
      However, it comes with a loss of stability with a respective sameness rate of $94.71$\% against $98.64$\%.
      Then, the custom canvas is more complex than the Morellian canvas as it includes several emojis, two strings including Swedish letters, many overlapping ellipses, a background with a color gradient, and a rotation of all these elements if the functionality is available.
      These improvements provide a higher normalized entropy of $0.420$, but a lower sameness rate of $92.16$\%.
      The main drawback of adding more complexity to the custom canvas is the temporal cost.
      It has a MCT of $257$ms against $37$ms for the Morellian canvas and $31$ms for the AmIUnique canvas.
      However, it represents less than $10$\% of the total median collection time of the fingerprints which is $2.92$~seconds, and less than $5$\% of the median loading time of a web page on a desktop browser which is $6.5$~seconds~\cite{WebPageMedianLoadingTime}.
      Finally, knowing the value of the custom canvas leaves less variability on the value of the two other canvases than the opposite.
      When knowing the value of the Morellian or AmIUnique canvas, the custom canvas has a respective NCE of $0.079$ and $0.103$.
      However, knowing the value of the custom canvas results in the Morellian canvas showing a NCE of $0.044$ and the AmIUnique canvas showing a NCE of $0.037$.
      To conclude, adding more instructions for the canvas drawing usually provides more distinctiveness, as each instruction can induce a difference between browsers.
      However, it comes at the cost of additional computation time and loss of stability as each additional instruction can constitute an instability factor.

    \subsubsection{WebGL canvas}
      The WebGL canvas is an image that is drawn using the canvas API within the \texttt{webgl} or \texttt{webgl2} contexts.
      These contexts use the WebGL library~\cite{WebGL} that leverages hardware accelerations to render and manipulate two-dimensional graphics and three-dimensional scenes.
      Canvas fingerprinting was first introduced by Mowery et al.~\cite{MS12} using the \texttt{webgl} context, but afterward most studies focused on the \texttt{2d} context~\cite{BKST16, BMPT16, AL17, VLRR18}.
      This can result from the unreliability of the WebGL context encountered by Laperdrix et al.~\cite{LRB16} for which Cao et al.~\cite{CLW17} proposed a remedy by setting specific parameters.

      Our WebGL canvas consists in sequential triangles with a color gradient and is simpler than our designed HTML5 canvas.
      It provides a normalized entropy of $0.263$ against $0.420$ for the custom HTML5 canvas.
      It is more stable that the other canvases with a sameness rate of $98.97$\%, and has a median collection time of $41$ms against $257$ms for the custom HTML5 canvas.
      Figure~\ref{fig:canvas-webgl-png} displays an example of our WebGL canvas.

    \subsubsection{Web Audio}
    \label{sec:web-audio-fingerprinting}
      Web Audio fingerprinting was discovered by Englehardt et al.~\cite{EN16} when assessing the use of web tracking methods on the web, and was studied thoroughly by Queiroz et al.~\cite{QF19}.
      It consists in processing audio signals in the browser and collecting the rendered result.
      Similarly to canvases, this processing relies on software and hardware components, and variations occur between different component stacks.
      It works by creating an \texttt{Audio Context}, which in our case is the \texttt{OfflineAudioContext}\footnote{
        The advantage of using OfflineAudioContext is the manipulation of audio signal without playing any sound on a genuine audio output peripheral.
      }, in which we manipulate \texttt{Audio Nodes}.
      The \texttt{Audio Nodes} are of three types.
      Source nodes generate an audio signal (e.g., from a microphone or an audio stream), destination nodes render the signal (e.g., playing it through speakers), and manipulation nodes manipulate the signal (e.g., increase its volume).
      These nodes are linked together to form a network that goes from source nodes to destination nodes and passes through manipulation nodes.
      We refer the reader to~\cite{QF19} for a broader description of the main audio nodes.

      We have three audio fingerprinting attributes.
      The \emph{audio FP simple}~(AFS) attribute relies on a simple process.
      The attributes \emph{audio FP advanced}~(AFA) and \emph{audio FP advanced frequency data}~(AFA-FD) rely on a more advanced process.
      Their concrete implementation is described in Appendix~\ref{app:attributes}.
      The most distinctive audio attribute is the AFA-FD that has a normalized entropy of $0.161$, followed by the AFS ($0.153$), and the AFA ($0.147$).
      They all have a sameness rate of approximately $95$\%.
      Their values are the string representations of floating-point numbers, hence they have a median size of $17$~bytes.
      The simple process has a median collection time (MCT) of $1.38$~seconds and the advanced process has a MCT of $1.64$~seconds.
      The AFA-FD is collected from the advanced version and has a MCT increased by $3$ms compared to the AFA.

%% file: 7-discussion.tex
\section{Discussion}
\label{sec:discussion}
  This section describes how browser fingerprinting can be integrated in a web authentication mechanism and discusses the attacks that are possible on such mechanism.

  \subsection{Browser fingerprinting-based authentication mechanism}
  \label{sec:browser-fingerprinting-based-authentication-mechanism}
    Browser fingerprinting can enhance an authentication mechanism by providing an additional barrier at a low usability and deployability cost.
    In this section, we provide an example of how it can concretely be implemented.
    The verifier controls a web platform on which her users are registered.
    Each registered account is associated to an identifier and to a set of authentication factors.
    They include a simple password and the fingerprints of the browsers used by the account owner.
    We emphasize that dynamic attributes can be integrated to the browser fingerprints to enforce a challenge-response mechanism that mitigates replay attacks~\cite{REKP19, LABN19}.

    \subsubsection{Enrollment}
      The enrollment consists for a user to create his account and to register the authentication factors.
      During this step, the user and the verifier agrees on the account identifier (e.g., username, email address, phone number) and on the authentication factors (e.g., password, email address, phone number) that are assigned to the user.
      The fingerprint of the browser in use during the enrollment is collected and stored as the first browser fingerprint of the user.
      To register an additional browser, the user is required to authenticate using other strong factors (e.g., a physical token, a one time password) before getting the fingerprint of the new browser registered.

    \subsubsection{Authentication}
      During each authentication, the user claims an account by providing the identifier and presenting the authentication factors.
      The verifier compares the presented authentication factors with the ones stored for this user.
      If they match, the user is deemed legitimate and is given access to the account.
      Otherwise, the access is denied to the user and the verifier can take preventive actions~\cite{GSD18} according to her policy (e.g., blocking the account).
      The verifier notably compares the collected browser fingerprint with the fingerprints of the browsers registered to the account.
      If the other factors match and the fingerprint of one of the registered browser matches the collected fingerprint, the fingerprint stored for this browser is updated to the newly collected one.
      By using the \texttt{User-Agent} HTTP header, it would also be possible to reduce the set of the stored fingerprints against which to compare the presented fingerprint (e.g., recognizing the family of the browser in use and getting the stored fingerprints of this family).

    \subsubsection{Account Recovery}
      It happens that legitimate users are not able to provide the authentication factors (e.g., a password is forgotten, a physical token is lost).
      When it occurs, users are given access to an account recovery mechanism~\cite{MGSD20} that leverages other authentication factors than the usual ones (e.g., face-to-face verification, email verification).
      Users cannot mistake their browser fingerprint, but it can become hard to recognize if too many changes are brought to the web environment of the browser.
      In this case, the user is asked to undergo the account recovery step and select the registered browser for which to update its fingerprint.

  \subsection{Attacks on the authentication mechanism}
    We discuss in this section the attacks that are possible on the browser fingerprinting-based authentication mechanism.
    We consider an attacker that can access the authentication page and seeks to impersonate as many legitimate users as possible.
    To impersonate a legitimate user, the attacker has to find the right combination of identifier and authentication factors.
    Below, we make assumptions about the attacker and describe the attacks that he can execute.

    \subsubsection{Assumptions about the attacker}
      We focus on the contribution of browser fingerprinting to the authentication mechanism and make the following assumptions.
      The attacker knows the identifiers of the targeted users and the authentication factors (e.g., their password) at the exception to the browser fingerprint\footnote{
        We discuss in Section~\ref{sec:replay-attack} the case where all the authentication factors are known to the attacker including the browser fingerprint.
      }.
      A concrete example is an attacker that obtained these pieces of information from a data leak or theft.
      The attacker cannot tamper with the authentication server of the verifier nor with the device or the browser of the user.
      The attacker cannot tamper nor eavesdrop the communications between the server of the verifier and the browser of the user.
      The attacker knows the attributes that are collected by the fingerprinting script.
      This knowledge can be obtained from static or dynamic analysis of the script~\cite{ASH19, BMEWZRL20, RTM21}.
      The attacker knows the domains of the attributes and the domain of the fingerprints.
      The latter is the Cartesian product of the former.
      This knowledge can stem from publicly accessible resources like statistics~\cite{CaratProjectStatistics}, documentation, or fingerprint datasets~\cite{HenningTillmannBFP, VLRR18}.
      It can also come from malicious sources like stolen fingerprints~\cite{MAR20}, phishing attacks~\cite{TLZBRIMCEMMPB17}, or a set of controlled browsers~\cite{PIPMIV19}.
      The attacker can control the values of the attributes that compose the fingerprint by using open source tools like Disguised Chromium Browser~\cite{BKST16} or Blink~\cite{LRB15}, commercial solutions like Multilogin~\cite{Multilogin} or AntiDetect~\cite{AntiDetect}, or by altering the network packets that contain the fingerprint with tools like BurpSuite~\cite{BurpSuite}.
      The attacker can submit a limited number of arbitrary fingerprints to try to impersonate a targeted user.
      We emphasize that the reach of the attacker depends on the number of attempts similarly to guessing attacks on passwords~\cite{BON12, WZWYH16}.
      In our online authentication context, the number of attempts depends on the rate limiting policy of the verifier~\cite{GSD18, LZLZL18}.
      After a given number of failed attempts, the verifier can lock the account and ask the account owner to update his authentication factors (e.g., change his password, update his fingerprint).

    \subsubsection{Brute force attack}
      The brute force attack consists in the attacker submitting randomly sampled fingerprints from the fingerprint domain.
      In practice, the attacker randomly picks the value of each attribute, forges a fingerprint by combining them, and submits this fingerprint.
      The attacker repeats this process until the account is locked out or after a number of attempts decided by herself.
      The reach of the brute force attack depends on the space of the possible fingerprints that the attacker manages to cover in the limited number of attempts.
      Two factors greatly limit this reach.
      First, the domain of the fingerprint grows exponentially with the domain of the composing attributes.
      Second, the fingerprints are sampled randomly from the possibility space, which can lead to highly improbable combination of attribute values due to contradicting or incompatible values.
      Contradicting values occur when two attributes that always provide the same value, like the \texttt{logicalXDPI} and \texttt{logicalYDPI} properties of the \texttt{screen} JavaScript object, are chosen with diverging values.
      Incompatible values occur when two attributes that are correlated are chosen with incompatible values, like one value that indicates a mobile device (e.g., a UserAgent typically observed on mobile devices) and the other value indicating a desktop device (e.g., a large screen size).
      On our experimental setup, these two factors are important and greatly reduce the reach of brute force attacks.
      Indeed, our dataset is composed of $262$ attributes (including the extracted ones) among which $20$\% have more than a thousand distinct values (see Section~\ref{sec:attribute-wise-analysis}).
      Moreover, our attributes tend to be highly correlated with each other: $49$ attributes have a null conditional entropy when the value of another attribute is known (see Section~\ref{sec:correlation-between-attributes}).
      Although the fingerprint verification mechanism can allow differences between the fingerprints, a randomly sampled fingerprint is expected to have a lower number of identical attributes compared to the fingerprint of the targeted user.
      For a given attribute, there is only one chance over the size of the attribute domain for the attribute to be identical, and there may be more chances for it to be similar.
      However, for the randomly sampled fingerprint to match the target fingerprint, there should be sufficient identical or similar attributes, whereas the chances for this to happen decreases with the number and domain size of the attributes.

    \subsubsection{Dictionary attack}
      The dictionary attack consists for an attacker to infer a distribution of the fingerprints (e.g., from stolen fingerprints~\cite{MAR20}) and to submit the most probable fingerprints.
      The reach of the dictionary attack depends on the correspondence between the fingerprint distribution among the protected users and the fingerprint distribution known by the attacker.
      We take the example of an attacker having the knowledge of a fingerprint distribution obtained from mobile browsers configured in English, but targeting users of desktop browsers configured in French.
      In this example, the fingerprints submitted by the attacker would have few chances to match as some attributes (e.g., the screen size, the browser language) would differ.
      The users that are impersonated are those for which one of the registered fingerprints matches with one of the most probable fingerprints given the knowledge of the attacker.
      The authors of the FPSelect study~\cite{AAL20} measured the reach of a dictionary attack on a sample of $30,000$ fingerprints of our dataset, considering an attacker that knows the exact distribution of the target fingerprints and a verification mechanism that allows differences between the compared fingerprints.
      They found out that $0.21$\% of the users are impersonated when the attacker is able to submit the $4$ most common fingerprints, which goes up to $0.51$\% for the $16$ most common fingerprints.
      These proportions are a higher bound on the proportion of the fingerprints that are shared respectively by the $4$ and the $16$ most common fingerprints due to the differences allowed by the verification mechanism.
      These proportions are slightly lower than the share of the most common passwords measured in previous studies.
      Thomas et al.~\cite{TLZBRIMCEMMPB17} analyzed nearly two billion passwords and found out that the $4$ most common passwords are shared by $0.72$\% of the users, which goes up to $1.01$\% for the $10$ most common passwords.
      Wang et al.~\cite{WZWYH16} worked on nine password datasets and found out that the ten most common passwords are shared by $0.79$\% to $10.44$\% of the users, with an average of $3.06$\% among the datasets.

    \subsubsection{Replay attack}
    \label{sec:replay-attack}
      The replay attack consists for an attacker to collect the fingerprint of a user (e.g., through a phishing attack) and to submit this fingerprint to impersonate the user.
      Recent works~\cite{REKP19, LABN19} proposed to use dynamic attributes -- notably the HTML5 canvas -- to design challenge-response mechanisms to thwart replay attacks.
      These challenge-response mechanisms leverage the value of the dynamic attributes which depends on the set of instructions that is used (e.g., the drawing instructions of the canvas).
      The challenge is then the set of instructions and the response is the value generated by the browser.
      By varying the challenge on each authentication, a fingerprint that would have been collected by the attacker would be invalid as it would be the response to a previous challenge.
      We refer to the work of Laperdrix et al.~\cite{LABN19} for a thorough analysis of the level of security provided by such challenge-response mechanism.
      Section~\ref{sec:focus-dynamic-attributes} analyzes the nine dynamic attributes of three different types that are included in our study.

    \subsubsection{Relay attack}
      The relay attack consists for an attacker to pose as a legitimate user to the verifier, to pose as the verifier to a targeted user, and to relay the communications from one to the other.
      A relay attack is divided in the three following steps.
      First, the attacker sets a phishing website that embarks an authentication page similar to the verifier.
      Then, the attacker lures a victim to visit this fake page.
      At the same time, the attacker connects to the authentication page of the verifier and forwards the fingerprinting script to the victim.
      The victim enters her credentials and gets her browser fingerprint collected.
      Finally, the attacker submits these pieces of information to the authentication page of the verifier.
      Amnesty International reports the usage of relay attacks in real life and shows that relay attacks are able to break two-factor authentication~\cite{AmnestyInternationalRelayAttack}.
      Laperdrix et al.~\cite{LABN19} acknowledge that a challenge-response mechanism is theoretically vulnerable to relay attacks.
      However, they discuss the practical implications of relay attacks and explain that relay attacks are costly for the attackers to execute in practice.
      Outside the web context, distance bounding protocols~\cite{ABBCHKKLMMPRSTTV19} were proposed to defend against relay attacks~\cite{FDC11}.
      We let the design of a distance bounding protocol that leverages the time taken by the browser and the server to communicate as a future work.

%% file: 8-related-work.tex
\section{Related works}
\label{sec:related-works}
  In this section, we present related works about the use of browser fingerprinting for authentication.
  In addition, Section~\ref{sec:dataset-differences} compares our dataset with the previously studied large-scale datasets~\cite{ECK10, LRB16, GLB18, PRGB20, LC20} and Section~\ref{sec:attribute-wise-analysis} compares the distinctiveness of our attributes with the distinctiveness reported in previous studies.

  We stress that most studies about browser fingerprinting focus on their use for identification.
  The use of browser fingerprinting for identification and for authentication differ in the objective when given a presented fingerprint~$f$.
  In identification, we seek to find the identity (e.g., an account, an advertisement profile) to which $f$ belongs among a pool of $N$~candidate identities.
  If $f$ is unrecognized, it is associated to a new identity that is then added to the candidate identities.
  Hence, in identification we operate a $1$~to~$N$ comparison.
  Depending on the matching method, the matching time can be proportional to~$N$ and become unrealistic in a large-scale setting~\cite{LC20}.
  On the contrary, in authentication the identity is already given (e.g., the claimed account) and we seek to verify that $f$ legitimately belongs to this identity.
  Hence, in authentication we operate a $1$~to~$1$ comparison\footnote{
    If users register $n$ browsers to their account, as described in Section~\ref{sec:browser-fingerprinting-based-authentication-mechanism}, $f$ is compared to the fingerprint of these $n$ browsers that are already identified.
    Users are expected to register fewer than ten devices~\cite{SMBE17}, hence $n$ is expected to be smaller than $N$ (e.g., a website having a thousand accounts registered).
    It would also be possible to find the right fingerprint among the $n$ possibilities by leveraging the UserAgent to recognize which browser the user is using.
  }.
  The studies about the use of browser fingerprinting for identification usually evaluate the threat posed to privacy by browser accessible information~\cite{ECK10, MS12, FE15, LRB16, SN17, GLB18, QF19}, propose counter-measures to avoid being tracked by this technique~\cite{BKST16, LBM17}, or measure its usage on the web~\cite{AEEJND14, EN16, BMEWZRL20, IES21}.

  The first works about the use of browser fingerprinting for authentication focused on its use for continuous authentication~\cite{UMFHSW13, PJ15, SPJ17}.
  Their objective is to verify that the authenticated session is not hijacked.
  These studies focus on the integration of browser fingerprinting in an authentication mechanism and provide few insights about the properties of the fingerprints (e.g., only~\cite{SPJ17} analyzes fingerprints and focuses on their classification efficacy).
  On the contrary, we identify several properties of authentication factors with which we evaluate real-life browser fingerprints, and explain the results by highlighting the contribution of single attributes.
  Moreover, these studies only consider a small fraction of the hundreds of available attributes and do not include the dynamic attributes that can be used in a challenge-response mechanism to thwart replay attacks.
  We include more than $200$ attributes -- including $9$ dynamic attributes -- for which we provide the implementation and the properties (e.g., number of distinct values, normalized entropy).

  Alaca et al.~\cite{AV16} provided a classification of the fingerprinting attributes according to properties that include the stability, the distinctiveness, and the resource usage.
  They qualitatively estimate these properties given the nature of the attributes.
  For some attributes, they provide the entropy measured in previous studies and acknowledge that further study is needed on this subject.
  We stress that comparing the attributes entropy between datasets of different sizes leads to comparability problems as explained by~\cite{LRB16} (e.g., the attribute of a dataset of $N$ fingerprints provides at most an entropy of $\log_2(N)$~bits).
  We provide the quantitative measure of these properties on our attributes from the analysis of real-life fingerprints, and also evaluate the accuracy of a simple verification mechanism.
  About the attributes considered by~\cite{AV16}, they include attributes that require interactions from the user (e.g., her location) or that are related to the network protocol (e.g., the TCP/IP stack fingerprinting that analyzes the response to specially crafted messages).
  Moreover, their classification leads to different attributes being grouped under a single designation, like the ``major software and hardware details'' that includes the attribute family of the JavaScript properties that we describe in Appendix~\ref{app:attributes}.
  The attributes of this family can show diverse properties, like the UserAgent that provides a normalized entropy of $0.394$ and a sameness rate of $0.98$\%, whereas the \texttt{screenX} property of the \texttt{window} object provides a normalized entropy of $0.125$ and a sameness rate of $0.93$\%.
  We show that more than $200$ attributes are easily accessible (i.e., they require few lines of JavaScript) and do not require any interaction from the user, which would reduce the usability (e.g., the user being asked permissions several times).
  Moreover, we provide the exhaustive list of the attributes with their concrete implementation and their properties.
  We also evaluate the properties of distinctiveness, stability, and resource usage on the complete fingerprints that combine as many attributes.

  Markert et al.~\cite{MGSD20} recently presented a \textit{work in progress} about the long-term analysis of fallback authentication methods.
  They plan to measure the recovery rate of the evaluated methods after an elapsed time of $6$, $12$, and $18$ months.
  These methods include browser fingerprinting, for which they acknowledge that ``not much about browser fingerprinting-based security systems is known''.

  Rochet et al.~\cite{REKP19} and Laperdrix et al.~\cite{LABN19} proposed challenge-response mechanisms that rely on dynamic attributes, especially the HTML5 canvas.
  The instructions provided to the canvas API are changed on each authentication, making the generated image vary on each fingerprinting.
  Trivial replay attacks then fails as the awaited canvas image changes each time.
  In this study, we propose the evaluation of nine dynamic attributes that include five HTML5 canvases, one WebGL canvas, and three audio fingerprinting methods.

%% file: 9-conclusion.tex
\section{Conclusion}
\label{sec:conclusion}
  In this study, we conduct the first large-scale empirical study of the properties of browser fingerprints when used for web authentication.
  We make the link between the digital fingerprints that distinguish Humans, and the browser fingerprints that distinguish browsers, to evaluate the latter according to properties inspired by biometric authentication factors.
  We formalize and evaluate the properties for the browser fingerprints to be usable and practical in a web authentication context.
  They include the distinctiveness of the fingerprints, their stability, their collection time, their size, the loss of efficacy among browser types, and the accuracy of a simple illustrative verification mechanism.
  We evaluate these properties on a real-life large-scale fingerprint dataset collected over a period of $6$~months that contains $4,145,408$~fingerprints composed of $216$~attributes.
  The attributes include nine dynamic attributes which are used in state-of-the-art authentication mechanisms to mitigate replay attacks.
  We thoroughly describe the preprocessing steps to prepare the dataset and the browser population that, contrary to most of the previous studies, is not biased towards technically-savvy users.
  We show that, considering time-partitioned datasets, our browser fingerprints provide a unicity rate above $81.3$\% which is stable over the $6$~months, and more than $94.7$\% of the fingerprints are shared by at most $8$~browsers.
  We observe a loss of distinctiveness from mobile browsers that show a lower unicity rate of $42$\%.
  About the stability and on average, more than $91$\% of the attributes are identical between two observations of the fingerprint of a browser, even when they are separated by nearly $6$~months.
  About the memory and the time consumption of our fingerprints, we show that they weigh a dozen of kilobytes and take a few seconds to collect.
  Our simple verification mechanism achieves an equal error rate of $0.61$\%.
  It comes from most of the consecutive fingerprints of a browser having at least $234$ identical attributes, whereas most of the fingerprints of different browsers have fewer.
  To better comprehend the results on the complete fingerprints, we evaluate the contribution of the attributes to each fingerprint property.
  We show that although the attributes show a lower distinctiveness compared to previous studies, $10$\% of the attributes provide a normalized entropy higher than $0.25$.
  Moreover, four attributes missing from the previous large-scale studies are part of the most distinctive of our attributes.
  They concern the size and position of elements of the browser interface.
  About the stability, $85$\% of the attributes stay identical between $99$\% of the pairs of consecutive fingerprints coming from the same browser.
  Few attributes consume a high amount of resources.
  Only $33$ attributes take more than $5$ms to collect and only $20$ attributes weigh more than $100$ bytes.
  When looking at the correlation between the attributes, we find that $49$ attributes can completely be inferred when the value of another attribute is known.
  We remark that the dynamic attributes are part of the most time-consuming attributes, but also among the most distinctive attributes.
  We show the importance of the set of instructions which influences both the distinctiveness and the stability of the generated values (e.g., the canvas image).
  We conclude that the browser fingerprints obtained from the combination of the studied browser population, and the large surface of fingerprinting attributes, carry the promise to strengthen web authentication mechanisms.

%% file: 10-acknowledgments.tex
\begin{acks}
  We would like to thank the anonymous reviewers for their valuable comments.
  We also thank Benoît Baudry and David Gross-Amblard for their valuable comments.
\end{acks}

%% file: 11-appendix.tex
\appendix

\section{Browser fingerprinting attributes}
\label{app:attributes}
  In this section, we describe the $216$~\emph{base attributes} that are included in our script and the $46$~\emph{extracted attributes} that are derived from the $9$~source attributes.
  We group the attributes in families and provide references to related studies.
  Their name is sufficient to retrieve the corresponding browser property.
  When needed, we provide a brief description of the method for reproducibility.
  We focus here on the description of the method and provide a \emph{complete list} of the attributes and their property in Appendix~\ref{app:attribute-list-and-property}.
  We denote \texttt{A.[B, C]} a property that is accessed either through \texttt{A.B} or \texttt{A.C}.
  If there is only one element inside the brackets, this element is optional.
  We denote \texttt{[...]} a part that is omitted but described in the corresponding attribute description.

  \subsection{JavaScript properties}
    Most attributes are \emph{properties} that are accessed through common \emph{Java\-Script objects}.
    The \texttt{navigator} object provides information on the browser (e.g., its version), its customization (e.g., the configured language), the underlying system (e.g., the operating system), and the supported functionalities (e.g., the list of available codecs).
    The \texttt{screen} object provides information on the screen size, the orientation, the pixel density, and the available space for the web page.
    The \texttt{window} object provides information on the window containing the web page, like its size or the support of storage mechanisms.
    The \texttt{document} object gives access to the page content and a few properties.
    The JavaScript properties are already included in previous studies~\cite{ECK10, LRB16, GLB18} or open source projects~\cite{FingerprintJS}, but are usually limited to fewer than $20$~properties.

  \subsection{HTTP headers}
    Our script collects $16$~attributes from \emph{HTTP headers}, most of which are already used in previous studies~\cite{ECK10, LRB16, GLB18}.
    Among these $16$~attributes, $15$~consist of the value of an explicitly specified header and the last attribute stores any remaining fields as pairs of name and value.

  \subsection{Enumeration or presence of browser components}
    One attribute family that provides a high diversity is the \emph{browser components}.
    The presence of some components can directly be accessed (e.g., the list of plugins~\footnote{
      Firefox browsers since version 29 require developers to probe the exact name of the wanted plugin and do not allow them to enumerate the list of plugins~\cite{FirefoxPluginsProbing}.
      This version was released in April 29, 2014, hence the Firefox browsers of our population are impacted by this limitation.
      As a result, the list of plugins of our Firefox browsers mostly contains a single entry that concerns the Flash plugin.
    }) whereas the presence of others have to be inferred (e.g., the installed fonts).
    The list of components that are given in this section have the components separated by a comma.

    \subsubsection{List attributes}
      Previous studies already identified the \emph{list of plugins} and the \emph{list of fonts} as highly distinctive~\cite{ECK10, LRB16}.
      We include these two attributes in our fingerprinting script.
      We enumerate the \emph{list of plugins} and the \emph{list of mime types} (i.e., the supported data format).
      The list of fonts is an asynchronous attribute.
      To collect it, we check the size of text boxes~\cite{FE15} to infer the presence of $66$~fonts.
      Additionally, we collect the \emph{list of speech synthesis voices} from the \texttt{speechSynthesis} property of the \texttt{window} object or one of its variant (e.g., prefixed with \texttt{webkit}).
      To do so, we use the \texttt{getVoices} function of this property.
      On some browsers, we have to wait for the \texttt{onvoiceschanged} event to be triggered before getting access to the list.
      This attribute is then asynchronous.

    \subsubsection{Support of video, audio, and streaming codecs}
      We infer the \emph{support of video codecs} by creating a \texttt{video} element and checking whether it can play a given type using the \texttt{can\-Play\-Type()} function.
      The \emph{support of audio codecs} is inferred by applying the same method on an \texttt{audio} element.
      We infer the \emph{support of streaming codecs} by calling the \texttt{is\-Type\-Supported()} function of the \texttt{window.[WebKit, moz, ms, $\varnothing$]MediaSource} object and checking the presence of both the audio and the video codecs.
      We apply the same method to infer the \emph{support of recording codecs} from the \texttt{Media\-Re\-cord\-er} object.

    \subsubsection{List of video codecs}
      The $15$ \emph{video codecs} for which we infer the presence are the following: vi\-deo/\-mp2t; co\-decs\-=\-"avc1.\-42\-E0\-1E,\-mp4a.\-40.\-2", vi\-deo/\-mp4; co\-decs\-=\-"avc1.\-42c00d", vi\-deo/\-mp4; co\-decs\-=\-"avc1.\-4D401E", vi\-deo/\-mp4; co\-decs\-=\-"mp4v.\-20.\-8", vi\-deo/\-mp4; co\-decs\-=\-"avc1.\-42E01E", vi\-deo/\-mp4; co\-decs\-=\-"avc1.\-42E01E, mp4a.\-40.\-2", vi\-deo/\-mp4; co\-decs\-=\-"hvc1.\-1.\-L0.\-0", vi\-deo/\-mp4; co\-decs\-=\-"hev1.\-\-1.\-\-L0.\-\-0", vi\-deo/\-ogg; co\-decs\-=\-"theora", vi\-deo/\-ogg; co\-decs\-=\-"vorbis", vi\-deo/\-webm; co\-decs\-=\-"vp8", vi\-deo/\-webm; co\-decs\-=\-"vp9", app\-li\-ca\-tion/\-dash+\-xml, app\-li\-ca\-tion/\-vnd.\-apple.\-mpeg\-URL, audio/\-mpeg\-url.

    \subsubsection{List of audio codecs}
      The $9$ \emph{audio codecs} for which we infer the presence are the following: au\-dio/\-wav; co\-decs\-=\-"1", au\-dio/\-mpeg, au\-dio/\-mp4; co\-decs\-=\-"mp4a.\-40.\-2", au\-dio/\-mp4; co\-decs\-=\-"ac-3", au\-dio/\-mp4; co\-decs\-=\-"ec-3", au\-dio/\-ogg; co\-decs\-=\-"vorbis", au\-dio/\-ogg; co\-decs\-=\-"opus", au\-dio/\-webm; co\-decs\-=\-"vorbis", au\-dio/\-webm; co\-decs\-=\-"opus".

    \subsubsection{List of detected fonts}
      The $66$ \emph{fonts} for which we infer the presence are the following: Andale Mono; AppleGothic; Arial; Arial Black; Arial Hebrew; Arial MT; Arial Narrow; Arial Rounded MT Bold; Arial Unicode MS; Bitstream Vera Sans Mono; Book Antiqua; Bookman Old Style; Calibri; Cambria; Cambria Math; Century; Century Gothic; Century Schoolbook; Comic Sans; Comic Sans MS; Consolas; Courier; Courier New; Garamond; Geneva; Georgia; Helvetica; Helvetica Neue; Impact; Lucida Bright; Lucida Calligraphy; Lucida Console; Lucida Fax; LUCIDA GRANDE; Lucida Handwriting; Lucida Sans; Lucida Sans Typewriter; Lucida Sans Unicode; Microsoft Sans Serif; Monaco; Monotype Corsiva; MS Gothic; MS Outlook; MS PGothic; MS Reference Sans Serif; MS Sans Serif; MS Serif; MYRIAD; MYRIAD PRO; Palatino; Palatino Linotype; Segoe Print; Segoe Script; Segoe UI; Segoe UI Light; Segoe UI Semibold; Segoe UI Symbol; Tahoma; Times; Times New Roman; Times New Roman PS; Trebuchet MS; Verdana; Wingdings; Wingdings 2; Wingdings 3.

  \subsection{Extension detection}
    The list of the installed \emph{browser extensions} cannot be directly accessed, but their presence can be inferred.
    We check the changes that are brought to the web page~\cite{SN17} by the $8$~extensions that are listed in Table~\ref{tab:extensions-dom-changes} and the availability of the web accessible resources~\cite{SVS17, KISP20} of the $8$~extensions that are listed in Table~\ref{tab:extensions-web-accessible-resources}.
    All these $16$~attributes are asynchronous.

    \subsubsection{Detection of an ad-blocker}
      We infer the presence of an \emph{ad-blocker} by creating an invisible dummy advertisement and checking whether it is removed or not.
      This attribute is then asynchronous.
      To detect the removal, we check whether the common methods used by ad-blockers are applied.
      They consist into removing the DOM element, setting its size to zero (e.g., using the \texttt{offsetWidth} or \texttt{clientHeight} property), setting its \texttt{visibility} to \texttt{hidden}, or setting its \texttt{display} to \texttt{none}.
      The dummy advertisement is a created division which has the \texttt{id} property set to ``\texttt{ad\-\_\-ads\-\_\-pub\-\_\-track}'', the \texttt{class} set to ``\texttt{ads .html?ad= /?view=ad text-ad textAd text\-\_\-ad text\-\_\-ads text-\-ads}'', and the \texttt{style} set to ``\texttt{width: 1px !imp\-or\-tant; height: 1px !\-imp\-ort\-ant; pos\-it\-ion: abs\-ol\-ute !\-imp\-ort\-ant; left: -1000\-px !\-imp\-ort\-ant; top: -1000px !\-imp\-ort\-ant;}''.

    \begin{table}
      \caption{
        Extensions detected by the changes they bring to the page content.
      }
      \label{tab:extensions-dom-changes}

      \centering
      \begin{tabular}{ll}
        \toprule
          \textbf{Extension} & \textbf{Page content change} \\
        \midrule
          Privowny           & W.privownyAddedListener[EXT] is supported \\
          UBlock             & D.head has \texttt{display: none !important;} and \texttt{:root} as style \\
          Pinterest          & D.body.data-pinterest-extension-installed is supported \\
          Grammarly          & D.body.data-gr-c-s-loaded is supported \\
          Adguard            & W.AG\_onLoad is supported \\
          Evernote           & Element with \texttt{style-1-cropbar-clipper} as \texttt{id} exists \\
          TOTL               & W.ytCinema is supported \\
          IE Tab             & W.ietab.getVersion() is supported \\
        \bottomrule
      \end{tabular}
    \end{table}

    \begin{table}
      \caption{
        Extensions detected by the availability of their web accessible resource.
        C~stands for chrome and R~stands for resource.
      }
      \label{tab:extensions-web-accessible-resources}

      \centering
      \begin{tabular}{ll}
        \toprule
          \textbf{Extension}  & \textbf{Web accessible resource} \\
        \midrule
          Firebug             & C://firebug/skin/firebugBig.png \\
          YahooToolbar        & R://635abd67-4fe9-1b23-4f01-e679fa7484c1/icon.png \\
          EasyScreenshot      & C://easyscreenshot/skin/icon16.png \\
          Ghostery            & R://firefox-at-ghostery-dot-com/data/images/ghosty-16px.png \\
          Kaspersky           & R://urla-at-kaspersky-dot-com/data/icon-16.png \\
          VideoDownloadHelper & R://b9db16a4-6edc-47ec-a1f4-b86292ed211d/data/images/icon-18.png \\
          GTranslate          & R://aff87fa2-a58e-4edd-b852-0a20203c1e17/icon.png \\
          Privowny            & C://privowny/content/icons/privowny\_extension\_logo.png \\
        \bottomrule
      \end{tabular}
    \end{table}

  \subsection{Size and color of web page elements}

    \subsubsection{Bounding boxes}
      The attributes related to the \emph{bounding boxes} concern a \texttt{div} element to which we append a \texttt{span} element.
      These attributes are asynchronous.
      The \texttt{div} element has his \texttt{style} property set to the values displayed in Table~\ref{tab:bounding-boxes-properties}.
      The \texttt{span} element contains a specifically crafted text that is provided below.
      The size of bounding boxes (i.e., the width and the height of the rectangles of the \texttt{div} and the \texttt{span} elements) are then collected using the \texttt{getClientRects} function.
      The text of the \texttt{span} element is ``\texttt{{\textbackslash}u\-a9\-c0 {\textbackslash}u\-26\-03 {\textbackslash}u\-20\-B9 {\textbackslash}u\-26\-04 {\textbackslash}u\-26\-9b {\textbackslash}u\-26\-24 {\textbackslash}u\-23\-B7 {\textbackslash}u\-26\-2c {\textbackslash}u\-26\-51 {\textbackslash}u\-26\-9d {\textbackslash}u\-06\-01 {\textbackslash}u\-06\-03 {\textbackslash}u\-aa\-c1 {\textbackslash}u\-06\-0e {\textbackslash}u\-06\-dd {\textbackslash}u\-d8\-3c{\textbackslash}u\-df\-e1 mmmmmmmmmmlil {\textbackslash}u\-10\-2a}''.
      The characters that start with ``\texttt{{\textbackslash}u}'' are special Unicode characters (e.g., emojis, letters of a non-latin alphabet).

      \begin{table}
        \caption{
          Properties of the \texttt{div} element that is measured to generate the value of the attributes related to the bounding boxes.
        }
        \label{tab:bounding-boxes-properties}

        \centering
        \begin{tabular}{ll}
          \toprule
            \textbf{Property} & \textbf{Value}                           \\
          \midrule
            position          & absolute                                 \\
            left              & -9999px                                  \\
            textAlign         & center                                   \\
            objectFit         & scale-down                               \\
            font              & 68px / 83px Helvetica, Arial, Sans-serif \\
            zoom              & 66\%                                     \\
            MozTransform      & scale(0.66)                              \\
            visibility        & hidden                                   \\
          \bottomrule
        \end{tabular}
      \end{table}

    \subsubsection{Width and position of a created div}
      The attribute named \emph{width and position} of a created \texttt{div} is the properties of \texttt{width} and \texttt{transform-origin} of a newly created \texttt{div} element which are obtained from the \texttt{get\-Computed\-Style} function.
      This attribute is asynchronous.
      This created \texttt{div} element is afterward used to infer the color of layout components as described below.

    \subsubsection{Colors of layout components}
      The attribute \emph{colors of layout components} is obtained by applying the color of several layout components (e.g., the scroll bar) to the created \texttt{div} element denoted \texttt{new\_div}, and getting the color back from the property \texttt{W.get\-Computed\-Style(new\-\_\-div).color}.
      Each of the tested component gets its color extracted this way and their colors are aggregated in this attribute.
      The color of each element is afterward extracted as a single attribute (see Section~\ref{sec:data-preprocessing}).
      They are displayed at the end of Table~\ref{tab:attributes-table} that lists the attributes and their properties.
      These attributes are asynchronous.

  \subsection{WebGL properties}
    Our script collects several properties from the WebGL API.
    To obtain them, we create a \texttt{canvas} element and get its WebGL context by calling \texttt{getContext()} using any of the following parameters: \texttt{webgl}, \texttt{webgl2}, \texttt{moz-webgl}, \texttt{experimental-webgl}, or \texttt{ex\-pe\-ri\-ment\-al-web\-gl\-2}.

    The property \texttt{MAX\-\_\-TEX\-TURE\-\_\-MAX\-\_\-ANI\-SO\-TRO\-PY\-\_\-EXT} is obtained from one of \texttt{[WEB\-KIT\-\_\-EXT\-\_\-, MOZ\-\_\-EXT\-\_\-, EXT\-\_\-]tex\-ture\-\_\-fil\-ter\-\_\-ani\-so\-tro\-pic}.
    To get the unmasked vendor and renderer, we first get an identifier named \texttt{id} from the unmasked property of the \texttt{get\-Extension('WEB\-GL\-\_\-de\-bug\-\_\-ren\-de\-rer\-\_\-in\-fo')} object, and then get the actual value by calling \texttt{get\-Parameter(id)}.
    The unmasked property of the renderer is named \texttt{UN\-MASK\-ED\_REN\-DER\-ER\_WEB\-GL} and the one of the vendor is named \texttt{UN\-MASK\-ED\_VEND\-OR\_WEB\-GL}.
    Finally, to get the \texttt{COM\-PRE\-SSED\-\_\-TEX\-TURE\-\_\-FOR\-MATS} property, we have to load the \texttt{[WEB\-KIT\-\_\-]WEBGL\-\_\-com\-pre\-ssed\-\_\-tex\-ture\-\_\-s3tc} extension first.

  \subsection{WebRTC fingerprinting}
    We include a WebRTC fingerprinting method similar to the method proposed by Takasu et al.~\cite{TSYI15}.
    The method consists in getting information about the Session Description Protocol of a generated WebRTC connection.
    Due to the variability of this information, we create two different connections and hold only the values that are identical between them.
    As this method leaks local IP addresses, we hash them directly on the client before sending this attribute.
    This attribute is asynchronous.

  \subsection{HTML5 canvases inspired by previous studies}
\begin{figure*}
      \minipage{0.49\textwidth}
        \setlength{\fboxsep}{0pt}  %
        \centering
        \fbox{\includegraphics[width=0.9\columnwidth]{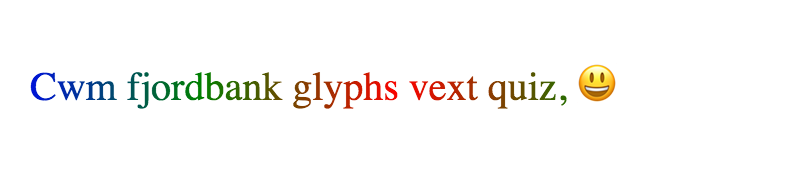}}
        \caption{
          A sample of the HTML5 canvas that is inspired by the AmIUnique~\cite{LRB16} study.
        }
        \label{fig:canvas-amiunique-png}
        \Description[
          A sample of the HTML5 canvas that is inspired by the AmIUnique~\cite{LRB16} study.
        ]{
          A sample of the HTML5 canvas that is inspired by the AmIUnique~\cite{LRB16} study.
          It consists in a pangram drawn with a color gradient next to a smiling face emoji.
        }
      \endminipage
      \hfill
      \minipage{0.49\textwidth}
        \centering
        \fbox{\includegraphics[width=0.9\columnwidth]{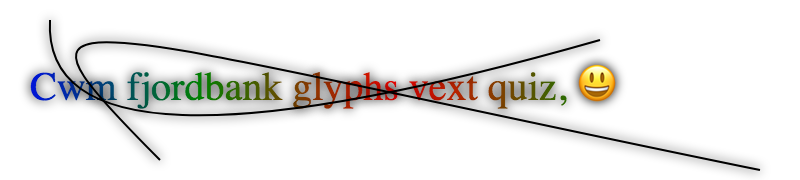}}
        \caption{
          A sample of the HTML5 canvas that is similar to the Morellian~\cite{LABN19} study.
        }
        \label{fig:canvas-morellian-png}
        \Description[
          A sample of the HTML5 canvas that is similar to the Morellian~\cite{LABN19} study.
        ]{
          A sample of the HTML5 canvas that is similar to the Morellian~\cite{LABN19} study.
          It consists in a pangram drawn with a color gradient next to a smiling face emoji and two curves drawn over these elements.
        }
      \endminipage
    \end{figure*}

    Our script includes a canvas inspired by the AmIUnique study~\cite{LRB16} in both PNG and JPEG formats, and an enhanced version similar to the Morellian study~\cite{LABN19}.
    Figure~\ref{fig:canvas-amiunique-png} and Figure~\ref{fig:canvas-morellian-png} respectively display an example of a canvas generated using these set of instructions.
    We refer to Section~\ref{sec:focus-dynamic-attributes} for a focus on the dynamic attributes.

  \subsection{Audio fingerprinting}
    In this section, we provide the concrete implementation and the network of \texttt{Audio\-Node} objects used within each audio fingerprinting methods.
    These methods are inspired by the work of Englehardt et al.~\cite{EN16}.
    They are designed to form complex networks of \texttt{Audio\-Node} objects to have more chances to induce fingerprintable behaviors.
    The three audio fingerprinting attributes are asynchronous.
    We refer to Section~\ref{sec:focus-dynamic-attributes} for a focus on the dynamic attributes.

    \subsubsection{Simple audio fingerprinting method}
\begin{figure}
        \centering
        \includegraphics[width=0.70\columnwidth]{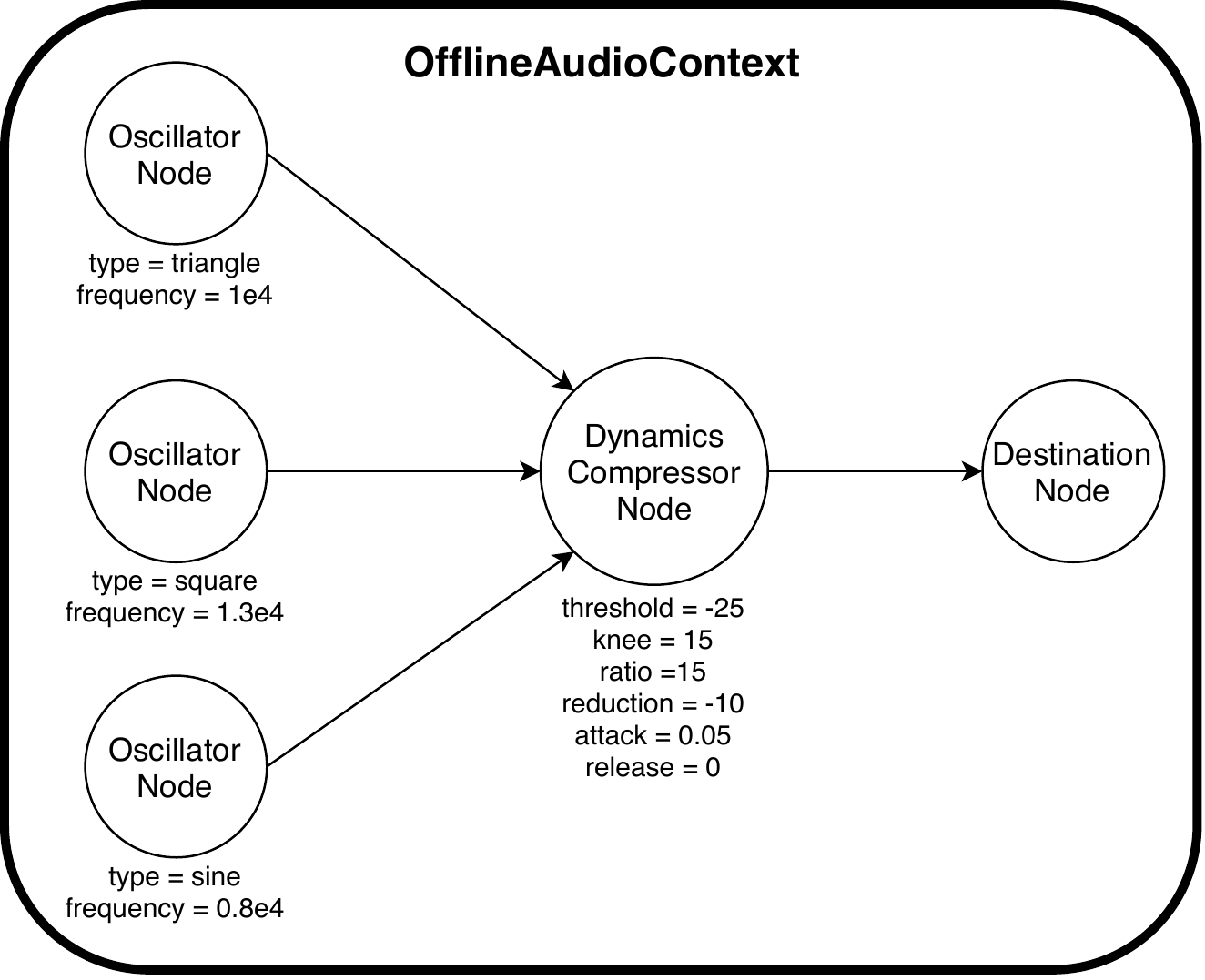}
        \caption{
          Architecture of the network of \texttt{Audio\-Node} objects for the simple audio fingerprinting method.
        }
        \label{fig:audio-nodes-architecture-simple-audio-fingerprinting}
        \Description[
          Architecture of the network of Audio\-Node objects for the simple audio fingerprinting method.
        ]{
          Architecture of the network of Audio\-Node objects for the simple audio fingerprinting method.
          It is composed of three Oscillator\-Node objects, one Dynamics\-Compressor\-Node, and one Audio\-Destination\-Node.
        }
      \end{figure}

      The \emph{simple process} consists of three \texttt{Oscillator\-Node} objects that generate a periodic wave, connected to a single \texttt{Dynamics\-Compressor\-Node}, and finishing to a \texttt{Audio\-Destination\-Node}.
      The architecture of the network of \texttt{Audio\-Node} objects for the simple process is depicted in Figure~\ref{fig:audio-nodes-architecture-simple-audio-fingerprinting} with the parameters set for each node.
      The \texttt{Oscillator\-Node} objects are started one after the other and overlap at some time.
      The sequence of events is the following: (1) the triangle oscillator node is started at ${t=0}$~seconds, (2) the square oscillator is started at ${t=0.10}$~seconds, (3) the triangle oscillator is stopped at ${t=0.20}$~seconds and the sine oscillator node is started, and finally (4) the square oscillator is stopped at ${t=0.25}$~seconds.
      When the rendering of the audio context is done, the \texttt{complete} event is triggered and gives access to a \texttt{renderedBuffer} that contains the audio data encoded as $32$~bits floating-point numbers.
      The \emph{audio FP simple}~(AFS) attribute is an integer computed as the sum of these numbers cast to absolute integers.

    \subsubsection{Advanced audio fingerprinting method}
\begin{figure}
        \centering
        \includegraphics[width=\columnwidth]{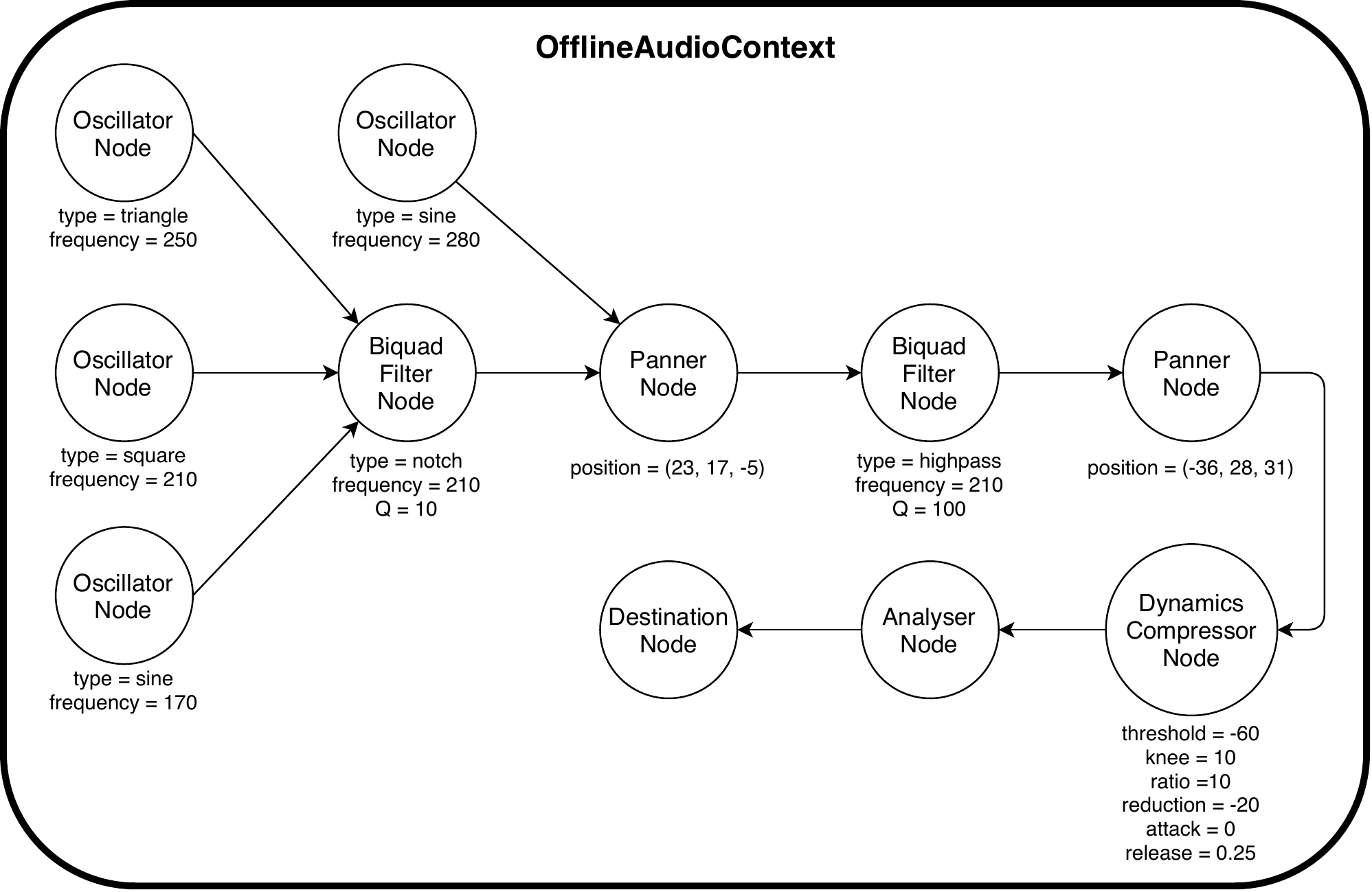}
        \caption{
          Architecture of the network of \texttt{Audio\-Node} objects for the advanced audio fingerprinting method.
        }
        \label{fig:audio-nodes-architecture-advanced-audio-fingerprinting}
        \Description[
          Architecture of the network of Audio\-Node objects for the advanced audio fingerprinting method.
        ]{
          Architecture of the network of Audio\-Node objects for the advanced audio fingerprinting method.
          It is composed of four Oscillator\-Node objects, two Biquad Filter Node objects, two Panner Node objects, one Dynamics\-Compressor\-Node, one Analyser\-Node, and one Audio\-Destination\-Node.
        }
      \end{figure}

      The \emph{advanced process} consists of four \texttt{Oscillator\-Node} objects, two \texttt{Biquad\-Filter\-Node} objects, two \texttt{Pan\-nerNode} objects, one \texttt{Dynamics\-Compressor\-Node}, one \texttt{Analyser\-Node}, and one \texttt{Audio\-Destination\-Node}.
      The architecture of the network of \texttt{Audio\-Node} objects for the advanced process is depicted in Figure~\ref{fig:audio-nodes-architecture-advanced-audio-fingerprinting} with the parameters set for each node.
      The \texttt{Oscillator\-Node} objects are started one after the other and overlap at some time.
      The sequence of events is the following: (1) the triangle oscillator node and the sine oscillator node with a frequency of $280$ are started at ${t=0}$~seconds, (2) the square oscillator is started at ${t=0.05}$~seconds, (3) the triangle oscillator is stopped at ${t=0.10}$~seconds, (4) the sine oscillator with a frequency of $170$ is stopped at ${t=0.15}$~seconds, (5) the square oscillator is stopped at ${t=0.20}$~seconds.
      When the rendering of the audio context is done, the \texttt{complete} event is triggered and gives access to a \texttt{renderedBuffer} that contains the audio data encoded as $32$~bits floating-point numbers.
      The \emph{audio FP advanced}~(AFA) attribute is an integer computed as the sum of these numbers cast to absolute integers.
      The \emph{audio FP advanced frequency data}~(AFA-FD) attribute is the sum of the frequency data obtained through the \texttt{getFloatFrequencyData} function of the \texttt{Analyser\-Node}.

\section{Anomalous collection times}
\label{app:anomalous-collection-times}
  In this section, we describe a side effect encountered by our script that results in some fingerprints and attributes showing a high collection time above $30$ seconds.
  These high collection times concern the sending of the fingerprint and the collection of the attributes that require the body to be set or that are collected asynchronously.
  In total, $27$ base attributes are impacted by these high collection times, from which we extract $45$ attributes.
  The single extracted attribute that is not concerned is the number of WebGL extensions.
  The remaining $189$ base attributes and one extracted attribute are not impacted.

  We measure the collection time of the fingerprints by the difference between two timestamps, one recorded at the starting of the script and the other one just before sending the fingerprint.
  Some fingerprints take a long time to collect that span from several hours to days.
  The high collection time of these fingerprints is explained below.
  Our fingerprinting script waits for the page body to load before collecting the attributes that require the body to be available, like the extension detection attributes.
  Due to a side effect, the script also waits for the body to be set before sending the fingerprint.
  We manage the waiting times using the \texttt{setTimeout} JavaScript function.
  The implementation of \texttt{setTimeout} in Chrome~\cite{BackgroundTimerAlignmentChrome} and Firefox~\cite{BackgroundTimerFirefox} throttles the time-out of the scripts that are running in inactive tabs to at least one second, and Firefox for Android increases up this threshold to $15$ minutes.
  The waiting for the page body to load and the throttling applied by major browser vendors to inactive tabs result in the high collection time of some fingerprints.
  Less than $1$\% of the entries of each device group (i.e., overall, mobiles, and desktops) are impacted by these high collection time and considered as outliers.

  \begin{table}
    \caption{
      The five attributes which collection time is missing in more than $1$\% of the overall, desktop, and mobile entries.
      The attributes are all asynchronous, and are ordered from the highest to the lowest proportion for the overall entries.
    }
    \label{tab:attributes-missing-collection-time}
    \centering
    \begin{tabular}{lccc}
      \toprule
        \textbf{Attribute} & \textbf{Overall} & \textbf{Desktop}
                           & \textbf{Mobile}                                \\
      \midrule
        WebRTC fingerprinting             &  24.04\%  &  26.10\%  & 20.31\% \\
        Audio FP advanced                 &  13.20\%  &  7.16\%   & 31.57\% \\
        Audio FP advanced frequency data  &  13.20\%  &  7.16\%   & 31.57\% \\
        Audio FP simple                   &  12.48\%  &  6.39\%   & 31.07\% \\
        List of speech synthesis voices   &  4.28\%   &  1.15\%   & 19.83\% \\
      \midrule
        Total entries                     & 5,714,738 & 4,610,640 & 808,318 \\
      \bottomrule
    \end{tabular}
  \end{table}

  We configured our fingerprinting script to collect the attributes in at most one second after the page body have been loaded.
  After this delay, the fingerprint is sent without waiting for all the attributes and the missing attributes do not have a collection time.
  All the attributes except five have less than $1$\% of the entries of each device group either having a high collection time (i.e., we deem the fingerprint an outlier) or missing the collection time.
  The five asynchronous attributes that miss collection time in more than $1$\% of the entries of each device group are presented in Table~\ref{tab:attributes-missing-collection-time}.
  The processing of these attributes take more time than the attributes which are a simple property.
  Moreover, they consist of complex processes that rely on components external to the browser (e.g., a real-time connection).
  As a result, the time limit can be reached before collecting the value and setting the collection time of these attributes.
  Mobile browsers show a higher proportion of missing collection time for the list of speech synthesis voices and the attributes related to the Web Audio API.
  This is due to the lack of support of these two APIs on mobile browsers at the time of the experiment~\cite{CanIUseSpeechSynthesis, CanIUseWebAudioAPI}.

\section{Keywords}
\label{app:keywords}
  In this section, we provide the keywords used to infer the browser family, the operating system, and whether the browser is a robot.
  We match these keywords on the \texttt{userAgent} JavaScript property which is first set to lower case.
  We manually compiled the keywords by searching for meaningful keywords inside the collected UserAgents.

  \subsection{Robot keywords}
    We ignore the fingerprints that come from robots (i.e., an automated tool and not a genuine visitor).
    To detect these fingerprints, we verify that the UserAgent does not contain the keywords, nor is set to the exact values, that are listed in Table~\ref{tab:blacklisted-robot-keywords}.

    \begin{table}
      \caption{
        The keywords and the exact UserAgent values that we use to detect robot browsers.
        The long values are cut at a blank space and displayed with indentations.
      }
      \label{tab:blacklisted-robot-keywords}
      \centering
      \begin{tabular}{l|l}
        \toprule
          \textbf{Blacklisted keyword} & \textbf{Blacklisted value} \\
        \midrule
          googlebot          & mozilla/4.0 (compatible; msie 7.0; windows nt 6.1; trident/7.0; slcc2; \\
          evaliant           & \hspace{0.5cm} .net clr 2.0.50727; .net clr 3.5.30729; .net clr 3.0.30729; \\
          bot.html           & \hspace{0.5cm} media center pc 6.0; .net4.0c; .net4.0e) \\
          voilabot           & mozilla/5.0 (x11; linux x86\_64) applewebkit/537.36 (khtml, like gecko) \\
          google web preview & \hspace{0.5cm} chrome/52.0.2743.116 safari/537.36 \\
          spider             & mozilla/5.0 (windows nt 6.3; rv:36.0) gecko/20100101 firefox/36.0 \\
          bingpreview        & mozilla/5.0 (macintosh; intel mac os x 10.10; rv:38.0) gecko/20100101 \\
                             & \hspace{0.5cm} firefox/38.0 \\
        \bottomrule
      \end{tabular}
    \end{table}

  \subsection{Device type}
    To infer the device type of a browser, we match keywords sequentially with the UserAgent of the browser.
    Table~\ref{tab:device-type-keywords} lists the keywords that we leverage to infer each device type.
    The set of keywords can overlap between two device types (e.g., the UserAgent of tablet browsers often contain keywords of mobile browsers like \texttt{mobile}).
    Due to this overlapping problem, we apply the following methodology.
    The \emph{mobile devices} are smartphones and do not include tablets.
    We check that their UserAgent contains a \texttt{mobile} keyword and no \texttt{tablet} nor \texttt{miscellaneous} keywords.
    To infer that a device is a \emph{tablet}, we check that its UserAgent contains a \texttt{tablet} keyword and no \texttt{miscellaneous} keywords.
    The \emph{miscellaneous devices} are game consoles and smart TVs.
    We check that their UserAgent contains a \texttt{miscellaneous} keyword.
    Finally, to infer that a device is a \emph{desktop} computer, we check that its UserAgent does not contain any of the \texttt{mobile},  \texttt{tablet}, or \texttt{miscellaneous} keywords.
    In Table~\ref{tab:device-type-keywords}, we omit a \texttt{miscellaneous} keyword due to its size: ``\texttt{opera/9.80 (linux i686; u; fr) presto/2.10.287 version/12.00 ; sc/ihd92 stb}''.

    \begin{table}
      \centering
      \minipage{0.48\textwidth}
        \caption{
          The keywords that we use to infer the device type of browsers.
        }
        \label{tab:device-type-keywords}
        \centering
        \begin{tabular}{l|l|l}
          \toprule
            \textbf{Mobile} & \textbf{Tablet} & \textbf{Miscellaneous} \\
          \midrule
            phone           & ipad            & wii                    \\
            mobile          & tablet          & playstation            \\
            android         & terra pad       & smart-tv               \\
            iphone          & tab             & smarttv                \\
            blackberry      &                 & googletv               \\
            wpdesktop       &                 & opera tv               \\
                            &                 & appletv                \\
                            &                 & nintendo               \\
                            &                 & xbox                   \\
          \bottomrule
        \end{tabular}
      \endminipage
      \hfill
      \centering
      \minipage{0.48\textwidth}
        \caption{
          The keywords that we use to infer the device type of browsers.
        }
        \label{tab:browser-family-keywords}
        \centering
        \begin{tabular}{lll}
          \toprule
            \textbf{Browser Family} & \textbf{Keywords} \\
          \midrule
            Firefox                 & Firefox           \\
            Edge                    & Edge              \\
            Internet Explorer       & MSIE, Trident/7.0 \\
            Samsung Internet        & SamsungBrowser    \\
            Chrome                  & Chrome            \\
            Safari                  & Safari            \\
          \bottomrule
        \end{tabular}
      \endminipage
    \end{table}

  \subsection{Browser and operating system families}
    Table~\ref{tab:browser-family-keywords} lists the keywords that we use to infer the family of a browser, and Table~\ref{tab:operating-system-family-keywords} lists the keywords that we use to infer the operating system family of a browser.
    As a keyword can be in the UserAgent of two different families, we check the keywords sequentially in the order presented in the tables, and classify a device in the first family for which a keyword matches.

    \begin{table}
      \caption{
        The keywords that we use to infer the operating system family of browsers.
      }
      \label{tab:operating-system-family-keywords}

      \centering
      \begin{tabular}{ll}
        \toprule
          \textbf{Operating System Family} & \textbf{Keywords} \\
        \midrule
          Windows~$10$  & windows nt 10.0 \\
          Windows~$7$   & windows nt 6.1 \\
          Other Windows & windows nt, windows 7, windows 98, windows 95, \\
                        & windows ce \\
          Mac~OS        & mac os x (but not ipad, nor iphone) \\
          Linux-based   & linux, cros, netbsd, freebsd, openbsd, fedora, ubuntu, \\
                        & mint \\
        \midrule
          Android       & android \\
          Windows Phone & windows phone \\
          iOS           & ipad, iphone (but not mac os x) \\
        \bottomrule
      \end{tabular}
    \end{table}

\section{Advanced verification mechanism}
\label{app:advanced-verification-mechanism}
  Section~\ref{sec:verification-mechanism-accuracy} describes a simple verification mechanism that checks that the number of identical attributes between the presented and the stored fingerprint is below a threshold.
  In this section, we present the results obtained using an \emph{advanced verification mechanism} that incorporates matching functions that authorize limited changes between the attribute values of the presented and the stored fingerprint.
  The methodology to obtain the datasets is the same as described in Section~\ref{sec:accuracy-of-the-simple-verification-mechanism}.

  \subsection{Attributes matching}
    The \emph{advanced verification mechanism} leverages matching functions for the comparison of the attributes.
    It counts the attributes that match between the two compared fingerprints, given the matching functions, and deems the evolution legitimate if this number is above a threshold.
    More formally, we seek to compare the stored fingerprint~$f$ to the presented fingerprint~$g$.
    To do so, we compare the values $f[a]$ and $g[a]$ of the attribute~$a$ for the fingerprints $f$ and $g$, using the matching function~$\approx^a$.
    The \emph{matching function}~$\approx^a$ of the attribute~$a$ verifies that the distance between $f[a]$ and $g[a]$ is below a threshold~$\theta^a$.
    Finally, we deem $g$ a legitimate evolution of $f$ if the total number of matching attributes between $f$ and $g$ is above a threshold~$\Theta$.

    Similarly to previous studies~\cite{ECK10, KCFLLL17, VLRR18}, we consider a distance measure that depends on the type of the attribute.
    We use the minimum edit distance~\cite{JM09MinimumEditDistance} for the textual attributes, the Jaccard distance~\cite{WWWLY16} for the set attributes, the absolute difference for the numerical attributes, and the identity function for the categorical attributes.
    The \emph{distance thresholds} of each attribute is obtained by training a Support Vector Machines~\cite{HDOPS98} model on the two classes of each month sample and extract the threshold from the resulting hyperplane.
    At the exception of the dynamic attributes that are required to be identical (i.e., the distance threshold is null) as they would contribute to a challenge-response mechanism~\cite{LABN19, REKP19}.

  \subsection{Distribution of matching attributes}
\begin{figure*}
      \minipage{0.49\columnwidth}
        \centering
        \includegraphics[width=0.9\columnwidth]{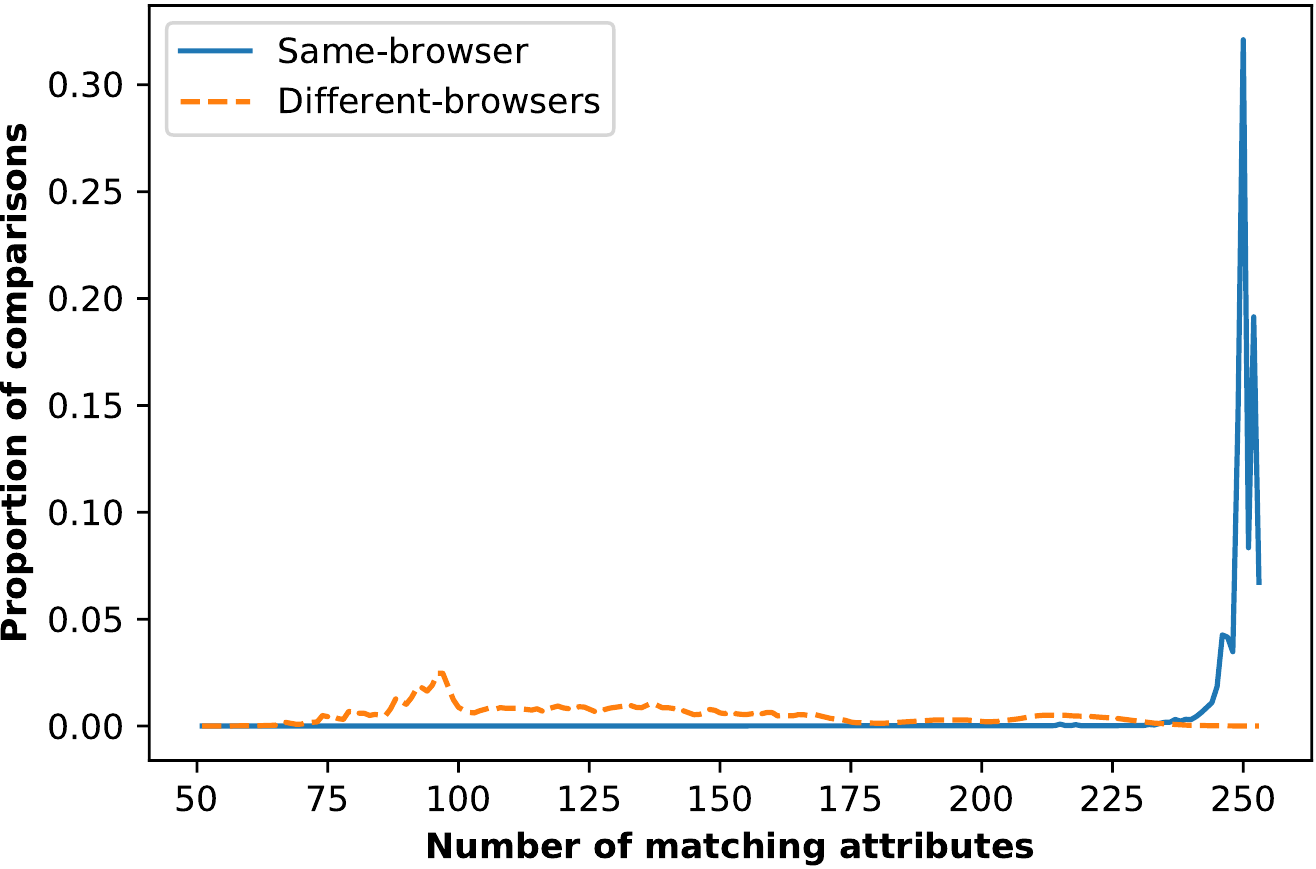}
        \caption{
          The number of matching attributes between the same-browser comparisons and the different-browsers comparisons.
        }
        \label{fig:matching-attributes-distribution-complete}
        \Description[
          The number of matching attributes between the same-browser comparisons and the different-browsers comparisons.
        ]{
          The number of matching attributes between the same-browser comparisons and the different-browsers comparisons.
          The two sets of comparisons are well separated, as $99$\% of the same-browser comparisons have at least $235$ matching attributes, and $99$\% of the different-browsers comparisons have fewer.
        }
      \endminipage
      \hfill
      \minipage{0.49\columnwidth}
        \centering
        \includegraphics[width=0.9\columnwidth]{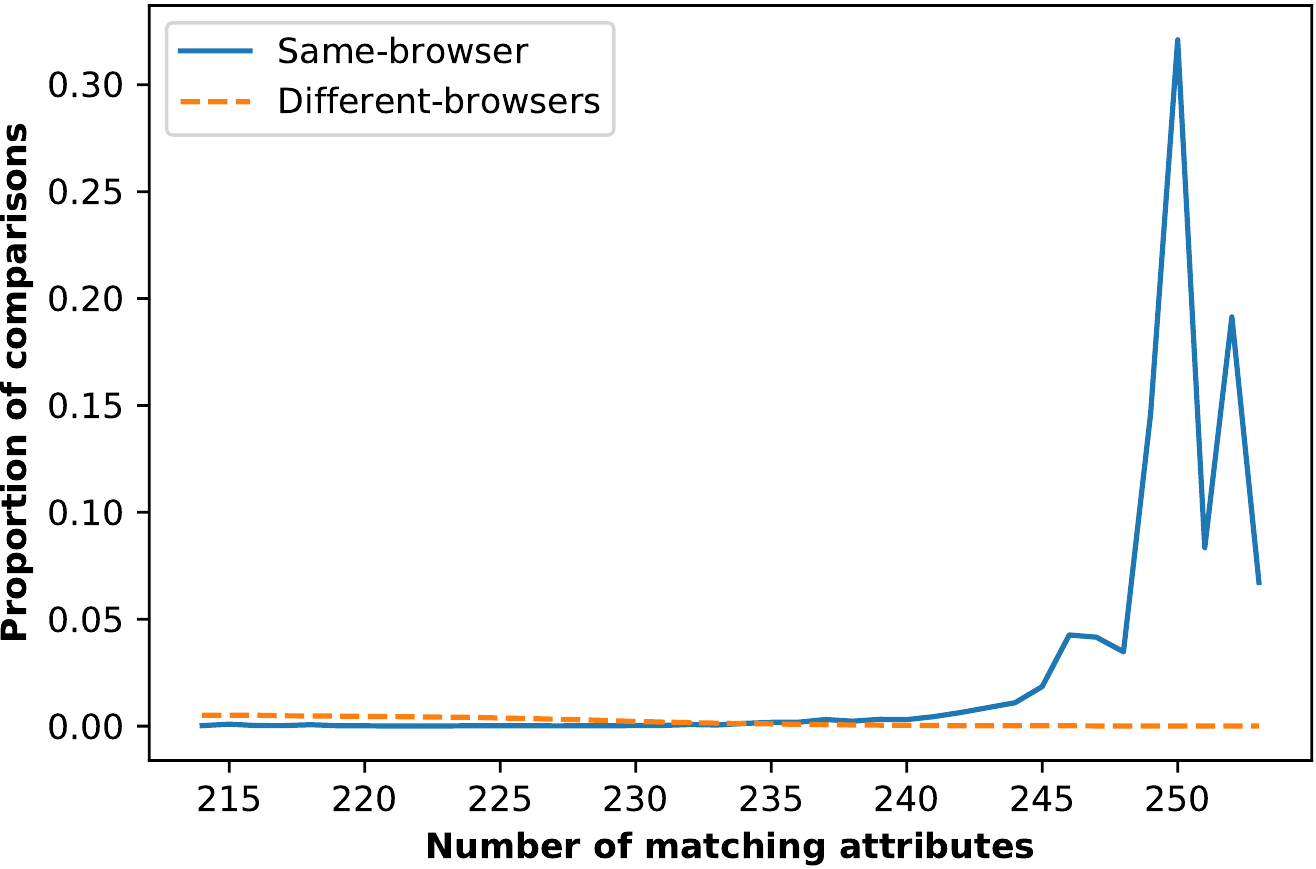}
        \caption{
          The number of matching attributes between the same-browser comparisons and the different-browsers comparisons, starting from $214$ matching attributes.
        }
        \label{fig:matching-attributes-distribution-zoom}
        \Description[
          The number of matching attributes between the same-browser comparisons and the different-browsers comparisons, starting from $214$ matching attributes.
        ]{
          The number of matching attributes between the same-browser comparisons and the different-browsers comparisons, starting from $214$ matching attributes.
          The two sets of comparisons are well separated, as $99$\% of the same-browser comparisons have at least $235$ matching attributes, and $99$\% of the different-browsers comparisons have fewer.
        }
      \endminipage
    \end{figure*}

    Figure~\ref{fig:matching-attributes-distribution-complete} displays the distribution of the matching attributes between the same-browser comparisons and the different-browsers comparisons, starting from $51$ matching attributes as there are no observed value below.
    Figure~\ref{fig:matching-attributes-distribution-zoom} presents a focus that starts from $214$ matching attributes, below which there are less than $0.001$ of the same-browser comparisons.
    We can observe that the two sets of comparisons are well separated, as $99$\% of the same-browser comparisons have at least $235$ matching attributes, and $99$\% of the different-browsers comparisons have fewer.
    The different-browsers comparisons have generally a fewer, and a more diverse, number of matching attributes compared to the same-browser comparisons.
    The different-browsers comparisons have between $51$ and $253$ matching attributes, with an average of $134.59$ attributes and a standard deviation of $43.25$ attributes.
    The same-browser comparisons have between $81$ and $253$ matching attributes, with an average of $249.45$ attributes and a standard deviation of $3.69$ attributes.

  \subsection{Distribution of match rates}
\begin{figure}
      \centering
      \includegraphics[width=0.70\columnwidth]{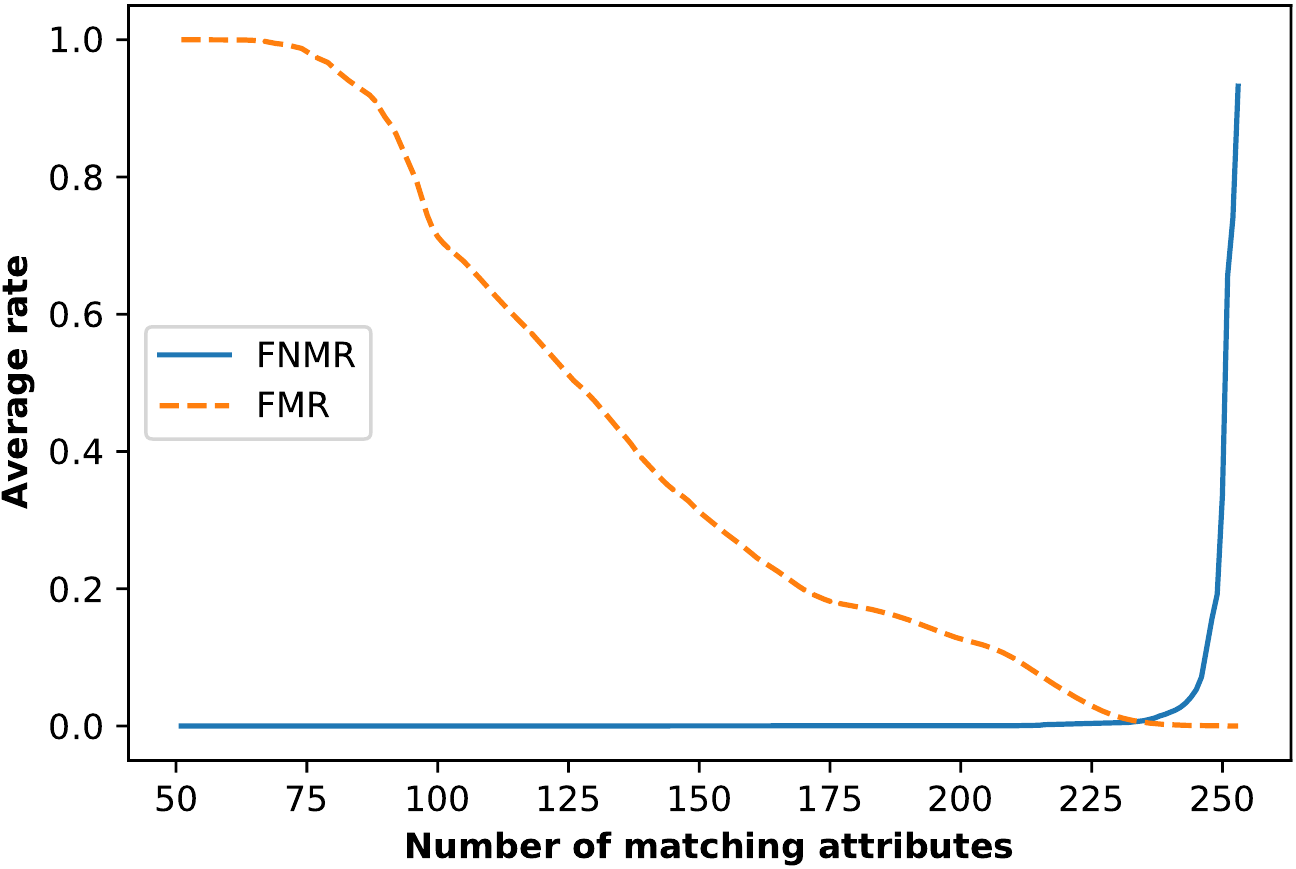}
      \caption{
        False match rate (FMR) and false non-match rate (FNMR) given the required number of matching attributes, averaged among the month six samples.
      }
      \label{fig:matching-fmr-fnmr-distribution}
      \Description[
        False match rate (FMR) and false non-match rate (FNMR) given the required number of matching attributes, averaged among the month six samples.
      ]{
        False match rate (FMR) and false non-match rate (FNMR) given the required number of matching attributes, averaged among the month six samples.
        The equal error rate, which is the rate where both the FMR and the FNMR are equal, is of $0.66$\% and is achieved for $234$ matching attributes.
      }
    \end{figure}

    Figure~\ref{fig:matching-fmr-fnmr-distribution} displays the false match rate (FMR) which is the proportion of the same-browser comparisons that are classified as different-browsers comparisons, and the false non-match rate (FNMR) which is the inverse.
    The displayed results are the average for each number of matching attributes among the six month-samples.
    As there are no same-browser comparisons that have less than $235$ matching attributes, the FNMR is null until this value.
    However, after exceeding this threshold, the FNMR increases as the same-browser comparisons begin to be classified as different-browsers comparisons.
    The equal error rate, which is the rate where both the FMR and the FNMR are equal, is of $0.66$\% and is achieved for $234$ matching attributes.

  \subsection{Comparison with identical matching}
    The matching functions of the advanced verification mechanism leads to more matching attributes than identical attributes between two fingerprints.
    The higher number of matching attributes happens for the same-browser comparisons, but also for the different-browsers comparisons.
    This reduces the False Non-Match Rate (FNMR), but increases the False-Match Rate (FMR).
    Due to the FMR being higher, the equal error rate is slightly higher for the advanced verification mechanism that leverages matching functions.

    Table~\ref{tab:comparison-identical-vs-matching-attributes} compares the results of the simple verification mechanism that leverages the identical attributes, and the advanced verification mechanism that leverages the matching attributes.
    Considering the matching functions only increases the average number of matching attributes for the same-browser comparisons by $0.81$, whereas the increase is greater for the different-browsers comparisons at $7.18$.
    The matching functions seem to contribute more to falsely linking different-browsers comparisons than same-browser comparisons.
    Due to this, the equal error rate is slightly higher for the advanced verification mechanism than for the simple one, respectively at $0.66$\% against $0.61$\%.
    Finally, we remark that the range of the identical attributes for the same-browser comparisons only goes up to $252$~attributes.
    This is explained by our deduplication step during the preprocessing of the dataset (see Section~\ref{sec:dataset-deduplication}) that removes the consecutive fingerprints that are identical.
    As a result, the same-browser comparisons have fewer than $253$ identical attributes and are forcibly different.
    As for the different-browsers comparisons, it always goes up to $253$~attributes as coincidence can make the fingerprints of different browsers match.

    \begin{table}
      \caption{
        Comparison between the simple verification mechanism that uses identical attributes and the advanced verification mechanism that uses matching attributes.
      }
      \label{tab:comparison-identical-vs-matching-attributes}
      \centering
      \begin{tabular}{lcc}
        \toprule
          \textbf{Result}              & \textbf{Simple} & \textbf{Advanced} \\
        \midrule
          Same-browser: range of identical or matching attributes
                                       & [72; 252]       & [81; 253]         \\
          Same-browser: average identical or matching attributes
                                       & 248.64          & 249.45            \\
          Same-browser: standard deviation  & 3.91            & 3.69              \\
        \midrule
          Different-browsers: range of identical or matching attributes
                                       & [34; 253]       & [51; 253]         \\
          Different-browsers: average identical or matching attributes
                                       & 127.41          & 134.59            \\
          Different-browsers: standard deviation & 44.06           & 43.25             \\
        \midrule
          Equal error rate             & 0.62\%          & 0.66\%            \\
          Threshold of identical or matching attributes
                                       & 232             & 234               \\
        \bottomrule
      \end{tabular}
    \end{table}

\section{Attributes list and property}
\label{app:attribute-list-and-property}
  In this section, we provide the complete list of our $262$ fingerprinting attributes and their property.
  Table~\ref{tab:attributes-table} lists our attributes with their number of distinct values seen during the experiment (Values), their normalized entropy (N. Ent.), their sameness rate (\% Same), their median size (Size), and their median collection time (Time).
  We refer to Section~\ref{sec:attribute-wise-analysis} for the distribution of the properties of the attributes: the number of distinct values, the normalized entropy, the minimum normalized conditional entropy, the sameness rate, the median collection time, and the median size.
  Table~\ref{tab:properties-distribution-information} provides the minimum, the average, the maximum, and the standard deviation of these properties among the attributes.

  \begin{table}
    \caption{
      The minimum, the average, the maximum, and the standard deviation (Std. dev.) of the distinct values, the normalized entropy, the minimum normalized conditional entropy, the median collection time in seconds, the median size in bytes, and the sameness rate of the attributes.
    }
    \label{tab:properties-distribution-information}
    \centering
    \begin{tabular}{lcccc}
      \toprule
        \textbf{Property}  &  \textbf{Minimum}  &  \textbf{Average}  &
        \textbf{Maximum}  &  \textbf{Std. dev.}  \\
      \midrule
        Distinct values                      &  1      &  7,633  &  671,254
                                             &  51,298  \\
        Normalized entropy                   &  0.000  &  0.090  &  0.420
                                             &  0.094   \\
        Minimum normalized conditional entropy &  0.000  &  0.008  &  0.181
                                             &  0.021   \\
        Sameness rate                        &  0.470  &  0.982  &  1.000
                                             &  0.069   \\
        Median collection time (seconds)     &  0.000  &  0.125  &  2.179
                                             &  0.425   \\
        Median size (bytes)                  &  1      &  23.51  &  502
                                             &  60.84   \\
      \bottomrule
    \end{tabular}
  \end{table}

  \subsection{Attributes list}
    To stay concise, we replace the name of common JavaScript objects or of API calls by abbreviations.
    We denote D the JavaScript \texttt{document} object, M the \texttt{Math} object, N the \texttt{navigator} object, S the \texttt{screen} object, and W the \texttt{window} object.
    Additionally, we denote A an initialized \texttt{Audio Context}, AA an initialized \texttt{AudioAnalyser}, and AD the \texttt{A.destination} property.
    Finally, we denote WG an initialized \texttt{WebGL Context}, WM the \texttt{WG.MAX\_} prefix, and WI the \texttt{WG.IMPLEMENTATION\_} prefix.

    Due to the diversity of JavaScript engines, some properties are accessible through different names (e.g., prefixed by \texttt{moz} for Firefox or \texttt{ms} for Internet Explorer).
    We denote \texttt{A.[B, C]} a property that is accessed either through \texttt{A.B} or \texttt{A.C}.
    If there is only one element inside the brackets, this element is optional.
    We denote \texttt{[...]} a part that is omitted but described in the corresponding attribute description.

  \begin{table}
    \caption{
      Browser fingerprinting attributes with their number of distinct values (Values), their normalized entropy (N. Ent.), their minimum normalized conditional entropy (MinNCE), their stability (\% Same), their median size (Size), and their median collection time (Time).
      The comparisons of the attribute with itself and with the source attributes from which we derive the extracted ones are ignored for the MinNCE result.
    }
    \label{tab:attributes-table}
    \begin{minipage}{\columnwidth}
      \begin{tabular}{lcccccc}
        \toprule
          \textbf{Attribute} & \textbf{Values} & \textbf{N. Ent.} &
          \textbf{MinNCE} & \textbf{\% Same} & \textbf{Size} & \textbf{Time} \\
        \midrule
          N.userAgent & 38,863 & 0.394 & 0.046 & 0.978 & 115 & 0.000 \\
          Listing of N & 1,660 & 0.207 & 0.009 & 0.989 & 502 & 0.001 \\
          Listing of screen & 82 & 0.129 & 0.000 & 0.999 & 209 & 0.000 \\
          N.language & 228 & 0.066 & 0.006 & 0.999 & 2 & 0.000 \\
          N.languages & 1,448 & 0.094 & 0.010 & 0.998 & 17 & 0.000 \\
          N.userLanguage & 124 & 0.036 & 0.001 & 1.000 & 1 & 0.000 \\
          N.systemLanguage & 115 & 0.037 & 0.001 & 1.000 & 1 & 0.000 \\
          N.browserLanguage & 52 & 0.036 & 0.000 & 1.000 & 1 & 0.000 \\
          N.platform & 32 & 0.068 & 0.000 & 1.000 & 5 & 0.000 \\
          N.appName & 5 & 0.003 & 0.000 & 1.000 & 8 & 0.000 \\
          N.appVersion & 37,310 & 0.342 & 0.000 & 0.984 & 107 & 0.000 \\
          N.appMinorVersion & 10 & 0.035 & 0.000 & 1.000 & 1 & 0.000 \\
          N.product & 2 & 0.003 & 0.000 & 1.000 & 5 & 0.000 \\
          N.productSub & 10 & 0.067 & 0.000 & 1.000 & 8 & 0.000 \\
          N.vendor & 21 & 0.064 & 0.000 & 1.000 & 1 & 0.000 \\
          N.vendorSub & 2 & 0.035 & 0.000 & 1.000 & 1 & 0.000 \\
          N.cookieEnabled & 1 & 0.000 & 0.000 & 1.000 & 4 & 0.000 \\
          N.cpuClass & 6 & 0.039 & 0.000 & 1.000 & 1 & 0.000 \\
          N.oscpu & 60 & 0.071 & 0.000 & 1.000 & 1 & 0.000 \\
          N.hardwareConcurrency & 28 & 0.086 & 0.025 & 0.999 & 1 & 0.000 \\
          N.buildID & 1,351 & 0.076 & 0.010 & 0.989 & 1 & 0.000 \\
          {[N.security, D.security[Policy]]} & 30 & 0.038 & 0.001 & 1.000 & 7 & 0.000 \\
          N.permissions & 3 & 0.045 & 0.000 & 1.000 & 1 & 0.000 \\
          W.Notification.permission & 5 & 0.043 & 0.001 & 0.999 & 7 & 0.000 \\
          W.Notification.maxActions & 3 & 0.041 & 0.000 & 1.000 & 1 & 0.000 \\
          N.[msM, m]axTouchPoints & 42 & 0.098 & 0.004 & 0.999 & 3 & 0.000 \\
          D.createEvent("TouchEvent") support & 3 & 0.032 & 0.000 & 1.000 & 1 & 0.000 \\
          W.ontouchstart support & 3 & 0.032 & 0.000 & 1.000 & 1 & 0.000 \\
          N.javaEnabled() & 4 & 0.045 & 0.008 & 0.997 & 1 & 0.000 \\
          N.taintEnabled() & 3 & 0.045 & 0.000 & 1.000 & 1 & 0.000 \\
          {[[N, W].doNotTrack, N.msDoNotTrack]} & 11 & 0.085 & 0.012 & 1.000 & 6 & 0.000 \\
          N.connection support & 3 & 0.022 & 0.000 & 1.000 & 1 & 0.000 \\
          N.connection.type & 12 & 0.028 & 0.003 & 0.992 & 1 & 0.000 \\
          N.connection.downlink & 91 & 0.032 & 0.003 & 0.995 & 1 & 0.000 \\
          N.[mozC, c]onnection.bandwidth & 6 & 0.023 & 0.000 & 1.000 & 3 & 0.000 \\
          N.mediaDevices support & 3 & 0.044 & 0.000 & 0.999 & 1 & 0.000 \\
          N.mediaDevices.getSupportedConstraints() & 12 & 0.090 & 0.000 & 0.997 & 144 & 0.000 \\
          W.Intl.Collator().resolvedOptions() & 311 & 0.097 & 0.000 & 0.999 & 115 & 0.005 \\
          W.Intl.DateTimeFormat().resolvedOptions() & 1,849 & 0.154 & 0.011 & 0.996 & 111 & 0.003 \\
          W.Intl.NumberFormat().resolvedOptions() & 260 & 0.070 & 0.000 & 0.999 & 138 & 0.001 \\
          W.Intl.v8BreakIterator().resolvedOptions() & 75 & 0.046 & 0.000 & 0.999 & 1 & 0.000 \\
          N.getGamepads() & 18 & 0.090 & 0.000 & 0.998 & 1 & 0.001 \\
        \bottomrule
      \end{tabular}
    \end{minipage}
  \end{table}
  \begin{table}
    \begin{minipage}{\columnwidth}
      \begin{tabular}{lcccccc}
        \toprule
          \textbf{Attribute} & \textbf{Values} & \textbf{N. Ent.} &
          \textbf{MinNCE} & \textbf{\% Same} & \textbf{Size} & \textbf{Time} \\
        \midrule
          W.InstallTrigger.enabled() & 4 & 0.037 & 0.000 & 1.000 & 1 & 0.000 \\
          W.InstallTrigger.updateEnabled() & 4 & 0.037 & 0.000 & 1.000 & 1 & 0.000 \\
          N.msManipulationViewsEnabled & 5 & 0.052 & 0.000 & 1.000 & 3 & 0.000 \\
          N.[msP, p]ointerEnabled & 9 & 0.051 & 0.000 & 1.000 & 3 & 0.000 \\
          D.msCapsLockWarningOff & 3 & 0.039 & 0.000 & 1.000 & 1 & 0.000 \\
          D.msCSSOMElementFloatMetrics & 4 & 0.039 & 0.000 & 1.000 & 1 & 0.000 \\
          N.[msW, w]ebdriver & 6 & 0.048 & 0.001 & 1.000 & 3 & 0.000 \\
          W.Debug.debuggerEnabled & 5 & 0.042 & 0.000 & 0.989 & 1 & 0.000 \\
          W.Debug.setNonUserCodeExceptions & 4 & 0.042 & 0.000 & 0.989 & 1 & 0.000 \\
          new Date(2016, 1, 1).getTimezoneOffset() & 60 & 0.008 & 0.001 & 0.999 & 2 & 0.000 \\
          Different Timezone at 01/01 and 06/01 & 3 & 0.005 & 0.002 & 0.999 & 1 & 0.000 \\
          S.width & 1,280 & 0.192 & 0.005 & 0.987 & 4 & 0.000 \\
          S.height & 1,016 & 0.188 & 0.015 & 0.987 & 3 & 0.000 \\
          W.screenX & 3,071 & 0.125 & 0.047 & 0.925 & 1 & 0.000 \\
          W.screenY & 1,181 & 0.126 & 0.049 & 0.925 & 1 & 0.000 \\
          S.availWidth & 1,746 & 0.202 & 0.016 & 0.985 & 4 & 0.000 \\
          S.availHeight & 1,353 & 0.268 & 0.058 & 0.984 & 3 & 0.000 \\
          S.availTop & 460 & 0.060 & 0.003 & 1.000 & 1 & 0.000 \\
          S.availLeft & 372 & 0.048 & 0.005 & 0.999 & 1 & 0.000 \\
          S.(pixelDepth, colorDepth) & 14 & 0.031 & 0.001 & 1.000 & 5 & 0.000 \\
          S.deviceXDPI & 249 & 0.073 & 0.000 & 0.993 & 1 & 0.000 \\
          S.deviceYDPI & 249 & 0.073 & 0.000 & 0.993 & 1 & 0.000 \\
          S.systemXDPI & 75 & 0.053 & 0.000 & 0.999 & 1 & 0.000 \\
          S.systemYDPI & 75 & 0.053 & 0.000 & 0.999 & 1 & 0.000 \\
          S.logicalXDPI & 6 & 0.039 & 0.000 & 1.000 & 1 & 0.000 \\
          S.logicalYDPI & 6 & 0.039 & 0.000 & 1.000 & 1 & 0.000 \\
          W.innerWidth & 2,572 & 0.263 & 0.023 & 0.957 & 4 & 0.000 \\
          W.innerHeight & 2,297 & 0.388 & 0.181 & 0.906 & 3 & 0.000 \\
          W.outerWidth & 2,481 & 0.293 & 0.074 & 0.909 & 4 & 0.000 \\
          W.outerHeight & 4,046 & 0.327 & 0.117 & 0.872 & 3 & 0.000 \\
          W.devicePixelRatio & 2,035 & 0.103 & 0.026 & 0.992 & 1 & 0.000 \\
          W.mozInnerScreenX & 3,682 & 0.065 & 0.011 & 0.991 & 1 & 0.000 \\
          W.mozInnerScreenY & 3,170 & 0.102 & 0.020 & 0.991 & 1 & 0.000 \\
          W.offscreenBuffering & 4 & 0.067 & 0.000 & 1.000 & 4 & 0.000 \\
          S.orientation & 3 & 0.044 & 0.000 & 1.000 & 1 & 0.000 \\
          S.[orientation.type, [moz, ms]Orientation] & 26 & 0.107 & 0.003 & 0.994 & 21 & 0.000 \\
          S.orientation.angle & 7 & 0.050 & 0.002 & 0.996 & 1 & 0.000 \\
          W.localStorage support & 4 & 0.001 & 0.001 & 1.000 & 1 & 0.001 \\
          W.sessionStorage support & 4 & 0.000 & 0.000 & 1.000 & 1 & 0.000 \\
          W.indexedDB support & 3 & 0.002 & 0.000 & 1.000 & 1 & 0.000 \\
          W.openDatabase support & 3 & 0.045 & 0.000 & 1.000 & 1 & 0.000 \\
          W.caches support & 3 & 0.045 & 0.000 & 1.000 & 1 & 0.000 \\
          M.tan(-1e300) & 15 & 0.087 & 0.002 & 1.000 & 19 & 0.000 \\
          M.tan(3.14159265359 * 0.3333 * 1e300) & 12 & 0.085 & 0.000 & 0.999 & 18 & 0.000 \\
          M.acos(0.000000000000001) & 4 & 0.058 & 0.000 & 1.000 & 18 & 0.000 \\
          M.acosh(1.000000000001) & 7 & 0.071 & 0.000 & 1.000 & 24 & 0.000 \\
        \bottomrule
      \end{tabular}
    \end{minipage}
  \end{table}
  \begin{table}
    \begin{minipage}{\columnwidth}
      \begin{tabular}{lcccccc}
        \toprule
          \textbf{Attribute} & \textbf{Values} & \textbf{N. Ent.} &
          \textbf{MinNCE} & \textbf{\% Same} & \textbf{Size} & \textbf{Time} \\
        \midrule
          M.asinh(0.00001) & 6 & 0.071 & 0.000 & 1.000 & 23 & 0.000 \\
          M.asinh(1e300) & 6 & 0.058 & 0.000 & 1.000 & 17 & 0.000 \\
          M.atan(2) & 3 & 0.018 & 0.000 & 1.000 & 18 & 0.000 \\
          M.atan2(0.01, 1000) & 3 & 0.045 & 0.000 & 1.000 & 23 & 0.000 \\
          M.atanh(0.0001) & 5 & 0.070 & 0.000 & 1.000 & 22 & 0.000 \\
          M.cosh(15) & 8 & 0.058 & 0.000 & 1.000 & 18 & 0.000 \\
          M.exp(-1e2) & 8 & 0.028 & 0.000 & 1.000 & 21 & 0.000 \\
          M.exp(1e2) & 9 & 0.028 & 0.000 & 1.000 & 22 & 0.000 \\
          M.LOG2E & 3 & 0.039 & 0.000 & 1.000 & 18 & 0.000 \\
          M.LOG10E & 3 & 0.000 & 0.000 & 1.000 & 18 & 0.000 \\
          M.E & 2 & 0.000 & 0.000 & 1.000 & 17 & 0.000 \\
          M.LN10 & 3 & 0.000 & 0.000 & 1.000 & 17 & 0.000 \\
          D.defaultCharset & 71 & 0.075 & 0.001 & 0.999 & 1 & 0.000 \\
          Width and height of fallback font text & 2,347 & 0.199 & nan & 0.998 & 11 & 0.099 \\
          W.[performance, console].jsHeapSizeLimit & 24 & 0.083 & 0.002 & 0.991 & 3 & 0.000 \\
          W.menubar.visible & 5 & 0.035 & 0.000 & 1.000 & 4 & 0.000 \\
          W.isSecureContext & 4 & 0.045 & 0.000 & 0.999 & 5 & 0.000 \\
          S.fontSmoothingEnabled & 4 & 0.042 & 0.001 & 1.000 & 1 & 0.000 \\
          new Date(0) & 1,846 & 0.118 & 0.004 & 0.998 & 82 & 0.004 \\
          new Date("0001-1-1") & 2,107 & 0.150 & 0.010 & 0.999 & 60 & 0.002 \\
          new Date(0) then setFullYear(0) & 2,376 & 0.136 & 0.013 & 0.998 & 61 & 0.001 \\
          Detection of an adblocker & 19 & 0.002 & 0.001 & 0.999 & 1 & 2.124 \\
          Firebug resource detection & 3 & 0.037 & 0.000 & 1.000 & 1 & 0.054 \\
          YahooToolbar resource detection & 3 & 0.037 & 0.000 & 1.000 & 1 & 0.056 \\
          EasyScreenshot resource detection & 3 & 0.037 & 0.000 & 1.000 & 1 & 0.056 \\
          Ghostery resource detection & 3 & 0.037 & 0.000 & 1.000 & 1 & 0.056 \\
          Kaspersky resource detection & 3 & 0.037 & 0.000 & 1.000 & 1 & 0.057 \\
          VideoDownloadHelper resource detection & 3 & 0.038 & 0.001 & 0.998 & 1 & 0.056 \\
          GTranslate resource detection & 3 & 0.037 & 0.000 & 1.000 & 1 & 0.058 \\
          Privowny resource detection & 2 & 0.037 & 0.000 & 1.000 & 1 & 0.057 \\
          Privowny page content change & 3 & 0.000 & 0.000 & 1.000 & 3 & 2.162 \\
          UBlock page content change & 4 & 0.000 & 0.000 & 1.000 & 1 & 2.151 \\
          Pinterest page content change & 10 & 0.001 & 0.001 & 0.999 & 1 & 2.151 \\
          Grammarly page content change & 3 & 0.000 & 0.000 & 1.000 & 1 & 2.152 \\
          Adguard page content change & 3 & 0.000 & 0.000 & 1.000 & 1 & 2.179 \\
          Evernote page content change & 3 & 0.000 & 0.000 & 1.000 & 1 & 2.156 \\
          TOTL page content change & 3 & 0.000 & 0.000 & 1.000 & 1 & 2.153 \\
          IE Tab page content change & 11 & 0.000 & 0.000 & 1.000 & 1 & 2.170 \\
          WebRTC fingerprinting & 671,254 & 0.294 & 0.144 & 0.765 & 1 & 0.771 \\
          WG.SHADING\_LANGUAGE\_VERSION & 23 & 0.103 & 0.000 & 0.996 & 18 & 0.001 \\
          WG.VERSION & 247 & 0.123 & 0.008 & 0.995 & 10 & 0.000 \\
          WG.VENDOR & 11 & 0.080 & 0.000 & 0.997 & 7 & 0.000 \\
          WG.RENDERER & 14 & 0.089 & 0.000 & 0.996 & 12 & 0.000 \\
          WG.ALIASED\_POINT\_SIZE\_RANGE & 42 & 0.129 & 0.006 & 0.996 & 5 & 0.000 \\
          WG.ALIASED\_LINE\_WIDTH\_RANGE & 30 & 0.077 & 0.003 & 0.996 & 3 & 0.000 \\
          WM.VIEWPORT\_DIMS & 13 & 0.107 & 0.009 & 0.995 & 11 & 0.000 \\
        \bottomrule
      \end{tabular}
    \end{minipage}
  \end{table}
  \begin{table}
    \begin{minipage}{\columnwidth}
      \begin{tabular}{lcccccc}
        \toprule
          \textbf{Attribute} & \textbf{Values} & \textbf{N. Ent.} &
          \textbf{MinNCE} & \textbf{\% Same} & \textbf{Size} & \textbf{Time} \\
        \midrule
          WG.SUBPIXEL\_BITS & 9 & 0.039 & 0.001 & 0.997 & 1 & 0.000 \\
          WG.SAMPLE\_BUFFERS & 5 & 0.039 & 0.000 & 0.996 & 1 & 0.000 \\
          WG.SAMPLES & 9 & 0.075 & 0.001 & 0.992 & 1 & 0.000 \\
          WG.COMPRESSED\_TEXTURE\_FORMATS & 3 & 0.034 & 0.000 & 0.998 & 23 & 0.000 \\
          WM.VERTEX\_UNIFORM\_VECTORS & 18 & 0.118 & 0.006 & 0.996 & 3 & 0.000 \\
          WM.COMBINED\_TEXTURE\_IMAGE\_UNITS & 19 & 0.079 & 0.003 & 0.996 & 2 & 0.000 \\
          WM.FRAGMENT\_UNIFORM\_VECTORS & 18 & 0.109 & 0.004 & 0.996 & 3 & 0.000 \\
          WM.CUBE\_MAP\_TEXTURE\_SIZE & 11 & 0.084 & 0.008 & 0.995 & 5 & 0.000 \\
          WG.STENCIL\_VALUE\_MASK & 8 & 0.050 & 0.000 & 0.996 & 10 & 0.000 \\
          WG.STENCIL\_WRITEMASK & 7 & 0.050 & 0.000 & 0.996 & 10 & 0.000 \\
          WG.STENCIL\_BACK\_VALUE\_MASK & 8 & 0.050 & 0.000 & 0.996 & 10 & 0.000 \\
          WG.STENCIL\_BACK\_WRITEMASK & 7 & 0.050 & 0.000 & 0.996 & 10 & 0.000 \\
          WM.TEXTURE\_SIZE & 10 & 0.081 & 0.009 & 0.995 & 5 & 0.000 \\
          WG.DEPTH\_BITS & 7 & 0.047 & 0.000 & 0.996 & 2 & 0.000 \\
          WM.VARYING\_VECTORS & 19 & 0.121 & 0.009 & 0.996 & 2 & 0.000 \\
          WI.COLOR\_READ\_FORMAT & 7 & 0.073 & 0.003 & 0.994 & 4 & 0.000 \\
          WM.RENDERBUFFER\_SIZE & 11 & 0.080 & 0.003 & 0.995 & 5 & 0.000 \\
          WG.STENCIL\_BITS & 5 & 0.016 & 0.000 & 0.997 & 1 & 0.000 \\
          WM.TEXTURE\_IMAGE\_UNITS & 7 & 0.033 & 0.000 & 0.997 & 2 & 0.000 \\
          WM.VERTEX\_ATTRIBS & 8 & 0.017 & 0.000 & 0.997 & 2 & 0.000 \\
          WM.VERTEX\_TEXTURE\_IMAGE\_UNITS & 9 & 0.057 & 0.001 & 0.996 & 2 & 0.000 \\
          WI.COLOR\_READ\_TYPE & 6 & 0.041 & 0.000 & 0.996 & 4 & 0.000 \\
          WM.TEXTURE\_MAX\_ANISOTROPY\_EXT & 9 & 0.029 & 0.000 & 0.997 & 2 & 0.001 \\
          WG.getContextAttributes() & 54 & 0.114 & 0.009 & 0.995 & 138 & 0.000 \\
          WG.getSupportedExtensions() & 535 & 0.209 & 0.027 & 0.990 & 401 & 0.008 \\
          WebGL vendor (unmasked) & 27 & 0.115 & 0.000 & 0.995 & 9 & 0.000 \\
          WebGL renderer (unmasked) & 3,786 & 0.268 & 0.073 & 0.991 & 20 & 0.000 \\
          WebGL precision format & 25 & 0.071 & 0.001 & 0.996 & 114 & 0.001 \\
          Our designed WebGL canvas & 1,158 & 0.263 & 0.023 & 0.990 & 64 & 0.041 \\
          Width and position of a created \texttt{div} & 17,832 & 0.324 & nan & 0.940 & 18 & 0.084 \\
          Colors of layout components & 7,707 & 0.153 & nan & 0.986 & 492 & 0.089 \\
          Size of bounding boxes of a created \texttt{div} & 16,396 & 0.369 & nan & 0.470 & 31 & 0.197 \\
          Presence of fonts & 17,960 & 0.305 & 0.110 & 0.996 & 198 & 0.450 \\
          Support of video codecs & 84 & 0.114 & 0.001 & 0.999 & 78 & 0.002 \\
          Support of audio codecs & 52 & 0.128 & 0.002 & 0.999 & 61 & 0.001 \\
          Support of streaming codecs & 50 & 0.132 & 0.010 & 0.999 & 133 & 0.002 \\
          Support of recording codecs & 7 & 0.069 & 0.000 & 0.999 & 140 & 0.001 \\
          List of speech synthesis voices & 3,967 & 0.204 & 0.034 & 0.945 & 250 & 0.546 \\
          N.plugins & 314,518 & 0.394 & 0.100 & 0.950 & 134 & 0.001 \\
          N.mimeTypes & 174,876 & 0.311 & 0.017 & 0.982 & 112 & 0.000 \\
          A.state & 5 & 0.082 & 0.000 & 0.999 & 7 & 0.000 \\
          A.sampleRate & 16 & 0.070 & 0.019 & 0.997 & 5 & 0.000 \\
          AD.channelCount & 5 & 0.036 & 0.000 & 1.000 & 1 & 0.000 \\
          AD.channelCountMode & 4 & 0.036 & 0.000 & 1.000 & 8 & 0.000 \\
          AD.channelInterpretation & 4 & 0.036 & 0.000 & 1.000 & 8 & 0.000 \\
          AD.maxChannelCount & 20 & 0.058 & 0.003 & 1.000 & 1 & 0.000 \\
        \bottomrule
      \end{tabular}
    \end{minipage}
  \end{table}
  \begin{table}
    \begin{minipage}{\columnwidth}
      \begin{tabular}{lcccccc}
        \toprule
          \textbf{Attribute} & \textbf{Values} & \textbf{N. Ent.} &
          \textbf{MinNCE} & \textbf{\% Same} & \textbf{Size} & \textbf{Time} \\
        \midrule
          AD.numberOfInputs & 3 & 0.035 & 0.000 & 1.000 & 1 & 0.000 \\
          AD.numberOfOutputs & 3 & 0.035 & 0.000 & 1.000 & 1 & 0.000 \\
          AA.channelCount & 5 & 0.067 & 0.000 & 1.000 & 1 & 0.001 \\
          AA.channelCountMode & 5 & 0.037 & 0.000 & 1.000 & 3 & 0.000 \\
          AA.channelInterpretation & 4 & 0.036 & 0.000 & 1.000 & 8 & 0.000 \\
          AA.numberOfInputs & 4 & 0.036 & 0.000 & 1.000 & 1 & 0.000 \\
          AA.numberOfOutputs & 4 & 0.036 & 0.000 & 1.000 & 1 & 0.000 \\
          AA.fftSize & 3 & 0.035 & 0.000 & 1.000 & 4 & 0.000 \\
          AA.frequencyBinCount & 3 & 0.035 & 0.000 & 1.000 & 4 & 0.000 \\
          AA.maxDecibels & 3 & 0.035 & 0.000 & 1.000 & 3 & 0.000 \\
          AA.minDecibels & 3 & 0.035 & 0.000 & 1.000 & 4 & 0.000 \\
          AA.smoothingTimeConstant & 4 & 0.046 & 0.000 & 0.998 & 3 & 0.000 \\
          Audio FP simple & 337 & 0.153 & 0.004 & 0.958 & 18 & 1.380 \\
          Audio FP advanced & 561 & 0.147 & 0.001 & 0.953 & 17 & 1.644 \\
          Audio FP advanced frequency data & 546 & 0.161 & 0.011 & 0.950 & 17 & 1.647 \\
          Our designed HTML5 canvas (PNG) & 269,874 & 0.420 & 0.021 & 0.922 & 64 & 0.257 \\
          Our designed HTML5 canvas (JPEG) & 205,005 & 0.399 & 0.001 & 0.936 & 64 & 0.262 \\
          HTML5 canvas inspired by AmIUnique (PNG) & 8,948 & 0.353 & 0.002 & 0.986 & 64 & 0.031 \\
          HTML5 canvas inspired by AmIUnique (JPEG) & 6,514 & 0.312 & 0.001 & 0.989 & 64 & 0.039 \\
          HTML5 canvas similar to Morellian (PNG) & 41,845 & 0.385 & 0.034 & 0.947 & 64 & 0.037 \\
          Accept HTTP header & 26 & 0.028 & 0.000 & 0.997 & 3 & 0.000 \\
          Accept-Encoding HTTP header & 30 & 0.019 & 0.002 & 1.000 & 13 & 0.000 \\
          Accept-Language HTTP header & 2,833 & 0.124 & 0.022 & 0.999 & 35 & 0.000 \\
          User-Agent HTTP header & 20,961 & 0.350 & 0.002 & 0.978 & 108 & 0.000 \\
          Accept-Charset HTTP header & 18 & 0.002 & 0.000 & 1.000 & 1 & 0.000 \\
          Cache-Control HTTP header & 47 & 0.055 & 0.023 & 0.706 & 1 & 0.000 \\
          Connection HTTP header & 2 & 0.000 & 0.000 & 1.000 & 5 & 0.000 \\
          TE HTTP header & 2 & 0.000 & 0.000 & 1.000 & 1 & 0.000 \\
          Upgrade-Insecure-Requests HTTP header & 2 & 0.000 & 0.000 & 1.000 & 1 & 0.000 \\
          X-WAP-Profile HTTP header & 4 & 0.000 & 0.000 & 1.000 & 1 & 0.000 \\
          X-Requested-With HTTP header & 151 & 0.004 & 0.000 & 1.000 & 1 & 0.000 \\
          X-ATT-DeviceId HTTP header & 1 & 0.000 & 0.000 & 1.000 & 1 & 0.000 \\
          X-UIDH HTTP header & 1 & 0.000 & 0.000 & 1.000 & 1 & 0.000 \\
          X-Network-Info HTTP header & 4 & 0.000 & 0.000 & 1.000 & 1 & 0.000 \\
          Via HTTP header & 4,272 & 0.007 & 0.003 & 0.999 & 1 & 0.000 \\
          Any remaining HTTP headers & 5,394 & 0.095 & 0.042 & 0.899 & 192 & 0.000 \\
          Number of bounding boxes & 15 & 0.062 & 0.008 & 0.998 & 1 & 0.197 \\
          Number of plugins & 54 & 0.147 & 0.000 & 0.984 & 1 & 0.001 \\
          Number of WebGL extensions & 28 & 0.176 & 0.000 & 0.991 & 2 & 0.008 \\
          Width and height of first bounding box & 12,937 & 0.350 & nan & 0.486 & 30 & 0.197 \\
          Width and height of second bounding box & 1,332 & 0.103 & nan & 0.941 & 1 & 0.197 \\
          Width and height of third bounding box & 772 & 0.076 & nan & 0.965 & 1 & 0.197 \\
          List of widths of bounding boxes & 6,690 & 0.299 & nan & 0.986 & 16 & 0.197 \\
          List of heights of bounding boxes & 2,222 & 0.264 & nan & 0.474 & 14 & 0.197 \\
          Width of first bounding box & 4,418 & 0.281 & 0.038 & 0.987 & 14 & 0.197 \\
          Height of first bounding box & 1,848 & 0.246 & 0.070 & 0.490 & 14 & 0.197 \\
        \bottomrule
      \end{tabular}
    \end{minipage}
  \end{table}
  \begin{table}
    \begin{minipage}{\columnwidth}
      \begin{tabular}{lcccccc}
        \toprule
          \textbf{Attribute} & \textbf{Values} & \textbf{N. Ent.} &
          \textbf{MinNCE} & \textbf{\% Same} & \textbf{Size} & \textbf{Time} \\
        \midrule
          Width of second bounding box & 471 & 0.085 & 0.002 & 0.998 & 1 & 0.197 \\
          Height of second bounding box & 398 & 0.088 & 0.007 & 0.941 & 1 & 0.197 \\
          Width of third bounding box & 224 & 0.060 & 0.004 & 0.999 & 1 & 0.197 \\
          Height of third bounding box & 343 & 0.064 & 0.000 & 0.966 & 1 & 0.197 \\
          Width of a created \texttt{div} & 15,473 & 0.316 & 0.007 & 0.940 & 6 & 0.084 \\
          Position of a created \texttt{div} & 16,375 & 0.316 & 0.008 & 0.942 & 11 & 0.084 \\
          Width of fallback font text & 1,029 & 0.184 & 0.024 & 0.998 & 5 & 0.099 \\
          Height of fallback font text & 1,159 & 0.164 & 0.010 & 0.998 & 5 & 0.099 \\
          Color of ActiveBorder element & 702 & 0.078 & 0.005 & 1.000 & 18 & 0.089 \\
          Color of ActiveCaption element & 475 & 0.073 & 0.002 & 1.000 & 18 & 0.089 \\
          Color of AppWorkspace element & 321 & 0.067 & 0.001 & 1.000 & 18 & 0.089 \\
          Color of Background element & 2,917 & 0.074 & 0.017 & 1.000 & 16 & 0.089 \\
          Color of ButtonFace element & 297 & 0.079 & 0.004 & 1.000 & 18 & 0.089 \\
          Color of ButtonHighlight element & 264 & 0.058 & 0.000 & 1.000 & 18 & 0.089 \\
          Color of ButtonShadow element & 343 & 0.076 & 0.001 & 1.000 & 18 & 0.089 \\
          Color of ButtonText element & 104 & 0.004 & 0.000 & 1.000 & 12 & 0.089 \\
          Color of CaptionText element & 123 & 0.014 & 0.000 & 1.000 & 12 & 0.089 \\
          Color of GrayText element & 333 & 0.071 & 0.004 & 1.000 & 18 & 0.089 \\
          Color of Highlight element & 1,088 & 0.097 & 0.016 & 0.987 & 17 & 0.089 \\
          Color of HighlightText element & 89 & 0.049 & 0.001 & 1.000 & 18 & 0.089 \\
          Color of InactiveBorder element & 334 & 0.060 & 0.000 & 1.000 & 18 & 0.089 \\
          Color of InactiveCaption element & 441 & 0.062 & 0.001 & 1.000 & 18 & 0.089 \\
          Color of InactiveCaptionText element & 265 & 0.088 & 0.006 & 0.999 & 15 & 0.089 \\
          Color of InfoBackground element & 239 & 0.057 & 0.000 & 1.000 & 18 & 0.089 \\
          Color of InfoText element & 96 & 0.003 & 0.000 & 1.000 & 12 & 0.089 \\
          Color of Menu element & 376 & 0.087 & 0.004 & 1.000 & 18 & 0.089 \\
          Color of MenuText element & 124 & 0.020 & 0.001 & 1.000 & 12 & 0.089 \\
          Color of Scrollbar element & 275 & 0.072 & 0.000 & 1.000 & 18 & 0.089 \\
          Color of ThreeDDarkShadow element & 75 & 0.071 & 0.001 & 1.000 & 18 & 0.089 \\
          Color of ThreeDFace element & 297 & 0.062 & 0.000 & 1.000 & 18 & 0.089 \\
          Color of ThreeDHighlight element & 261 & 0.048 & 0.000 & 1.000 & 18 & 0.089 \\
          Color of ThreeDLightShadow element & 280 & 0.074 & 0.001 & 1.000 & 18 & 0.089 \\
          Color of ThreeDShadow element & 339 & 0.075 & 0.000 & 1.000 & 18 & 0.089 \\
          Color of Window element & 329 & 0.019 & 0.001 & 1.000 & 18 & 0.089 \\
          Color of WindowFrame element & 140 & 0.069 & 0.000 & 1.000 & 18 & 0.089 \\
          Color of WindowText element & 107 & 0.004 & 0.000 & 1.000 & 12 & 0.089 \\
        \bottomrule
      \end{tabular}
    \end{minipage}
  \end{table}